\documentclass[aps,prl,reprint,superscriptaddress]{revtex4-2}

\usepackage{upgreek,mathtools}
\usepackage{color}
\usepackage{graphicx, svg}
\usepackage{nccmath}
\usepackage{bm}
\usepackage{hyperref}
\hypersetup{colorlinks,breaklinks,
            urlcolor=[rgb]{0,0,0.36},
            linkcolor=[rgb]{0,0,0.36},
            citecolor=[rgb]{0,0,0.36}}
\usepackage[mathlines]{lineno}

\usepackage{dcolumn}
\usepackage{mathtools}
\usepackage{dsfont, amsfonts, bbold}
\usepackage{comment}
\usepackage{physics}
\usepackage{xcolor}
\usepackage{adjustbox}
\usepackage{tabularray}

\newcommand{\dbar}{d\hspace*{-0.08em}\bar{}\hspace*{0.1em}}

\newcommand{\sign}{\operatorname{sgn}}
\newcommand{\imag}{\operatorname{Im}}

\newcommand{\ex}[1]{\left\langle #1 \right\rangle}

\newcommand{\id}{\mathbb{1}}

\newcommand{\pauli}{\tau}

\newcommand{\OO}{\mathcal{O}}

\newcommand{\neighbor}{\delta}
\newcommand{\neighboraux}{\epsilon}

\newcommand{\mgn}{\hat{\alpha}}

\newcommand{\nambu}{\hat{\Psi}}

\newcommand{\mpr}{\mathbb D}

\newcommand{\hpr}{G}

\newcommand{\msf}{\mathbb A}

\newcommand{\msfsc}{A}

\newcommand{\hsf}{B}

\newcommand{\mse}{\mathbb{\Pi}}

\newcommand{\hse}{\Sigma}

\newcommand{\og}{\text{O}}

\newcommand{\rpa}{\text{RPA}}

\newcommand{\scba}{\text{SCBA}}

\newcommand{\hnn}{\text{NN}}

\newcommand{\qrot}{\mathbf{R}}

\newcommand{\qafm}{\mathbf{Q}_\text{AFM}}

\newcommand{\km}{\mathbf k}
\newcommand{\p}{\mathbf p}
\newcommand{\q}{\mathbf q}
\newcommand{\Qm}{\mathbf Q}

\newcommand{\e}{\varepsilon}

\newcommand{\holepf}{\alpha_h}

\newcommand{\magpfl}{\alpha_m^{(1)}}
\newcommand{\magpfnl}{\alpha_m^{(2)}}

\newcommand{\swscale}{\Omega}

\newcommand{\oco}{r}

\newcommand{\bsv}{\Gamma}

\newcommand{\bsvf}{X}

\newcommand{\rampr}{\zeta}

\newcommand{\mb}{L}

\newcommand{\mmvc}{\mathcal{C}}
\newcommand{\mmvk}{\mathcal{K}}
\newcommand{\mmvq}{\mathcal{Q}}
\newcommand{\mmva}{\mathcal{A}}
\newcommand{\mmvs}{\mathcal{S}}

\newcommand{\lfn}{L}

\newcommand{\ramanbm}{\Lambda}

\newcommand{\hcomb}{\rm{c}}

\newcommand{\ETH}{Institute for Theoretical Physics, ETH Z\"urich, 8093 Z\"urich, Switzerland}
\newcommand{\MPQ}{Max-Planck-Institut f\"{u}r Quantenoptik, 85748 Garching, Germany}
\newcommand{\MCQST}{Munich Center for Quantum Science and Technology, 80799 Munich, Germany}
\newcommand{\LMU}{Fakult\"{a}t f\"{u}r Physik, Ludwig-Maximilians-Universit\"{a}t, 80799 Munich, Germany}

\newcommand{\radu}[1]{\textcolor{blue}{[RA] #1}}

\newcommand{\silent}[1]{}

\newcommand{\fwfigsize}{0.75} 

\begin{document}

\title{Universal magnetic energy scale in the doped Fermi-Hubbard model}

\author{Radu~Andrei}
\affiliation{\ETH}

\author{Ivan~Morera}
\affiliation{\ETH}

\author{Jonathan~B.~Curtis}
\affiliation{\ETH}

\author{Immanuel~Bloch}
\affiliation{\MPQ}
\affiliation{\MCQST}
\affiliation{\LMU}

\author{Eugene~Demler}
\affiliation{\ETH}

\date{\today}

\begin{abstract}
Magnetic correlations of doped Mott insulators hold the key to the unusual characteristics of many quantum materials. Recent experiments with ultracold atoms in optical lattices have provided new information about the magnetic properties of the Fermi-Hubbard model on a square lattice. We demonstrate that recent measurements \silent{in Refs [X], [Y] }indicate that a single doping-dependent energy scale determines both static correlations and dynamical response of these systems. To understand these experimental findings, we employ a self-consistent formalism to describe the coupling between antiferromagnetic magnons and doped holes, and we uncover the emergence of a universal magnetic energy scale at finite doping, which we denote by $J^*$. We present the single- and two-magnon spectral properties at finite doping and discuss the appearance of a bimagnon peak in lattice-modulation spectroscopy, at frequencies set by $J^*$. Furthermore, we argue that this same energy scale sets the onset of pseudogap phenomena, leading to the hypothesis $k_BT^* = c J^*$, with $c$ an order one number. We identify another low-energy scale emerging from our analysis of magnetic excitations, and argue that it controls the stability of N\'{e}el order at the lowest temperatures, ultimately driving a transition to an incommensurate spin-density-wave at finite doping. We discuss the relation between this low-energy scale and the nature of fermionic quasiparticles. Our analysis suggests that stability of the commensurate antiferromagentic phase at finite doping can be controlled experimentally by introducing additional quasiparticle broadening via disorder or low-frequency noise.

\end{abstract}

\maketitle

{\it Introduction}\textemdash In strongly correlated magnetic insulators, the motion of itinerant carriers is intimately coupled to collective magnetic excitations ~\cite{Auerbach1998,ScalapinoReviewHighTc,KeimerReviewHighTc}. This coupling profoundly reshapes many-body correlations and is commonly expected to underlie the enigmatic properties of strongly correlated electron systems, including the pseudogap and strange metal regimes of high-$T_c$ cuprates~\cite{Lee_review_hightc,Tom_review_pseudogap,Chubukov.1997,Lee.1992,Wen.1996}, heavy fermion systems~\cite{Gegenwart_review_heavy_fermion,Qimiao_review_heavy_fermion}, layered organic materials~\cite{Powell_review_organic}, and moir\'e systems~\cite{Young_review_moire,MacDonald_review_moire}. Whether these phenomena are material-specific or represent a universal outcome of doping a strongly correlated insulator remains a topic of intense debate.

The Fermi-Hubbard model provides a minimal framework for describing doped magnetic insulators. While the dynamics of itinerant carriers in this model have been extensively studied, leading to the conclusion that strongly renormalized magnetic polarons emerge at low doping~\cite{Bulaevskii1968,Brinkman1970,Trugman1988,Schmitt1988,Shraiman1988,Sachdev1989,Kane1989,Dagotto1989,Trugman1990,Auerbach1991,Martinez.1991,Manousakis1991,Horsch1991,Khaliullin.1993,Sherman.1994,Vojta.1996,Grusdt2018,Bohrdt2020}, the fate of magnetic correlations in the presence of doping remains unclear. The development of controllable and tunable quantum simulators~\cite{Bloch2012_review, Blatt2012_review, Browaeys2020_review}, such as ultracold atoms in optical lattices, has opened new avenues for tackling this question. In particular, single-site resolution has enabled direct imaging of the spatial structure of magnetic polarons~\cite{Koepsell2019,OL_Prichard2024_Nagaoka,OL_Lebrat2024_Nagaoka} and the evolution of many-body correlations upon doping~\cite{Koepsell_doping_2021,Gull.2013,Werner.2007lvq,Kyung.2006}. Combined with recent breakthroughs enabling experiments to reach cryogenic temperatures~\cite{Xu_cryogenic_2025}, these advances provide a fresh perspective on the emergence of unconventional phases upon doping. Notably, two recent studies~\cite{chalopin_observation_2026,kendrick2025pseudogapfermihubbardquantumsimulator} have investigated the onset of the pseudogap state and its connection to spin correlations~\cite{Millis.1990}, highlighting that a detailed understanding of magnetism in doped systems is essential for elucidating the nature of the pseudogap.

In this work, we identify the emergence of a universal magnetic energy scale in the doped Fermi–Hubbard model. Denoted by $J^*$, this scale decreases linearly with doping, and arises from the competition between antiferromagnetic superexchange and frustration induced by dopant motion. Additionally, it controls both the thermodynamic and dynamical magnetic properties of the system, for all but the lowest frequencies. First, we show that $J^*$ controls the equilibrium long-distance decay of spin–spin correlation functions, recently explored in~\cite{chalopin_observation_2026}. Second, we demonstrate that the same energy scale dictates the single- and two-magnon spectral functions, giving rise to a universal dynamical response in lattice-modulation spectroscopy experiments~\cite{kendrick2025pseudogapfermihubbardquantumsimulator}. Within our formalism, we thus establish that these two seemingly disparate probes—one accessing equilibrium properties and the other dynamical responses—are governed by the same underlying energy scale $J^*$. 
Moreover, we address the fate of long-range antiferromagnetic N{\'e}el order at ultra-low temperatures. Our framework reveals a separation of energy scales, whereby a distinct, low-energy $J_\rho$ controls the stability of magnetic order at finite doping. Unlike $J^*$, this latter scale depends sensitively on the quasiparticle nature of the dopants, with their lifetime determining the critical doping at which N{\'e}el order melts. Finally, we argue that $J^*$ sets a natural frequency cutoff for optical Raman-type probes: below this scale, contributions from charge degrees of freedom are suppressed, since hole motion inevitably disturbs the antiferromagnetic background through which it propagates. This leads to the hypothesis $k_BT^* = c J^*$, and strongly suggests that short-range magnetic correlations play a crucial role in shaping the experimental signatures of the pseudogap regime.

\section*{Universal Scaling}
In the absence of doping, and in the strong-coupling regime $U/t\gg 1$, the magnetic correlations and low-energy dynamics of the Hubbard model are essentially described by the superexchange energy $J = 4t^2/U$.
On the one hand, the spin-stiffness $\rho \propto J$ governs the thermodynamic properties of the system; for example, at any finite temperature $T$, the spin correlations exhibit an exponential decay at long distances, characterized by the correlation length $\xi(T) \propto e^{2\pi \rho/T}$~\cite{Chakravarty.1989}.
On the other hand, high-energy magnetic excitations also reside at frequencies set by $J$. These modes in turn govern the short-range magnetic correlations, which can be accessed by probes such as Raman \cite{Blumberg1996,Devereaux.2007} or inelastic neutron~\cite{Coldea.2001,Headings.2010,Yamada.1989,Hayden.1991} scattering. The former typically shows a bimagnon resonant absorption feature at frequency $\omega_{\rm 2M}(0) \approx 3.3J/\hbar$ \cite{Canali.1992}.

For an undoped system, it is indeed expected that all of these properties should be controlled by $J$, as it is the only energy scale in the problem. However, upon doping the Hubbard model, an {\it a priori} new scale $t$ enters the low-energy description. In this work, we show that $J$ and $t$ combine to yield an universal doping-dependent scale $J^*(\delta)$, which governs the thermodynamic as well as dynamical magnetic properties in the underdoped regime:
\begin{equation}
    \frac{J^*(\delta)}{J_0}=1 - a \; \frac{U\delta}{t},
\label{Eq:Universal_Scaling}
\end{equation}
with $a$ an $\mathcal{O}(1)$ prefactor \footnote{The slope $a$ in Eq. \eqref{Eq:Universal_Scaling} does have a weak dependence on the Hamiltonian parameters $t$ and $U$, entering at orders $t/U$ and higher; in the strong-coupling regime $t/U \ll 1$ this is however unimportant.}, and $J_0 \equiv (1 + r_0) J$ the bare superexchange together with the so-called Oguchi correction $r_0\sim 0.157$, which encodes the effect of quantum zero-point fluctuations on the single-magnon excitation spectrum in antiferromagnets \cite{Oguchi.1960}. For clarity, the principal magnetic energy scales discussed in this work are summarized in Table~\ref{tab:j_definitions_sm} of the SM.

\begin{figure}
    \centering
    \includegraphics[width=0.48\textwidth]{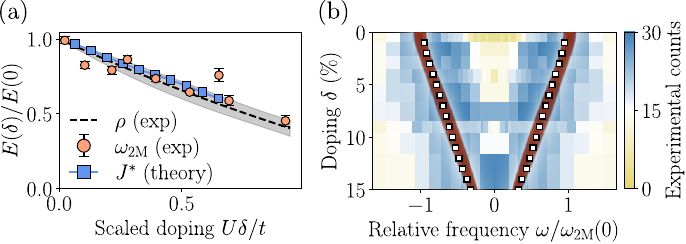}
    \caption{Universal renormalization of magnetic energy scales as a function of doping. (a)~Comparison between equilibrium measurements \cite{chalopin_observation_2026} of the finite-temperature spin stiffness $\rho$ (dashed line, error bounds shaded in grey) at $U/t = 6.5$; dynamical probing \cite{kendrick2025pseudogapfermihubbardquantumsimulator} of high-energy magnon dispersion via the bimagnon peak frequency $\omega_{\rm 2M}$ ($U/t = 7$, circles); and theoretical calculation of $J^*$ ($U/t = 6.5$, squares). Scaling the doping axis by $U/t$ leads to collapse of data, cf. Eq.~\eqref{Eq:Universal_Scaling}. (b)~Experimental Raman-type bimagnon spectra (blue-yellow colorplot, see Fig.~\ref{fig:assembled_3}a for measurement protocol), at $U/t = 7$, as a function of doping $\delta$ and frequency $\omega / \omega_{\rm 2M}(0)$, compared to theoretical calculations of $J^*(\delta) / J_0$ (white squares) and $T=0$ bimagnon Raman spectrum (dark red color; width in the frequency direction represents peak FWHM). Experimental data reproduced with permission, based on Fig.~3f of Ref.~\cite{chalopin_observation_2026}, and Fig.~3b of Ref.~\cite{kendrick2025pseudogapfermihubbardquantumsimulator}; see SM for details.}
    \label{fig:assembled_1}
\end{figure}

To demonstrate the emergence of this universal scale, we consider data from two recent quantum simulation experiments, probing different aspects of the doped Fermi-Hubbard model, see  Fig.~\ref{fig:assembled_1}(a). The first one extracts the spin-stiffness $\rho$ from the temperature dependence of the correlation length $\xi(T)$, and shows that $\rho$ follows a linear dependence on doping~\cite{chalopin_observation_2026}. In contrast, the second experiment uses lattice modulation spectroscopy to probe the dynamics of the Fermi-Hubbard model~\cite{kendrick2025pseudogapfermihubbardquantumsimulator}. Interestingly, it reveals the presence of a resonant absorption peak which closely mimics the bimagnon signature in the undoped, Heisenberg case, but at a renormalized frequency $\omega_{\rm 2M}(\delta)$. This is emphasized further in Fig.~\ref{fig:assembled_1}(b), where we overlay the experimental absorption spectrum from Ref.~\cite{kendrick2025pseudogapfermihubbardquantumsimulator} with the theoretical predictions for the bimagnon peak position and spread derived in this manuscript. We remark that the redshift of the bimagnon peak upon doping is in close analogy to $B_{1g}$ Raman scattering in underdoped cuprates \cite{sugai_bimagnon_2003}. 

In order to meaningfully compare doping-induced modifications of the spin-stiffness $\rho$ and bimagnon peak frequency $\omega_{\rm 2M}$ -- two energy scales which are vastly different in the absence of doping -- in Fig.~\ref{fig:assembled_1}(a) we divide out the zero-doping values, and plot $\rho(\delta) / \rho(0)$ as well as $\omega_{\rm 2M}(\delta) / \omega_{\rm 2M}(0)$. Despite probing dramatically different frequency regimes, using different experimental protocols, and even different values of $U/t$, we see that both data sets are excellently described by the same renormalized superexchange interaction $J^*(\delta)$, which depends linearly on the scaled doping $\delta_{\rm eff} = U/t \times \delta$, cf. Eq.~\eqref{Eq:Universal_Scaling}.

We will now show that these phenomenological observations can be explained using a single microscopic treatment, describing the self-consistent interactions between doped holes and magnons in the Fermi-Hubbard model. This allows us to predict without any free parameters the scaling of $J^*(\delta)$ in terms of $\delta_{\rm eff}$, finding excellent agreement with the two different experimental observations. Furthermore, we use this theoretical analysis to make additional predictions for the evolution of magnetic properties of the Fermi-Hubbard model with doping, and discuss how to relate our findings to the general picture of the pseudogap regime.

\section*{Theoretical Framework}

\begin{figure}
    \centering
    \includegraphics[width=0.48\textwidth]{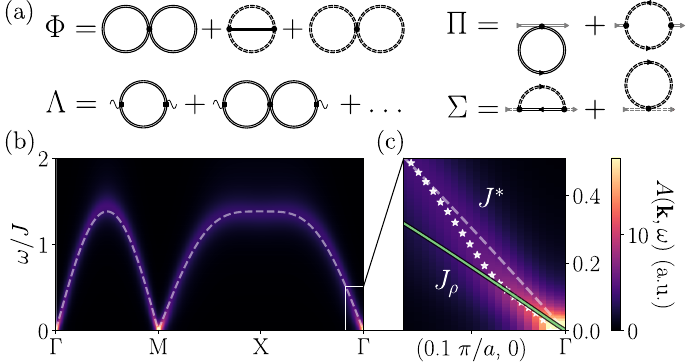}
    \caption{Magnon spectrum renormalization due to dopants. (a)~Luttinger-Ward functional $\Phi[\mpr, \hpr]$, magnon and holon self-energies $\mse$ and $\hse$ respectively, and bimagnon Raman response $\Lambda$. (b)~Typical single-magnon spectral function at $U/t = 8$ and finite doping $\delta = 8.5\%$, together with the renormalized Heisenberg dispersion $\omega_{\mathbf k} = 2 J^* \epsilon_{\mathbf k}$ (dashed line). Note the excellent agreement at all but the lowest energies, as well as the significant spectral broadening at the top of the magnon band. (c)~Detail of dispersion around the gapless $\Gamma$ point, where deviations from $J^*$ (dashed white line) are described by a distinct energy scale $J_\rho$ (solid green line). For clarity, white stars mark the location of the spectral function peak at each momentum.}
    \label{fig:assembled_2}
\end{figure}

\begin{figure}
    \centering
    \includegraphics[width=0.48\textwidth]{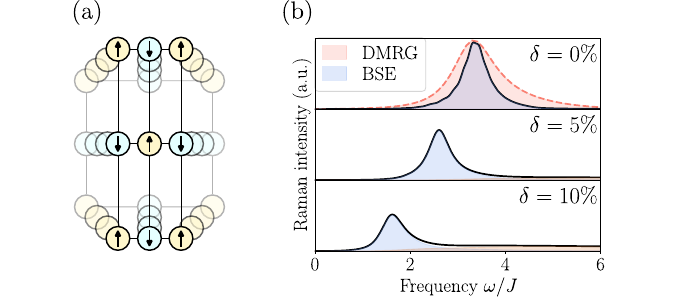}
    \caption{Quantum simulation of Raman scattering. (a)~Lattice modulation protocol for $d-$wave spectroscopy experiments~\cite{kendrick2025pseudogapfermihubbardquantumsimulator}. With respect to the standard square configuration, hosting an isotropic hopping amplitude $t$, the lattice is compressed in the $x$ direction and stretched along $y$; this yields a $d$-wave modulation of $t$, analogous to optical Raman experiments in the $B_{1g}$ channel. (b)~Theoretical Raman intensity of bimagnon (blue) and estimated charge background (brown) contributions, at three different dopings. At half-filling, the peak position is in excellent agreement with DMRG+TDVP calculations (pink). See SM for computation details.}
    \label{fig:assembled_3}
\end{figure}

We consider the Fermi-Hubbard model on a 2D square lattice, given by the Hamiltonian 
\begin{align}
    \hat{H} &= - t \sum_{\langle j,k \rangle, \sigma} \left( \hat{c}^{\dagger}_{j\sigma} \hat{c}_{k\sigma} + \rm{h.c.} \right) + U \sum_{j} \hat{n}_{j\uparrow}\hat{n}_{j\downarrow} ,
\end{align}
where $\hat{c}_{j\sigma}$ and $\hat{n}_{j\sigma}$ are the electron annihilation and number operators on site $j$, with spin $\sigma \in \{\uparrow,\downarrow\}$; $\langle j, k\rangle$ denotes summation over nearest neighbors $j$ and $k$; the hopping amplitude is $t$, and $U$ represents the on-site repulsion. We focus on the strongly correlated regime $U/t\gg 1$, where the low-energy behavior of the Fermi–Hubbard model can be mapped onto the extended $t-J$ model, including the so-called three-site terms~\cite{Troyer1995}. We employ a zero-temperature, real-time diagrammatic formalism combined with a Luttinger-Ward functional approach (see Fig.~\ref{fig:assembled_2}a), to develop a self-consistent description of spin waves coupled to the motion of doped holes. Within this framework, we solve the Dyson equation to obtain the single-particle properties, as well as the Bethe–Salpeter equation (BSE) to compute the two-magnon response probed in Raman-type and lattice-modulation spectroscopy experiments. For details of the mapping from the $t-J$ model to the diagrammatic formalism, including the auxiliary-particle representation and the unitary transformation to the staggered rotation frame, we refer the reader to the Supplementary Material. Additionally, to verify the applicability of our diagrammatic technique, we perform large-scale unbiased tensor network simulations for the zero-doping case ($S=1/2$ Heisenberg model). In particular, we study the dynamical bimagnon response to lattice modulations using a density-matrix renormalization group (DMRG) approach, combined with the time-dependent variational principle (TDVP). The excellent agreement between the two methods confirms the validity of our diagrammatic approach, see Fig.~\ref{fig:assembled_3}(b).

\section*{Kinetic reduction of spin exchange}

The renormalization of the magnetic energy scale in Eq.~\eqref{Eq:Universal_Scaling} fundamentally arises from the competition between antiferromagnetic superexchange at half-filling and the spin reshuffling caused by hole motion, in accordance with the Nagaoka–Thouless mechanism \cite{Nagaoka1966,Thouless1965,NagyLovas2017}. Interestingly, such a scaling was previously proposed in the context of kinetic magnetism \cite{Haerter-Shastry,Morera_High_temperature} at large $U/t$, and has been confirmed for the triangular lattice \cite{TMD_Ciorciaro2023,OL_Lebrat2024_Nagaoka,OL_Prichard2024_Nagaoka}. More recently, similar renormalizations have been observed in the dynamical spin structure factor near full polarization at intermediate $U/t$ \cite{Prichard_MagnonPolaron}, as well as in ultrafast dynamics following photodoping of an insulator \cite{Radovskaia2026}. Our work demonstrates that this renormalization remains relevant all the way down to the regime $U/t\sim 7$, where it universally governs the finite-doping magnetic properties.

It is not obvious \emph{a priori} that the magnon dispersion, in the presence of doped holes, should closely resemble that of the Heisenberg model at half filling -- and, in particular, that it would be described by a single energy scale $J^*$. For example, while the delocalization of charge carriers throughout the lattice generally suppresses long-range antiferromagnetic order, it is established \silent{(\radu{citations?})}that robust short-range correlations persist even at finite doping. One might therefore expect more pronounced modifications of the low-energy magnon spectrum, compared to the high-energy part. To investigate the shape of single-magnon spectra in the presence of dopants, we employ our self-consistent approach to calculate the transverse dynamical spin structure factor,
\begin{equation}
    A(\mathbf{k}, \omega) = \textrm{Re}\int dt \;  \Theta(t) \; \left\langle \hat{S}^+_{\mathbf{k}}(t) \hat{S}^-_{\mathbf{k}}(0) \right\rangle \; e^{i\omega t},
\end{equation}
which is directly accessible in ultracold atom experiments, by using a two-photon Raman-type process \cite{Prichard_MagnonPolaron}. In magnetically ordered states, $A(\mathbf{k}, \omega)$ is primarily given by the single-magnon spectral function, thus providing a direct probe for the dispersion of these excitations. 

Figure~\ref{fig:assembled_2}(b) depicts a typical finite-doping result for the one-magnon spectral function, along a 1D cut through the Brillouin zone. As emphasized in panel~(c), a narrow region surrounding the gapless $\Gamma$ and M points indeed exhibits a distinct behavior compared to the rest of the spectrum, which we will later analyze in detail. Remarkably, however, the majority of the magnon dispersion is well described by a single renormalized exchange scale $J^*$, as highlighted in panel~(b). In order to understand the origin of this apparently simple description, we must examine the spectral properties of doped holes.

The dynamics of dopants in a quantum antiferromagnet have been the subject of extensive theoretical and numerical investigations~\cite{Bulaevskii1968,Brinkman1970,Trugman1988,Schmitt1988,Shraiman1988,Sachdev1989,Kane1989,Dagotto1989,Trugman1990,Auerbach1991,Martinez.1991,Manousakis1991,Horsch1991,Khaliullin.1993,Sherman.1994,Vojta.1996,Grusdt2018,Bohrdt2020,Chubukov.1997,Werner.2007lvq,Kyung.2006,Lee.1992,Wen.1996,Millis.1990,Gull.2013}. A key outcome of these studies — and of our approach (see Fig.~\ref{fig:sm_spectral_functions} in the SM for representative spectral functions) — is that spin excitations dress a doped hole to form a magnetic polaron, whose dispersion exhibits minima at momenta $(\pm \pi/2, \pm \pi/2)$. Away from these points, the hole spectral function broadens substantially, reflecting processes in which the polaron decays to the aforementioned minima via emission of magnons. At even higher energies, broad `string-like' excitations appear across the entire Brillouin zone. The key takeaway is that, aside from the bottom of the polaron band, the hole spectral function is predominantly incoherent. This observation is essential for understanding the magnon spectrum renormalization, as it motivates the consideration of a simplified model based on purely incoherent holes, remarkably reproducing the qualitative scaling of Eq.~\eqref{Eq:Universal_Scaling}. Here, we highlight the underlying physical mechanism, while full calculation details can be found in the SM.

Within our self-consistent formalism, doped holes modify the magnon spectrum through an RPA self-energy correction, taking the form of a polarization bubble $i\int_{\mathbf q, \epsilon} G_{\mathbf q}(\epsilon) \; G_{\mathbf k + \mathbf q}(\omega + \epsilon)$, see Fig.~\ref{fig:assembled_2}(a). For external momenta $\mathbf k$ and frequencies $\omega$ that are not too small, the two hole spectral functions entering the bubble are sufficiently shifted in energy such that their sharp, coherent features do not overlap, since the polaron bandwidth is given by $J$ and thus narrow; on the other hand, the incoherent parts of the hole spectral function extend over a broad range of frequencies ($\sim 8t$), yielding a dominant contribution to the polarization bubble which is only weakly dependent on $\mathbf{k}$. To leading order in doping $\delta$, the bubble therefore scales as $g(\omega) \times \delta / t$, with $g(\omega)$ a dimensionless function of order unity. Substituting this result into the magnon propagator shows that the net effect of the holes is to induce a renormalized exchange interaction,
\begin{equation} \label{eq:anderson_toy_model_main}
    J^*(\omega) = J_{\delta} - 2 g(\omega) \times t \delta .
\end{equation}
Here, $J_{\delta} \equiv (1 - \delta)^2 [1 + r(\delta)] J$ encompasses both a trivial reduction $(1-\delta)^2$ of the local magnetic moments upon doping, as well as the the finite-doping Oguchi correction $r(\delta)$. Meanwhile, the second term captures the kinetic renormalization of $J^*$ induced by the motion of the doped holes. We emphasize the close similarity of \eqref{eq:anderson_toy_model_main} to the typical Anderson-Shastry expression encountered in treatments of kinetic magnetism. For $\omega \ll t$, one may replace $g(\omega)$ by the positive constant $g(0)$; the approximation is reasonable even in the middle of the magnon spectrum. This immediately leads to a single effective exchange $J^*$, which obeys the universal scaling, Eq.~\eqref{Eq:Universal_Scaling} \footnote{Ignoring the difference between $J_\delta$ and $J_0$, as well as the $U/t$ dependence of $g(0)$, yields Eq. \eqref{Eq:Universal_Scaling} with a constant slope $a = g(0) / 2(1+r_0)$. Taking these two effects into account will introduce corrections to the slope $a$ at order $t/U$ and above, as mentioned before.}. On the other hand, for $\omega \sim t$, relevant for the top of the magnon band, the main effect is that $g(\omega)$ acquires a substantial imaginary part, which leads to the considerable broadening visible in Fig.~\ref{fig:assembled_2}(b). While these arguments successfully capture the qualitative structure of the magnon spectra, they do not provide a controlled way to compute the value of $g(0)$, or equivalently the prefactor $a$ in Eq.~\eqref{Eq:Universal_Scaling}. Quantitative predictions still require the full self-consistent solution of the coupled hole–magnon problem, which we achieve numerically.

\section*{Two-magnon Raman scattering}

Having analyzed the behavior of single magnons in doped systems, we turn our attention to the two-magnon Raman scattering process~\cite{Devereaux.2007}, as well as its quantum simulation in ultracold atom systems. By modulating the hopping amplitude $t$ with a $d$-wave spatial profile (Fig.~\ref{fig:assembled_3}a), the spin degrees of freedom are perturbed with the Loudon-Fleury operator \cite{Fleury.1968},
\begin{equation} \label{eq:lf_vertex_main}
    \hat{H}_\text{LF} \propto \sum_{j} \vec S_{j} \cdot (\vec S_{j + \hat y} + \vec S_{j - \hat y} - \vec S_{j + \hat x} - \vec S_{j - \hat x}).
\end{equation}
In the solid state, this modulation is typically realized by applying an external electric field in the context of Raman scattering, while in optical lattices the hopping amplitude can be directly controlled by adjusting the potential barrier between neighboring sites~\cite{kendrick2025pseudogapfermihubbardquantumsimulator}. In both cases, the applied perturbation \eqref{eq:lf_vertex_main} creates local pairs of spin flips~\cite{Fleury.1968,Devereaux.2007}, which makes bimagnon Raman scattering an invaluable tool for probing the dynamics of short-range magnetic correlations. Owing to the local nature of these excitation pairs, the interaction between magnons also plays a crucial role in determining the Raman spectrum. In undoped systems, the magnetic Raman signal can be understood as follows~\cite{Canali.1992}: an applied perturbation at zero total momentum will create $(\mathbf k, -\mathbf k)$ magnon pairs. Since the single-magnon density of states is peaked at the top of the band (energy $2J_0$), one may expect a sharp Raman peak at $4 J_0$; in fact, the attractive magnon-magnon interaction broadens the resonance, and redshifts it to $2.9 J_0 \simeq 3.3 J$. 

Using a ladder approximation to the two-magnon BSE (Fig. \ref{fig:assembled_2}a, see SM for details), we compute the spin contribution to Raman scattering intensity at arbitrary doping; typical curves are shown in Fig.~\ref{fig:assembled_3}(b). At zero doping, we benchmark the BSE result against DMRG+TDVP simulations and find excellent agreement in the peak position. Upon increasing $\delta$, the bimagnon peak redshifts and loses spectral weight, the latter effect also being a consequence of the trivial spin depletion $(1-\delta)^2$ introduced earlier. Finally, as highlighted in Fig.~\ref{fig:assembled_1}(b), the relative redshift of the bimagnon peak closely follows that of $J^*$, indicating that this dynamical observable is governed by the same universal energy scale that controls equilibrium spin probes.

\section*{Long-wavelength magnons}

While the effect of dopants on high-energy magnons is dominated by the incoherent part of the hole spectral function, and is therefore largely insensitive to precise details of the magnetic polaron dispersion and lifetime, the situation is markedly different for low-energy magnon modes. Expanding the magnon self-energy $\mathbb{\Pi}_{\mathbf k}(\omega)$ at low $(\mathbf{k}, \omega)$ yields an expression for the spin stiffness that differs qualitatively from Eq. \eqref{eq:anderson_toy_model_main},
\begin{equation} \label{eq:j_rho_main}
    J_{\rho} = \sqrt{[J_\delta - t A(\delta)][J_\delta - t B(\delta)]},
\end{equation}
where $J_\delta$ is the same as in \eqref{eq:anderson_toy_model_main}, and $A, B > 0$ are obtained from integrating the diagrammatic bubble $i\int_{ \epsilon} G_{\mathbf q}^2(\epsilon)$ over momentum, with certain $\mathbf q-$dependent prefactors (see SM). In practice, we find that $A \ll B$, implying that the long-wavelength spin stiffness -- and, consequently, the stability of antiferromagnetic order at $T=0$ -- is controlled by the competition between $J_\delta$ and the kinetic term $t B(\delta)$. Crucially, because the hole bubble $i\int_{ \epsilon} G_{\mathbf q}^2(\epsilon)$ involves no relative shifts in frequency or momentum, the coefficient $B(\delta)$ is now directly sensitive to sharp features in the magnetic polaron dispersion.

If the hole spectral function displays quasiparticle peaks at the dispersion minima, it can be shown that $B(\delta)$ is proportional to the density of states at the Fermi level. Then, the AFM stability condition $J_\delta - t B(\delta) > 0$ translates to a Stoner-type criterion~\cite{Shraiman.1989}, which is easily violated owing to the large effective mass of the magnetic polaron. Such a line of reasoning leads to the conclusion that, at any infinitesimal doping, AFM order is unstable~\cite{Auerbach.1991,Shraiman.1989} towards e.g. spin-spiral states. On the other hand, a number of theoretical and numerical studies have found a finite critical doping $\delta_{\rm AFM}>0$~\cite{Vojta.1996,Khaliullin.1993,Sherman.1994}, which is also supported by experimental investigations~\cite{Yamada.1998}; the fate of long-range order in doped antiferromagnets has remained a controversial question.

In order to clarify some of these disagreements, as well as to reflect experimental reality, we additionally introduce a polaron broadening $\eta_F>0$ which makes its lifetime finite even at the bottom of the band. With this modification, the characteristic low-doping behavior $B(\delta) \propto \delta / \sqrt{\eta_F}$ emerges, which in turn stabilizes AFM order up to $\delta_{\rm AFM} \propto \sqrt{\eta_F}$. Moreover, for dopings slightly below this critical value, we find $J_\rho \propto \sqrt{\delta_{\rm AFM} - \delta}$. We also remark that, in our approach with $U/t = 8$, very small broadenings $\eta_F / t \simeq 0.05$ are sufficient to yield the typical values $\delta_{\rm AFM} \simeq 4-5\%$. Regardless of whether \cite{Bohrdt2020, monte_carlo_single_hole_z, mischchenko_delta_function_peak} or not \cite{phase_string_1, phase_string_2} spectral function $\delta-$peaks with finite weight $Z>0$ arise from the idealized Fermi-Hubbard or $t-J$ models, nonzero broadening is expected in any experimental setting, making $\eta_F$ an important control parameter to consider.

\begin{figure}
    \centering
    \includegraphics[width=0.48\textwidth]{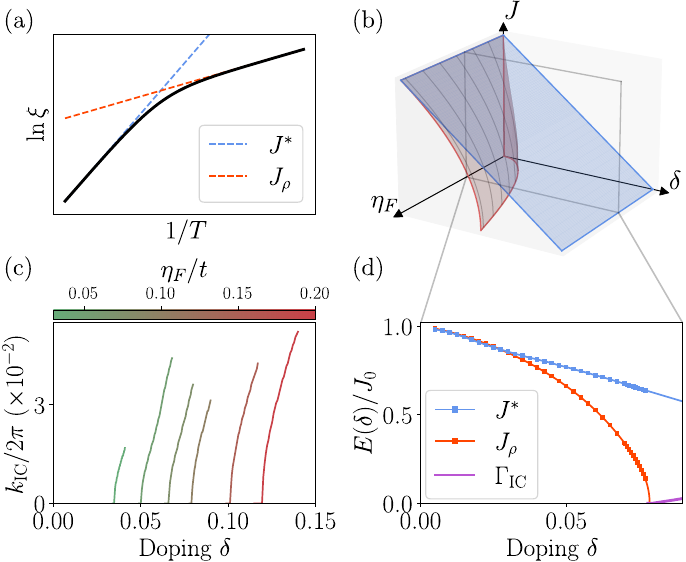}
    \caption{Implications of a second magnetic energy scale $J_\rho$, and instability towards incommensurate magnetic order. (a)~Schematic behavior of spin-spin correlation length versus inverse temperature, in the weakly-doped 2D Fermi-Hubbard model. At low enough temperatures, a crossover between $J^*$ and $J_\rho$ behaviors is expected. (b)~Doping dependence of $J^*$ (blue) and $J_\rho$ (orange), at different values of the inverse polaronic lifetime $\eta_F$. While $J^*$ is insensitive to $\eta_F$ and depends linearly on doping, the stiffness $J_\rho$ shows the characteristic $\sqrt{\delta_{\rm AFM} - \delta}$ shape, with $\delta_{\rm AFM}$ sensitive to $\eta_F$. The striking similarity of $\eta_F > 0$ cuts with the underdoped region of cuprate phase diagrams leads to the hypothesis that $T_N \propto J_\rho$ while $T^* \propto J^*$. (c)~Estimated doping dependence of incommensuration $k_{\rm IC} \equiv |(\pi, \pi) - \tilde{\mathbf Q}|$, for dopings $\delta > \delta_{\rm AFM}$ beyond AFM instability, at six different values of $\eta_F$. (d)~Doping dependence of magnetic energy scales $J^*$ and $J_\rho$, illustrated at $\eta_F/t = 0.1$. Squares denote values extracted from self-consistent numerical solutions, lines are extrapolations based on Eqs.~\eqref{Eq:Universal_Scaling} and \eqref{eq:j_rho_main}. Purple line represents the rate $\Gamma_{\rm IC}$ of instability towards the incommensurate order.} 
    \label{fig:assembled_4}
\end{figure}

The consequences of having a distinct scale $J_\rho < J^*$ for low-energy magnons are presented in the top part of Fig.~\ref{fig:assembled_4}. Panel~(a) schematically depicts the expected temperature dependence of the spin-spin correlation length $\xi$, which in the 2D Heisenberg model obeys $\ln  \xi \propto J/T$~\cite{Chakravarty.1989}. For the relatively high temperatures $T \sim J$ currently accessible in quantum simulators, we have seen that $J^*$ is the only relevant energy scale; however, upon further cooling which should be accessible in the near future, we expect that the slope will tend to a lower value set by $J_\rho$. For example, with $U/t = 8$ and $\eta_F/t \simeq 0.15 - 0.2$, we expect that for doping levels $\delta \simeq 8-10\%$, the ratio $J_\rho / J^* \simeq 0.65$ will be small enough to make the two energy scales experimentally distinguishable on temperature scales $T/t \simeq 0.1$ and below.\silent{Source: TS 1770890293 R8.} As shown in Fig.~\ref{fig:assembled_4}(b), further control over the relation between $J^*$ and $J_\rho$ is offered by the broadening $\eta_F$; by tuning the amount of static disorder present in ultracold atom arrays (e.g. from an optical speckle pattern), or the low-frequency noise, one could adjust the value of $J_\rho$ and even the stability of long-range AFM order. On the other hand, we remark that the high-temperature magnetic properties, governed by $J^*$, are completely insensitive to $\eta_F$.

\section*{Instability of the AFM phase towards incommensurate order}

For doping values beyond the stability threshold of AFM order, $\delta > \delta_{\rm AFM}$, we find no self-consistent solution for the hole and magnon propagators. 
This indicates an instability of the AFM state at wavevector $\mathbf{Q} = (\pi, \pi)$, signaling a doping-induced transition to an incommensurate phase characterized by a wavevector $\tilde{\mathbf{Q}}$. The goal of this section is to estimate, through the study of this instability, the doping dependence of $\tilde{\mathbf Q}$, as well as the typical energy (and thus, temperature) scale at which signatures of incommensurate magnetic order should be observable.  

To understand the commensurate-to-incommensurate (C-IC) magnetic transition qualitatively, we can extend the low-momentum expansion of $\mathbb{\Pi}_{\mathbf k}(\omega)$ of the previous section to fourth order, yielding
\begin{equation}
\omega_k^2 = -\mathcal{A}\left(\delta - \delta_{\mathrm{AFM}}\right) k^2 + \mathcal{K} \, k^4,
\label{Eq:GL_IC}
\end{equation}
where $\mathcal{A}$ and $\mathcal{K}$ are smooth, positive functions of doping. For dopings above the transition point, $\delta > \delta_{\mathrm{AFM}}$, the magnon spectrum exhibits an instability at small momenta; however, the quartic term stabilizes the dispersion at larger $k$. This results in a characteristic wavevector $k_{\mathrm{IC}} \propto \sqrt{\delta - \delta_{\mathrm{AFM}}}$ at which the instability rate is maximal, $\Gamma_{\rm IC} \equiv \textrm{Im}(\omega_{k_{\rm IC}})\propto \delta-\delta_{\rm AFM}$. As the strongest instability dictates the new state towards which the system tends, we identify $k_{\rm IC}$ as the difference between the new ordering wavevector $\tilde{\mathbf{Q}}$ and the N\'eel one $\mathbf Q$. We explicitly verified that the instability is always strongest for momenta parallel to a crystal axis, i.e. we obtain $\tilde{\mathbf{Q}} = (\pi \pm k_{\rm IC}, \pi)$ and respectively $(\pi, \pi \pm k_{\rm IC})$. Furthermore, the instability rate is proportional to the energy density difference between the unstable and stable phases \cite{pekker_babadi_instabilities, Yuzbashyan_instability, Barankov_instability, abrikosov2012methods}, meaning that $\Gamma_{\rm IC}$ is expected to set the temperature scale at which signatures of incommensurate order become experimentally visible.

While the qualitative conclusions given by Eq.~\eqref{Eq:GL_IC} are correct, we go beyond this approximation by expanding the magnon self-energy $\Pi_{\mathbf{k}}(\omega)$ to second order in frequency, and determining the magnon dispersion numerically to all orders in $\km$ (see SM for details). In Fig.~\ref{fig:assembled_4}(c) and (d), we present the incommensurate wavevector $k_{\mathrm{IC}}$ and the instability rate $\Gamma_{\mathrm{IC}}$ as functions of doping, respectively. The results of the full numerical solution are in qualitative agreement with the low-momentum expansion, supporting its validity. Additionally, we have verified that by taking the clean limit $\eta_F\rightarrow 0$ the (C-IC) transition point reaches zero doping, in agreement with previous zero-temperature calculations~\cite{WhiteScalapinoStripes1998,WhiteScalapino2003,Hager2005,ShiweiStripes2010,Bo-XiaoStripes2017,WhiteHybridStripes2017,HuangStripes2018,Thom4leg2020,Wietek.2021,ShiweiAbsence2020}. While our analysis focuses on the incommensurate magnetic properties above $\delta_{\rm AFM}$, the accompanying charge distribution and possible stripe formation are left as a topic for future studies.

\section*{Implications for magnetic response in the pseudogap regime}

One puzzling feature of the pseudogap state is that its signatures are present in both charge and magnetic responses, which raises numerous questions about the underlying mechanism. Here, we argue that the magnetic scale $J^*$ can govern responses to both of these seemingly disparate probes. On the one hand, the uniform spin susceptibility $\chi(T)$ experimentally displays a maximum at the pseudogap temperature $T^*(\delta)$. In the undoped case, the spin susceptibility of the 2D Heisenberg model is theoretically known to have a maximum, at a temperature given by $k_B T^*_0 = c J_0$, with $c \sim \mathcal{O}(1)$ \cite{miyashita_thermodynamic_1988,gomez-santos_monte_1989,makivic_two-dimensional_1991,kim_low_1998,okabe_quantum_1988}. Since we have argued that most of the magnon spectrum in the doped case is practically a scaled-down version of the Heisenberg one, we expect that the same maximum will be seen at finite doping, namely when $k_B T^*(\delta) = c J^*(\delta)$. This hypothesis is strongly supported by the linear dependence of $J^*$ with $\delta$, Eq.~\eqref{Eq:Universal_Scaling}, which for $U/t = 8$ intercepts the doping axis at $\delta^* = t/aU \sim 20\%$, in good agreement with typical cuprate values \cite{pseudogap_temperature_doping_dependence}. Moreover, for layered materials, any interlayer exchange coupling $J_\perp \ll J_\rho$ will stabilize N\'{e}el order up to some finite temperature proportional to the in-plane spin stiffness, $T_N \propto J_\rho$. Comparing $\eta_F>0$ cuts of Fig.~\ref{fig:assembled_4}(b) with the underdoped region of typical cuprate phase diagrams, we find excellent qualitative agreement.

On the other hand, with regard to charge response, the typical signature of the pseudogap regime is spectral weight suppression at low frequency. In our approach, the kinetic term of the Hubbard model gets mapped at leading order to the hopping of a hole with the simultaneous creation of a local magnon,
\begin{equation}
     \hat{H}_\text{kin} \to \sqrt{2S} \; {t \over 2} \; \sum_{n, \neighbor} \left[ \hat{h}^\dag_{n + \neighbor} \hat{h}_n  \; (\hat{b}_{n + \neighbor} + \hat{b}^{\dagger}_{n})  + \rm{h.c.} \right] .
\end{equation}
In the momentum-space Hamiltonian, the matrix element for this process vanishes as the magnon momentum $\mathbf q \to 0$, while the magnon density of states is moreover reduced at low energies. As hole hopping primarily occurs via the creation or annihilation of a short-wavelength magnon, one naturally obtains a lower energy cutoff for charge response on the order of $J^*$.

To illustrate this mechanism, in Fig.~\ref{fig:assembled_3}(b) we estimate the charge background contribution to Raman scattering, in the absence of vertex corrections (see SM for details). While the hole bubble yields the expected flat response over a frequency window of approximately $8t$, convolution with the additional magnon propagator indeed suppresses this weight below a few $J^*$. Similar mechanisms should give rise to pseudogap signatures in the other probes sensitive to motion of charges, suggesting that $J^*$ may indeed govern a wide range of experimental observables in the pseudogap regime.

\section*{Conclusions and outlook}
Motivated by recent ultracold atom experiments exploring the pseudogap phenomenology, we demonstrate that the doped Fermi–Hubbard model exhibits a universal energy scale $J^*$ with a simple doping dependence \eqref{Eq:Universal_Scaling}, which governs both equilibrium and dynamical magnetic responses. To elucidate the origin of this scale, we develop a diagrammatic framework based on the Luttinger–Ward functional and obtain self-consistent solutions of the coupled hole–magnon problem in the two-dimensional Fermi–Hubbard model. In addition, we predict a secondary scale $J_\rho< J^*$, which controls the lowest-temperature physics and determines the stability of antiferromagnetic order at $T = 0$ against the formation of an incommensurate spin-density-wave state. We show that the critical doping $\delta_{\rm AFM}$ for the commensurate-to-incommensurate transition is intimately linked to the magnetic polaron lifetime, providing a new tuning knob for the phase diagram of the Fermi-Hubbard model. Finally, we argue that the magnetic susceptibility should exhibit a maximum at temperatures $T^* \propto J^*$, suggesting that the pseudogap behavior arises from strong short-range antiferromagnetic correlations—a prediction that can be tested in near-term ultracold-atom experiments.

\section*{Acknowledgments}
\begin{acknowledgments}
The authors would like to acknowledge helpful discussions with Andrei Bernevig, Annabele Bohrdt, Petar Bojovi\'{c}, Thomas Chalopin, Tom Devereaux, Titus Franz, Antoine Georges, Markus Greiner, Alex G\'{o}mez Salvador, Fabian Grusdt, Anant Kale, Lev Kendrick, Ehsan Khatami, Gil Refael, Duilio de Santis, Si Wang, and Aaron Young. RA, IM, JBC and ED acknowledge support from the SNSF project 200021$\_$212899, the Swiss State Secretariat for Education, Research and Innovation (contract number UeM019-1),
the SNSF Sinergia grant CRSII--222792,  NCCR SPIN, a National Centre of Competence in Research, funded by the Swiss National Science Foundation (grant number 225153). 
\end{acknowledgments}

\newpage
\clearpage
\appendix

\onecolumngrid

\section*{Supplementary Material}

\setcounter{equation}{0}
\renewcommand{\theequation}{S\arabic{equation}}

\section{Notation for magnetic energy scales}

Here, we summarize all the different magnetic energy scales defined throughout the paper, together with their intuitive interpretations.

\begin{table}[ht]
    \centering
    \begin{tblr}{
        width = \textwidth,   
        colspec = { | c | Q[c, wd=3cm] | X[c] | }, 
        rows = {m}   
    }
    \hline
     \textbf{Scale} & \textbf{Definition} & \textbf{Interpretation} \\ \hline \hline
        $J$ & $4 t^2 / U$ & Bare superexchange coupling in the $t-J$ model. \emph{Not directly accessible in experiments.} \\\hline
        $J_0$ & $(1+r_0) J$ & Effective superexchange in the Heisenberg (undoped) case. Includes renormalization due to quantum zero-point fluctuations in the AFM state (Oguchi correction), $r_0 \sim 0.157$. Note that $J_0$ rather than $J$ sets the Heisenberg single-magnon spectrum, via $\omega_{\km} (\delta = 0) = 2 J_0 \e_{\km}$. Therefore, $J_0$ is the scale \emph{directly accessible in experiments} such as inelastic neutron scattering, or temperature dependence of spin-spin correlation length. \\\hline
        $J_\delta$ & $(1 - \delta)^2 [1 + r(\delta)] J$ & Doping-induced renormalization of exchange, including the effects of local moment depletion $(1-\delta)^2$, as well as the finite-doping Oguchi correction $r(\delta)$. Crucially, the definition of $J_\delta$ \emph{ignores kinetic effects}, meaning that it does not fully capture magnetic response in doped systems, and thus $J_\delta$ by itself is \emph{not accessible experimentally}. \\\hline
        $J^*(\delta)$ & $\omega_{\km} (\delta) \approx 2 J^*(\delta) \; \e_{\km}$ for high-energy magnons & Universal magnetic energy scale in doped systems, governing a vast array of one- and two-magnon \emph{experimental signatures}. Controls \emph{short-range magnetic correlations}, and obeys the universal scaling, Eq. \eqref{Eq:Universal_Scaling}. Obtained numerically by extracting a single energy scale from the upper part of the magnon spectrum, Fig.~\ref{fig:sm_j_measures}. Understood analytically to be dominated by kinetic effects, see Eq.~\eqref{eq:anderson_toy_model_main}, as well as \eqref{eq:anderson_toy_model} and \eqref{eq:anderson_toy_model_divided} in the SM.\\\hline
        $J_\rho(\delta)$ & $\omega_{\km}(\delta) \approx \sqrt{2} \; J_{\rho}(\delta) \; |\km|$ for low-energy magnons & Secondary magnetic energy scale in doped systems, controlling \emph{long-range magnetic correlations}. Unlike $J^*$, $J_\rho$ is sensitive to details of the polaron dispersion, and yields a square-root doping dependence, see Eqs.~\eqref{eq:j_rho_full_expr} and \eqref{eq:j_rho_full_expr_simplified}. Thought to control the stability of long-range AFM order in 2D or layered systems, for $T=0$.\\\hline
    \end{tblr}
    \caption{Magnetic energy scales defined in this work and their interpretations.}
    \label{tab:j_definitions_sm}
\end{table}

\section{Comparison to experimental data}

In Figure~\ref{fig:assembled_1}(a), we compare our theoretical calculation for $J^*(\delta)$ against two recent experiments, probing different magnetic observables in doped systems: spin stiffness $\rho$ \cite{chalopin2024probingmagneticoriginpseudogap} and bimagnon peak position $\omega_{\rm 2M}$ in Raman-type spectra \cite{kendrick2025pseudogapfermihubbardquantumsimulator}. In undoped systems, both $\rho$ and $\omega_{\rm 2M}$ are directly proportional to $J$, albeit with very different numerical prefactors. These prefactors have been extensively studied in the past, and their numerical values are well-understood for the Heisenberg case. In order to meaningfully compare doping behavior of $\rho$ and $\omega_{\rm 2M}$, we divide out their zero-doping values, and focus on the \emph{relative scaling} with $\delta$. Therefore, our objects of interest are $\rho(\delta) / \rho(0)$ and respectively $\omega_{\rm 2M}(\delta) / \omega_{\rm 2M}(0)$. As the experimental data in Refs.~\cite{chalopin2024probingmagneticoriginpseudogap, kendrick2025pseudogapfermihubbardquantumsimulator} are not presented in this way, we describe here the procedure by which we obtain the values depicted in Fig.~\ref{fig:assembled_1}(a).\\

\textbf{Spin stiffness.} We rely on the data presented in Fig.~3f of Ref.~\cite{chalopin2024probingmagneticoriginpseudogap}. Although denoted by $\Theta(\delta)$ in the original work, we use the notation $\rho(\delta)$ for simplicity. As done in the original work, we fit a quadratic polynomial $\rho(\delta) \approx \rho_0 + \rho_1 \delta + \rho_2 \delta^2$ to the experimental measurements of $(\rho, \delta)$ pairs. Dividing out the zero-doping value, and plugging in $\delta_{\rm eff} = U/t \times \delta$, we arrive at
\begin{equation} \label{eq:expt_comp_details_stiffness}
    {\rho(\delta) \over \rho(0)} \approx 1 + {\rho_1 \over \rho_0} {t \over U} \; \delta_{\rm eff} + {\rho_2 \over \rho_0} \left( {t \over U} \right)^2 \delta_{\rm eff}^2.
\end{equation}
The two resulting coefficients $t \rho_1 / U \rho_0$ and $t^2 \rho_2 / U^2 \rho_0$ are obtained by combining the result of the quadratic fit, described previously, with the Hamiltonian parameter ratio $U/t = 6.5 \pm 0.5$, as given in the original work. All error bounds are propagated using the \emph{uncertainties} package in python. The result, in the form given by Eq.~\eqref{eq:expt_comp_details_stiffness}, is the curve labeled $\rho$ in Fig.~\ref{fig:assembled_1}(a).\\

\textbf{Bimagnon peak position.} We use the Raman-type spectra shown in Fig.~3b of Ref.~\cite{kendrick2025pseudogapfermihubbardquantumsimulator}. For every doping value in the dataset, we fit to the experimental spectrum a symmetric double-Lorentzian curve plus background,
\begin{equation}
    f(\omega) = C_1  \left[ {\gamma^2 \over (\omega - \omega_{\rm 2M})^2 + \gamma^2} + {\gamma^2 \over (\omega + \omega_{\rm 2M})^2 + \gamma^2} \right] + C_0,
\end{equation}
from which we extract the parameters $\omega_{\rm 2M}$, $\gamma$, $C_0$, and $C_1$. From the list of $(\omega_{\rm 2M}, \delta)$ pairs, we divide all frequencies the value of $\omega_{\rm 2M}(0)$, multiply all dopings by $U/t = 7$, and plot the results as the circles labeled $\omega_{\rm 2M}$ in Fig.~\ref{fig:assembled_1}(a).

\section{Extended $t-J$ model and Raman vertices}

In this section, we briefly review the derivation of the extended $t-J$ model as a low-energy effective description of the Hubbard one, at strong interactions $U \gg t$. Starting from a spatio-temporal modulation of the hopping amplitude $t$, as may be implemented in optical lattices, we extract the corresponding perturbation within the low-energy manifold. Finally, we argue that it is analogous to optical Raman experiments in solid-state systems.\\

The starting point is the Hamiltonian
\begin{equation}
    \hat{H}(\tau) = \hat{T}  + \hat{H}_{\rm int} + \hat{V}(\tau), 
\end{equation}
where $\tau$ denotes time, as $t$ is reserved for hopping amplitude. The Hubbard term is $\hat{H}_{\rm int} = U \sum_{j} \hat{n}_{j\uparrow}\hat{n}_{j\downarrow}$, with $U>0$ the on-site repulsion strength; the electron number operator for site $j$ and spin $\sigma$ is $\hat{n}_{j \sigma} = \hat{c}^\dagger_{j \sigma} \hat{c}_{j \sigma}$ ; and the time-dependent perturbation $\hat{V}(\tau)$ encodes the optical lattice modulation, which will only be considered in eq. \eqref{eq:perturbation_definition} after introducing the Schrieffer-Wolff transformation. Meanwhile, the kinetic term $\hat{T} = - t \sum_{\langle j,k \rangle \sigma} \left( \hat{c}^{\dagger}_{j\sigma} \hat{c}_{k\sigma} + \rm{h.c.} \right)$ can be separated into operators that preserve, decrease, and respectively increase the double occupancy. With $\neighbor \in \{\pm \hat x, \pm \hat y\}$ denoting nearest-neighbor directions, we write $\hat{T} = - t \sum_{\neighbor} \left( \hat{T}_0^{\neighbor} + \hat{T}_{-1}^{\neighbor} + \hat{T}_1^{\neighbor} \right)$, where the individual terms are
\begin{subequations}
\begin{align}
     \hat{T}_0^{\neighbor} &\equiv \sum_{\mathbf{j}, \sigma} \left[ n_{\mathbf{j}, \bar \sigma} c^\dag_{\mathbf{j}, \sigma} c_{\mathbf{j} + \neighbor, \sigma} n_{\mathbf{j} + \neighbor, \bar \sigma} + \left( 1 - n_{\mathbf{j}, \bar \sigma} \right) c^\dag_{\mathbf{j}, \sigma} c_{\mathbf{j} + \neighbor, \sigma} \left(1 - n_{\mathbf{j} + \neighbor, \bar \sigma} \right) \right], \\
	\hat{T}_{-1}^{\neighbor} &\equiv \sum_{\mathbf{j}, \sigma} \left( 1 - n_{\mathbf{j}, \bar \sigma} \right) c^\dag_{\mathbf{j}, \sigma} c_{\mathbf{j} + \neighbor, \sigma} n_{\mathbf{j} + \neighbor, \bar \sigma}, \\
	\hat{T}_{1}^{\neighbor} &\equiv \sum_{\mathbf{j}, \sigma} n_{\mathbf{j}, \bar \sigma} c^\dag_{\mathbf{j}, \sigma} c_{\mathbf{j} + \neighbor, \sigma} \left( 1 - n_{\mathbf{j} + \neighbor, \bar \sigma} \right).
\end{align}
\end{subequations}
Here, $\bar \sigma$ denotes the spin opposite to $\sigma$. The split is advantageous, since we may write $[\hat{H}_\text{int}, \hat{T}_{\alpha}^{\neighbor}] = \alpha U \hat{T}_{\alpha}^{\neighbor}$ and $[\hat{T}_{\alpha}^{\neighbor}]^\dag = \hat{T}_{-\alpha}^{-\neighbor}$ for $\alpha \in \{-1,0,1\}$. We now change the basis via the transformation $\hat{U} = e^{i \hat{S}}$, where $\hat{S}$ is picked to cancel, at leading order, the coupling between different double-occupancy sectors. To second order in $t/U$, and ignoring $\hat{V}(\tau)$, the transformed Hamiltonian is
\begin{equation} \label{eq:transformed_hamiltonian_schrieffer_wolff}
    \hat{H}_{\rm eff} = \hat{U}^\dag \hat{H} \hat{U} \approx \hat{H}_{\rm int} + \hat{T} + i [\hat{H}_{\rm int}, \hat{S}] + i [\hat{T}, \hat{S}] - {1 \over 2} \left[ [\hat{H}_{\rm int}, \hat{S}], \hat{S} \right] + \dots,
\end{equation}
and imposing $-t \left( \hat{T}_{-1} + \hat{T}_{1} \right) + i [\hat{H}_{\rm int}, \hat{S}] = 0$, we use the aforementioned commutators to find the solution
\begin{equation} \label{eq:schrieffer_wolff_solution}
    \hat{S} = -i\; {t \over U} \left( \hat{T}_{1} - \hat{T}_{-1} \right).
\end{equation}
Without loss of generality, we will focus on the hole-doped case. Thus, to extract the low-energy theory, the remaining terms in the rotated Hamiltonian \eqref{eq:transformed_hamiltonian_schrieffer_wolff} must be projected into the sector of no double occupancies. With $\hat{P}$ the projector achieving this, we can plug \eqref{eq:schrieffer_wolff_solution} into \eqref{eq:transformed_hamiltonian_schrieffer_wolff} to arrive at
\begin{equation}
    \hat{H}_{\rm eff} \to \hat{P} \left( - t \hat{T}_0 + {t^2 \over  U} \sum_{\neighbor, \neighboraux} [\hat{T}_{1}^{\neighbor}, \hat{T}_{-1}^{-\neighboraux}] \right) \hat{P}.
\end{equation}
The first term $- t \; \hat{P} \hat{T}_0 \hat{P}$ represents the usual restricted hopping of the $t-J$ model, while the $\neighboraux = \neighbor$ component of the second one gives superexchange:
\begin{equation}
    {t^2 \over  U} \; \hat{P} \left( \sum_{\neighbor} [\hat{T}_{1}^{\neighbor}, \hat{T}_{-1}^{-\neighbor}] \right) \hat{P} = {4 t^2 \over U} \sum_{\langle j, k \rangle} \left( \mathbf{S}_{j} \cdot \mathbf{S}_{k} - \frac{1}{4} \hat{n}_{j} \hat{n}_{k} \right),
\end{equation}
where the spin operators are defined by $\mathbf{S}_{j} = \frac{1}{2} \; \hat{c}^{\dagger}_{j, \alpha} \boldsymbol{\sigma}^{\alpha \beta} \hat{c}_{j, \beta}$, while the total number operators are $\hat{n}_{j} = \hat{n}_{j, \uparrow} + \hat{n}_{j, \downarrow}$. The remaining $\neighboraux \neq \neighbor$ contributions, often neglected in spite of appearing at the same order in $t/U$ as the superexchange, yield the so-called 3-site correlated hopping terms:
\begin{equation}
    {t^2 \over  U} \; \hat{P} \left( \sum_{\neighbor \neq \neighboraux} [\hat{T}_{1}^{\neighbor}, \hat{T}_{-1}^{-\neighboraux}] \right) \hat{P} = - {t^2 \over  U} \; \hat{P} \Bigg[ \sum_{\substack{\mathbf{j}, \sigma \\ \neighbor \neq \neighboraux}} \left( \hat{c}^\dag_{\mathbf{j}, \sigma} \;  \hat{c}_{\mathbf{j} - \neighboraux, \bar \sigma}^\dag \; \hat{c}_{\mathbf{j} - \neighboraux, \bar \sigma} \; \hat{c}_{\mathbf{j} + \neighbor - \neighboraux, \sigma} - \hat{c}^\dag_{\mathbf{j}, \sigma} \;  \hat{c}_{\mathbf{j} - \neighboraux, \bar \sigma}^\dag \; \hat{c}_{\mathbf{j} - \neighboraux, \sigma}  \; \hat{c}_{\mathbf{j} + \neighbor - \neighboraux, \bar \sigma} \right) \Bigg] \hat{P}.
\end{equation}
Combining these three contributions, one recovers the extended $t-J$ model, Eq. \eqref{eq:extended_t_j_hamiltonian}. We now turn our attention to the perturbation $\hat{V}(\tau)$; for spatially uniform modulations of the hopping amplitude, it can be expressed as
\begin{equation} \label{eq:perturbation_definition}
    \hat{V}(\tau) = - \Delta t \sum_{\neighbor} f^{\neighbor}(\tau) \left( \hat{T}_0^{\neighbor} + \hat{T}_{-1}^{\neighbor} + \hat{T}_1^{\neighbor} \right).
\end{equation}
In the case of ultracold atom arrays, this is directly achieved by modulating the height of the energy barrier between neighboring optical lattice sites. For example, modulating the $\hat{x}-$ and $\hat{y}-$hoppings with equal intensity $\Delta t$ and frequency $\omega$, one may achieve either an $s-$wave driving when the two directions are modulated in phase,
\begin{equation} \label{eq:perturbation_form_factors_s_wave}
    f^{\pm \hat{x}}(\tau) = f^{\pm \hat{y}}(\tau) = \sin (\omega \tau),
\end{equation}
or otherwise a $d-$wave perturbation if the modulations are perfectly out of phase:
\begin{equation} \label{eq:perturbation_form_factors_d_wave}
    f^{\pm \hat{x}}(\tau) = -f^{\pm \hat{y}}(\tau) = \sin (\omega \tau).
\end{equation}
This latter approach will allow us to access the bimagnon peak. We also note that for optical Raman experiments in the solid state, one would couple the probe to the system via Peierls coupling $t \to t e^{i e A \cdot \neighbor}$, which at first order gives form factors $f^{\neighbor}_{(1)} \propto i A \cdot \neighbor$, at second order $f^{\neighbor}_{(2)} \propto  (A \cdot \neighbor)^2$, etc.\\

As with the original kinetic Hamiltonian, the perturbation $\hat{V}(\tau)$ has a contribution which does not connect subspaces of different double-occupancy, and therefore can be directly projected as $ - \Delta t  \; \hat{P} \left(\sum_{\neighbor} f^{\neighbor} (\tau) \; \hat{T}_0^{\neighbor} \right) \hat{P}$. The rest is plugged into the Schrieffer-Wolff transformation. Combining both, and omitting the explicit $\tau$ dependence on $\hat{V}_{\rm eff}$ as well as on the form factors for brevity, the end result is
\begin{align}
    \hat{V}_{\rm eff} &\to  -\Delta t \; \hat{P} \Bigg[  \sum_{\neighbor} f^{\neighbor} \; \sum_{\mathbf{j}, \sigma} \hat{c}^\dag_{\mathbf{j}, \sigma} \hat{c}_{\mathbf{j} + \neighbor, \sigma} - {2 t \over U} \sum_{\neighbor} \left( f^{\neighbor} + f^{-\neighbor}\right) \sum_{\mathbf j} \left( \mathbf{S}_{\mathbf j} \cdot \mathbf{S}_{\mathbf j + \neighbor} - {1 \over 4} \hat{n}_{\mathbf j} \hat{n}_{\mathbf j + \neighbor} \right) \nonumber \\
    & \qquad \qquad + {t \over U} \sum_{\neighbor \neq \neighboraux } \left( f^{\neighbor} + f^{-\neighboraux}\right) \sum_{\mathbf{j}, \sigma} \left( \hat{c}^\dag_{\mathbf{j}, \sigma} \;  \hat{c}_{\mathbf{j} - \neighboraux, \bar \sigma}^\dag \; \hat{c}_{\mathbf{j} - \neighboraux, \bar \sigma} \; \hat{c}_{\mathbf{j} + \neighbor - \neighboraux, \sigma} - \hat{c}^\dag_{\mathbf{j}, \sigma} \; \hat{c}_{\mathbf{j} - \neighboraux, \bar \sigma}^\dag \; \hat{c}_{\mathbf{j} - \neighboraux, \sigma}  \;  \hat{c}_{\mathbf{j} + \neighbor - \neighboraux, \bar \sigma} \right) \Bigg] \hat{P}, \label{eq:perturbation_term_effective_Hamiltonian}
\end{align}
from which one may extract the charge and spin Raman vertices, as we will do in the next section.

\newcommand{\jl}{\mathbf{j}}

\section{Separating spin and charge degrees of freedom}

The goal of this section is to introduce an auxiliary-boson transformation which makes the distinction between spin and charge at the operator level. Then, we write the Hamiltonian governing the evolution of these coupled degrees of freedom, as well as the corresponding Raman vertices.\\

We begin by considering the extended $t-J$ model on a 2D square lattice,
\begin{align} 
    \hat{H} = &-t \sum_{\jl, \neighbor, \sigma} \hat{P}_{\jl} \; \hat{c}^{\dagger}_{\jl, \sigma} \hat{c}_{\jl + \neighbor, \sigma} \; \hat{P}_{\jl + \neighbor} + {J \over 2} \sum_{\jl, \neighbor} \left( \mathbf{S}_{\jl} \cdot \mathbf{S}_{\jl + \neighbor} - \frac{1}{4} \hat{n}_{\jl} \hat{n}_{\jl + \neighbor} \right) \nonumber \\
    &- {J \over 4} \sum_{\substack{\jl, \sigma \\ \neighbor \neq \neighboraux}} \hat{P}_{\jl} \left( \hat{c}^\dag_{\jl, \sigma} \;  \hat{c}_{\jl - \neighboraux, \bar \sigma}^\dag \; \hat{c}_{\jl - \neighboraux, \bar \sigma} \; \hat{c}_{\jl + \neighbor - \neighboraux, \sigma} - \hat{c}^\dag_{\jl, \sigma} \;  \hat{c}_{\jl - \neighboraux, \bar \sigma}^\dag \; \hat{c}_{\jl - \neighboraux, \sigma}  \; \hat{c}_{\jl + \neighbor - \neighboraux, \bar \sigma} \right) \hat{P}_{\jl + \neighbor - \neighboraux}, \label{eq:extended_t_j_hamiltonian}
\end{align}
where we introduce the superexchange coupling $J=4t^2/U$, and the on-site projector $\hat{P}_{\jl}=(1 - \hat{n}_{\jl, \uparrow} \hat{n}_{\jl, \downarrow})$. Since we are working with a bipartite lattice, define the sublattice-rotated fermion operators
\begin{equation}
    \hat{f}_{\jl, \sigma} \equiv \begin{cases}
    \hat{c}_{\jl, \sigma} & \jl \in A,\\
    \hat{c}_{\jl, \overline{\sigma}} & \jl \in B,
    \end{cases}
\end{equation}
such that an AFM arrangement in the original basis translates to an FM one in the new frame. The original Hamiltonian \eqref{eq:extended_t_j_hamiltonian} now becomes
\begin{align}
    \hat{H} = &- t \sum_{\jl, \neighbor, \sigma} \hat{P}_{\jl} \; \hat{f}^{\dagger}_{\jl, \sigma} \hat{f}_{\jl + \neighbor, \overline{\sigma}} \; \hat{P}_{\jl + \neighbor} - {J \over 2} \sum_{\jl, \neighbor} \left[ \tilde{S}^z_{\jl} \tilde{S}^z_{\jl + \neighbor} - {1 \over 2} \left( \tilde{S}^{+}_{\jl} \tilde{S}^{+}_{\jl + \neighbor} + \tilde{S}^{-}_{\jl} \tilde{S}^{-}_{\jl + \neighbor}\right) + \frac{1}{4} \; \hat{n}_{\jl} \hat{n}_{\jl + \neighbor} \right] \nonumber \\
    &- {J \over 4} \sum_{\substack{\jl, \sigma \\ \neighbor \neq \neighboraux}} \hat{P}_{\jl} \left( \hat{f}^\dag_{\jl, \sigma} \;  \hat{f}_{\jl - \neighboraux, \sigma}^\dag \; \hat{f}_{\jl - \neighboraux, \sigma} \; \hat{f}_{\jl + \neighbor - \neighboraux, \sigma} - \hat{f}^\dag_{\jl, \sigma} \;  \hat{f}_{\jl - \neighboraux, \sigma}^\dag \; \hat{f}_{\jl - \neighboraux, \bar \sigma}  \; \hat{f}_{\jl + \neighbor - \neighboraux, \bar \sigma} \right) \hat{P}_{\jl + \neighbor - \neighboraux}, \label{eq:rotated_hamiltonian}
\end{align}
where the rotated spin operators are analogously defined through $\tilde{\mathbf{S}}_{\jl} = \frac{1}{2} \; \hat{f}^{\dagger}_{\jl, \alpha} \boldsymbol{\sigma}^{\alpha \beta} \hat{f}_{\jl, \beta}$.\\

We aim to separate the charge and spin degrees of freedom by introducing the auxiliary-boson transformation 
$\hat{f}^{\dagger}_{\jl, \sigma} = \hat{b}^{\dagger}_{\jl, \sigma} \hat{h}_{\jl}$. Here, $\hat{h}_{\jl}$ ($\hat{h}^\dag_{\jl}$) is a spinless fermionic operator annihilating (creating) a hole at site $\jl$, while the bosonic spin-wave operator $\hat{b}_{\jl, \sigma}$ ($\hat{b}^\dagger_{\jl, \sigma}$) represents the annihilation (creation) of a Schwinger boson. The auxiliary-boson mapping is faithful, provided that the local constraint
\begin{equation} \label{eq:local_constraint}
    \sum_{\sigma} \hat{b}^\dag_{\jl, \sigma} \hat{b}_{\jl, \sigma} + \hat{h}^\dag_{\jl} \hat{h}_{\jl} = 2S
\end{equation}
is always satisfied at every site $\jl$. Since we are describing the AFM ordered state (corresponding to FM in the rotated frame), we can condense one of the boson species. In the rotated frame, take the order along the positive $\hat{z}$ direction, and carry out the replacement   
\begin{equation} \label{eq:condensation_with_holes}
    \hat{b}_{\jl, \uparrow}, \; \hat{b}^\dag_{\jl, \uparrow} \to \sqrt{2S - \hat{b}^\dag_{\jl, \downarrow} \hat{b}_{\jl, \downarrow} - \hat{h}^\dag_{\jl} \hat{h}_{\jl}}.
\end{equation}
After performing this procedure, the only remaining boson operators are $\hat{b}^\dag_{\jl, \downarrow}, \hat{b}_{\jl, \downarrow}$. For convenience, we will drop the spin index, and denote these simply by $\hat{b}^\dag_{\jl}, \hat{b}_{\jl}$ from now on. Using the local constraint \eqref{eq:local_constraint}, as well as $\left( \hat{h}^\dagger_{\jl} \right)^2 = 0$, the $\hat{z}-$component of the rotated spin operator can be directly expressed as
\begin{equation}
    \tilde{S}_{\jl}^z = \hat{h}_{\jl} \hat{h}_{\jl}^\dag \left(S - \hat{b}^\dag_{\jl} \hat{b}_{\jl} \right), 
\end{equation}
while the transverse components can be expanded by
\begin{align}
    \tilde{S}_{\jl}^+ &= \sqrt{2S} \; \hat{h}_{\jl} \hat{h}_{\jl}^\dag \; \sqrt{1 - {\hat{b}^\dag_{\jl} \hat{b}_{\jl} \over 2S} - {\hat{h}^\dag_{\jl} \hat{h}_{\jl} \over 2S}} \; \hat{b}_{\jl} \nonumber \\
    &= \sqrt{2S} \; \hat{h}_{\jl} \hat{h}_{\jl}^\dag \; \sqrt{1 - {\hat{b}^\dag_{\jl} \hat{b}_{\jl} \over 2S}} \; \hat{b}_{\jl} \nonumber \\
    &= \sqrt{2S} \; \hat{h}_{\jl} \hat{h}_{\jl}^\dag \left[ \hat{b}_{\jl} - {1 \over 4S} \; \hat{b}^\dag_{\jl} \hat{b}_{\jl} \hat{b}_{\jl} + \OO(S^{-2}) \right],
\end{align}
and respectively $\tilde{S}_{\jl}^- = [\tilde{S}_{\jl}^+]^\dag$. Going from the first line to the second, the hole number operator $\hat{h}^\dag_{\jl} \hat{h}_{\jl}$ inside the square root can be eliminated by expanding in a series, and observing that the prefactor of $\hat{h}_{\jl} \hat{h}_{\jl}^\dag$ annihilates all resulting terms, except for the zeroth-order one. The second step is a typical expansion in the magnon density $\hat{b}^\dag_{\jl} \hat{b}_{\jl} / 2S$; in the Hamiltonian, we will only keep terms up to quartic order in $\hat{b}_{\jl}$ operators.\\

Plugging everything in, the superexchange term in \eqref{eq:rotated_hamiltonian} can be expressed order-by-order in $S$ as 
\begin{equation}
    \hat{H}_\text{exchange} = \hat{H}_\text{holes}^{\hcomb} + \hat{H}_\text{LSWT}^{\hcomb} + \hat{H}_\text{int}^{\hcomb} + \OO(S^{-1}),
\end{equation}
where the individual terms are
\begin{subequations} \label{eq:exchange_components_full}
\begin{align}
    \hat{H}_\text{holes}^{\hcomb} &=  - J S^2 \; \sum_{\jl, \neighbor} \hat{h}_{\jl} \hat{h}_{\jl}^\dag \hat{h}_{\jl + \neighbor} \hat{h}_{\jl + \neighbor}^\dag, \\
    \hat{H}_\text{LSWT}^{\hcomb} &= {J S \over 2} \sum_{\jl, \neighbor} \hat{h}_{\jl} \hat{h}_{\jl}^\dag \hat{h}_{\jl + \neighbor} \hat{h}_{\jl + \neighbor}^\dag \left( \hat{b}^\dag_{\jl} \hat{b}_{\jl} + \hat{b}^\dag_{\jl + \neighbor} \hat{b}_{\jl + \neighbor} +  \hat{b}_{\jl} \hat{b}_{\jl + \neighbor} + \hat{b}^\dag_{\jl} \hat{b}^\dag_{\jl + \neighbor} \right), \\
    \hat{H}_\text{int}^{\hcomb} &= - {J \over 2} \sum_{\jl, \neighbor} \hat{h}_{\jl} \hat{h}_{\jl}^\dag \hat{h}_{\jl + \neighbor} \hat{h}_{\jl + \neighbor}^\dag \Bigg[ \hat{b}^\dag_{\jl + \neighbor} \hat{b}^\dag_{\jl} \hat{b}_{\jl}  \hat{b}_{\jl + \neighbor} + {1 \over 4} \bigg( \hat{b}^\dag_{\jl + \neighbor} \hat{b}_{\jl + \neighbor} \hat{b}_{\jl + \neighbor} \hat{b}_{\jl} + \hat{b}^\dag_{\jl} \hat{b}_{\jl} \hat{b}_{\jl}  \hat{b}_{\jl + \neighbor} + \hat{b}^\dag_{\jl}  \hat{b}^\dag_{\jl + \neighbor} \hat{b}^\dag_{\jl + \neighbor} \hat{b}_{\jl + \neighbor} +  \hat{b}^\dag_{\jl + \neighbor} \hat{b}^\dag_{\jl} \hat{b}^\dag_{\jl} \hat{b}_{\jl} \bigg) \Bigg].
\end{align}  
\end{subequations}
Assuming translational invariance for any expectation values involving only hole operators, we can effectively decouple the hole and magnon degrees of freedom \footnote{This step is not strictly necessary: one may keep the full exchange terms \eqref{eq:exchange_components_full} and use them to write diagrams. When the legs corresponding to holes, originating from a single vertex, are contracted with each other, the result is equivalent to taking the approximate form \eqref{eq:exchange_components_decoupled}. On the other hand, having multiple such vertices and joining their hole legs to each other would give rise to higher-order terms, which we will ignore.} in all terms of \eqref{eq:exchange_components_full}. Defining the quantities
\begin{subequations}
\begin{align}
    \holepf &\equiv \left\langle \hat{h}_{\jl} \hat{h}_{\jl}^\dag \hat{h}_{\jl + \neighbor} \hat{h}_{\jl + \neighbor}^\dag \right\rangle, \\
    \magpfl &\equiv \left\langle \hat{b}^\dag_{\jl} \hat{b}_{\jl} + \hat{b}^\dag_{\jl + \neighbor} \hat{b}_{\jl + \neighbor} +  \hat{b}_{\jl} \hat{b}_{\jl + \neighbor} + \hat{b}^\dag_{\jl} \hat{b}^\dag_{\jl + \neighbor} \right\rangle, \\
    \magpfnl &\equiv \left\langle \hat{b}^\dag_{\jl + \neighbor} \hat{b}^\dag_{\jl} \hat{b}_{\jl}  \hat{b}_{\jl + \neighbor} + {1 \over 4} \bigg( \hat{b}^\dag_{\jl + \neighbor} \hat{b}_{\jl + \neighbor} \hat{b}_{\jl + \neighbor} \hat{b}_{\jl} + \hat{b}^\dag_{\jl} \hat{b}_{\jl} \hat{b}_{\jl}  \hat{b}_{\jl + \neighbor} + \hat{b}^\dag_{\jl}  \hat{b}^\dag_{\jl + \neighbor} \hat{b}^\dag_{\jl + \neighbor} \hat{b}_{\jl + \neighbor} +  \hat{b}^\dag_{\jl + \neighbor} \hat{b}^\dag_{\jl} \hat{b}^\dag_{\jl} \hat{b}_{\jl} \bigg) \right\rangle,
\end{align}
\end{subequations}
we find up to an irrelevant constant term
\begin{equation}
    \hat{H}_\text{exchange} \approx \hat{H}_\text{holes} + \hat{H}_\text{LSWT} + \hat{H}_\text{int} + \text{constant} + \OO(S^{-1}),
\end{equation}
where the components now read
\begin{subequations} \label{eq:exchange_components_decoupled}
\begin{align}
    \hat{H}_\text{holes} &=  - J S^2 \left( 1 - {1 \over 2S} \; \magpfl + {1 \over 2S^2} \; \magpfnl \right) \; \sum_{\jl, \neighbor} \hat{h}_{\jl} \hat{h}_{\jl}^\dag \hat{h}_{\jl + \neighbor} \hat{h}_{\jl + \neighbor}^\dag, \label{eq:effective_nn_hole_realspace} \\
    \hat{H}_\text{LSWT} &= {\holepf J S \over 2} \sum_{\jl, \neighbor} \left( \hat{b}^\dag_{\jl} \hat{b}_{\jl} + \hat{b}^\dag_{\jl + \neighbor} \hat{b}_{\jl + \neighbor} +  \hat{b}_{\jl} \hat{b}_{\jl + \neighbor} + \hat{b}^\dag_{\jl} \hat{b}^\dag_{\jl + \neighbor} \right), \\
    \hat{H}_\text{int} &= - {\holepf J \over 2} \sum_{\jl, \neighbor} \Bigg[ \hat{b}^\dag_{\jl + \neighbor} \hat{b}^\dag_{\jl} \hat{b}_{\jl}  \hat{b}_{\jl + \neighbor} + {1 \over 4} \bigg( \hat{b}^\dag_{\jl + \neighbor} \hat{b}_{\jl + \neighbor} \hat{b}_{\jl + \neighbor} \hat{b}_{\jl} + \hat{b}^\dag_{\jl} \hat{b}_{\jl} \hat{b}_{\jl}  \hat{b}_{\jl + \neighbor} + \hat{b}^\dag_{\jl}  \hat{b}^\dag_{\jl + \neighbor} \hat{b}^\dag_{\jl + \neighbor} \hat{b}_{\jl + \neighbor} +  \hat{b}^\dag_{\jl + \neighbor} \hat{b}^\dag_{\jl} \hat{b}^\dag_{\jl} \hat{b}_{\jl} \bigg) \Bigg].
\end{align}
\end{subequations}
The last two terms resemble the usual spin-wave Hamiltonian that one obtains from a Holstein-Primakoff approach in the absence of doping. We remark that $\holepf = (1-\delta)^2$ encodes the trivial reduction in average local magnetic moments, due to the probability $\delta$ of finding a hole on any given site. In momentum space, the magnetic terms in \eqref{eq:exchange_components_decoupled} read
\begin{subequations} \label{eq:simp_ham_magnons}  
\begin{align}
    \hat{H}_\text{LSWT} &= {\holepf z J S \over 2} \sum_{\km} \left[ \hat{b}^\dag_{\km} \hat{b}_{\km} + \hat{b}^\dag_{-\km} \hat{b}_{-\km} +  \gamma_{\km} \left( \hat{b}_{-\km} \hat{b}_{\km} + \hat{b}^\dag_{\km} \hat{b}^\dag_{-\km} \right) \right], \label{eq:hamiltonian_lswt}\\
    \hat{H}_\text{int} &= - {\holepf z J \over 2 N} \sum_{\km, \p, \q} \left[ \gamma_\q \; \hat{b}^\dag_{\km + \q} \hat{b}^\dag_{\p - \q} \hat{b}_{\p} \hat{b}_{\km} + {1 \over 2} \left( \gamma_{\km} \; \hat{b}^\dag_{\km + \q} \hat{b}_{\q - \p} \hat{b}_{\p} \hat{b}_{\km} + \gamma_{\km + \q} \; \hat{b}^\dag_{\km + \q}  \hat{b}^\dag_{\p - \q} \hat{b}^\dag_{-\p} \hat{b}_{\km} \right) \right], \label{eq:hamiltonian_magnon_int}
\end{align}
\end{subequations}
where $z=4$ is the coordination number, and $\gamma_\km = {1 \over z} \sum_{\neighbor} e^{i \km \neighbor}$ the square-lattice form factor. Meanwhile, the nearest-neighbor hole interaction \eqref{eq:effective_nn_hole_realspace} is
\begin{equation} \label{eq:simp_ham_holes}
    \hat{H}_\text{holes} =  - z J S^2 \left( 1 - {1 \over 2S} \; \magpfl + {1 \over 2S^2} \; \magpfnl \right) \left[N - 2 \sum_{\km} \hat{h}^\dag_{\km} \hat{h}_{\km} + {1 \over N} \sum_{\km, \q, \p} \gamma_{\q} \; \hat{h}^\dag_{\km + \q} \hat{h}^\dag_{\p - \q} \hat{h}_{\p} \hat{h}_{\km} \right],
\end{equation}
so that it contains an (irrelevant) overall constant, a chemical potential correction, as well as an actual normal-ordered interaction term. For brevity, we will define the following quantity,
\begin{equation}
    \lambda \equiv 1 - {1 \over 2S} \; \magpfl + {1 \over 2S^2} \; \magpfnl,
\end{equation}
which however contributes only to a renormalization of the hole chemical potential, and is therefore irrelevant for our current discussion, where we always fix doping.\\

Shifting focus to the kinetic term of \eqref{eq:rotated_hamiltonian}, note that the typical contribution $\hat{f}^{\dagger}_{\jl, \sigma} \hat{f}_{\jl + \neighbor, \overline{\sigma}} = \hat{b}^{\dagger}_{\jl, \sigma} \hat{h}_{\jl} \hat{h}^\dag_{\jl + \neighbor} \hat{b}_{\jl + \neighbor, \overline{\sigma}}$ involves a spin flip with respect to the rotated frame. Upon condensing via \eqref{eq:condensation_with_holes}, the leading-order term will therefore contain a single magnon operator. Specifically, it yields
\begin{equation}
    \hat{H}_\text{kin} = \sqrt{2S} \; {t \over 2} \; \sum_{\jl, \neighbor} \left[ \hat{h}^\dag_{\jl + \neighbor} \hat{h}_{\jl}  \; (\hat{b}_{\jl + \neighbor} + \hat{b}^{\dagger}_{\jl})  + \hat{h}^\dag_{\jl} \hat{h}_{\jl + \neighbor}  \; ( \hat{b}^\dag_{\jl + \neighbor} + \hat{b}_{\jl} ) \right] + \OO(S^{-1/2}),
\end{equation}
or equivalently in momentum space
\begin{equation} \label{eq:simp_ham_coupling}
    \hat{H}_\text{kin} = \sqrt{2S \over N} \; z t \; \sum_{\km, \q} \hat{h}^\dag_{\km + \q} \hat{h}_{\km} \left( \gamma_{\km} \hat{b}_{\q} + \gamma_{\km + \q} \hat{b}^\dag_{-\q}  \right) + \OO(S^{-1/2}).
\end{equation}
This term therefore describes the process through which a hole scatters by annihilating or creating a magnon. On the other hand, we discard higher-order terms in $S$, which couple the hole to continua of $3$ magnons, $5$ magnons, etc.\\ 

Finally, let us consider the 3-site terms in the extended $t-J$ model. Their main effect will be to give the holes a bare dispersion, which the usual kinetic and superexchange terms fail to do, as evidenced by \eqref{eq:simp_ham_holes} and \eqref{eq:simp_ham_coupling}. Beyond this, they will also introduce higher-order couplings between magnons and holes (i.e. 2 holes scattering on 2 magnons, and above), which we will neglect in the same spirit as when we took only the leading term in \eqref{eq:simp_ham_coupling}. The summand in \eqref{eq:rotated_hamiltonian} becomes
\begin{equation}
    \hat{f}^\dag_{\mathbf{j}, \sigma} \;  \hat{f}_{\mathbf{j} - \neighboraux, \sigma}^\dag \left( \hat{f}_{\mathbf{j} - \neighboraux, \sigma} \; \hat{f}_{\mathbf{j} + \neighbor - \neighboraux, \sigma} - \hat{f}_{\mathbf{j} - \neighboraux, \bar \sigma}  \;  \hat{f}_{\mathbf{j} + \neighbor - \neighboraux, \bar \sigma} \right) = \hat{b}^\dag_{\mathbf{j}, \sigma} \hat{b}_{\mathbf{j} - \neighboraux, \sigma}^\dag \hat{h}_{\mathbf{j}} \hat{h}_{\mathbf{j} - \neighboraux} \hat{h}_{\mathbf{j} - \neighboraux}^\dag \hat{h}_{\mathbf{j} + \neighbor - \neighboraux}^\dag \left( \hat{b}_{\mathbf{j} - \neighboraux, \sigma}  \hat{b}_{\mathbf{j} + \neighbor - \neighboraux, \sigma} - \hat{b}_{\mathbf{j} - \neighboraux, \bar \sigma}   \hat{b}_{\mathbf{j} + \neighbor - \neighboraux, \bar \sigma} \right).
\end{equation}
Abbreviating the spatial indices of the magnon operators by $\{1,2,3,4\}$, we see that upon condensation, they yield
\begin{align}
    \sum_{\sigma} \hat{b}_{1\sigma}^\dag \hat{b}_{2\sigma}^\dag \left( \hat{b}_{3\sigma} \hat{b}_{4\sigma} - \hat{b}_{3 \bar \sigma} \hat{b}_{4 \bar \sigma}\right) \to (2S)^2 - S \left[ 2 \hat{b}_1^\dag \hat{b}_2^\dag + 2 \hat{b}_3 \hat{b}_4 + \sum_{\alpha = 1}^4 \hat{b}_\alpha^\dag \hat{b}_{\alpha} \right] + \mathcal{O}(S^0),
\end{align}
and we are interested in the first term, which gives rise to an effective hole Hamiltonian
\begin{align}
    \hat{H}_{3-\rm{site}} &= - J S^2 \sum_{\jl, \neighbor \neq \neighboraux} \hat{h}_{\mathbf{j}} \hat{h}_{\mathbf{j} - \neighboraux} \hat{h}_{\mathbf{j} - \neighboraux}^\dag \hat{h}_{\mathbf{j} + \neighbor - \neighboraux}^\dag = J S^2 \sum_{\jl, \neighbor \neq \neighboraux} \hat{h}_{\mathbf{j} + \neighbor - \neighboraux}^\dag \hat{h}_{\mathbf{j}} - J S^2 \sum_{\jl, \neighbor \neq \neighboraux} \hat{h}_{\mathbf{j} + \neighbor - \neighboraux}^\dag \hat{h}_{\mathbf{j} - \neighboraux}^\dag \hat{h}_{\mathbf{j} - \neighboraux} \hat{h}_{\mathbf{j}}.
\end{align}
In momentum space, this reads
\begin{equation} \label{eq:hamiltonian_holes_corr_hop}
    \hat{H}_{3-\rm{site}} = 4 J S^2 \sum_{\km} \chi_{\km} \; \hat{h}_{\km}^\dag \hat{h}_{\km} - {4 J S^2 \over N} \sum_{\km, \p, \q} \xi_{\km, \q} \; \hat{h}^\dag_{\km + \q} \hat{h}^\dag_{\p - \q} \hat{h}_{\p} \hat{h}_{\km},
\end{equation}
where we have defined the form factors
\begin{subequations}
\begin{align}
    \chi_{\km} &= {1 \over 4} \sum_{\neighbor \neq \neighboraux} e^{i \km (\neighbor - \neighboraux)} = {\cos (2 k_x) + 4 \cos (k_x) \cos (k_y) + \cos (2 k_y) \over 2}, \label{eq:three_site_dispersion_def} \\
    \xi_{\km, \q} &= {1 \over 4} \sum_{\neighbor \neq \neighboraux} e^{i \km (\neighbor - \neighboraux)} e^{i \q \neighbor} \label{eq:three_site_interaction_form_def}.
\end{align}  
\end{subequations}
As promised, \eqref{eq:hamiltonian_holes_corr_hop} contains a bare dispersion $4 J S^2 \chi_{\km}$, but it also contributes to the interaction between holes. To sum up, the Hamiltonian that we will work with is the combination of terms \eqref{eq:simp_ham_magnons}, \eqref{eq:simp_ham_holes}, \eqref{eq:simp_ham_coupling}, and \eqref{eq:hamiltonian_holes_corr_hop}:
\begin{equation} \label{eq:simp_ham_total}
    \hat{H} \approx \hat{H}_\text{LSWT} + \hat{H}_\text{int} + \hat{H}_\text{holes} + \hat{H}_{3-\rm{site}} +\hat{H}_\text{kin} + \dots,
\end{equation}
where the $\dots$ denote constant as well as higher-order terms, which will be ignored.\\

For completeness, we also present the Raman vertices resulting from applying the auxiliary-boson transformation to the perturbation \eqref{eq:perturbation_term_effective_Hamiltonian}. As the operators involved in $\hat{V}_{\rm eff}$ are identical to those in the Hamiltonian, while only the form factors differ, we will skip the detailed derivations. Furthermore, we specialize to the $d-$wave form factors discussed in \eqref{eq:perturbation_form_factors_d_wave}. The kinetic part of $\hat{V}_{\rm eff}$, analogously to \eqref{eq:simp_ham_coupling}, will contain two hole operators and a magnon one,
\begin{equation} \label{eq:raman_vetex_kinetic}
    \hat{V}_{\rm kin} = \sqrt{2S \over N} \; z \Delta t \sum_{\mathbf k, \mathbf q} \hat{h}^\dag_{\mathbf k + \mathbf q} \hat{h}_{\mathbf{k}} \left[ \tilde{\gamma}_{\mathbf k} \hat{b}_{\mathbf q} + \tilde{\gamma}_{\mathbf k + \mathbf q} \hat{b}^\dag_{-\mathbf{q}} \right].
\end{equation}
Crucially, the presence of the $\hat{b}$ operators suppresses low-frequency response. Indeed, a hole-only contribution of the type $\hat{h}^\dag_{\mathbf k + \mathbf q} \hat{h}_{\mathbf{k}}$ could yield finite weight at low frequencies; but when the magnons are included, we are confronted with their vanishing density of states at the bottom of the band. This implies that the magnon itself must carry a finite frequency in order to obtain a sizable overall response, which means that said total response will be significantly suppressed as $\omega \to 0$. As discussed in the main text, this strongly reminds of pseudogap behavior.\\

The spin exchange term in $\hat{V}_{\rm eff}$ is of Loudon-Fleury form, and under the $d-$wave driving it maps to
\begin{equation} \label{eq:raman_vetex_bimagnon}
    \hat{V}_{\rm 2M} = \alpha_h \; (2S) \; {2 z t \Delta t \over U} \sum_{\mathbf k} \tilde{\gamma}_{\mathbf k} \left( \hat{b}_{\mathbf k} \hat{b}_{-\mathbf k} + \hat{b}^\dag_{\mathbf k} \hat{b}^\dag_{-\mathbf k} \right),
\end{equation}
mirroring the optical Raman vertex for the $B_{1g}$ symmetry channel in the solid state. Finally, the lowest-order contribution from the 3-site term is
\begin{equation} \label{eq:raman_vetex_3_site}
    \hat{V}_{3-\rm{site}} = (2S)^2 \; {8 t \Delta t \over U} \sum_{\mathbf k} \tilde{\chi}_{\mathbf k} \; \hat{h}^\dag_{\mathbf k} \hat{h}_{\mathbf k},
\end{equation}
where we defined the form factor
\begin{equation}
    \tilde{\chi}_{\mathbf k} = {\cos (2 k_x) - \cos (2 k_y) \over 2}.
\end{equation}
We remark, however, that the lack of any momentum transfer in \eqref{eq:raman_vetex_3_site} means that this vertex will not contribute to our estimation of the charge Raman response in the following sections.

\section{Single-particle properties: Green's functions, symmetries, and diagrammatic approach}

In this section, we define the single-particle Green's functions that form the basis of our analysis. We highlight their symmetry properties arising from the assumption of AFM order, as well as the point group of the square lattice. Then, we introduce the Luttinger-Ward functional and describe the self-consistent procedure used in calculating single-particle propagators and self-energies.

\subsection{Propagator definitions}

The linear spin-wave Hamiltonian \eqref{eq:hamiltonian_lswt} can be simplified by defining the Nambu spinor
\begin{equation}
    \nambu_{\km} = \begin{pmatrix}
        \hat{b}_{\km} \\ \hat{b}^\dag_{-\km}
    \end{pmatrix},
\end{equation}
in terms of which we have
\begin{equation} \label{eq:hamiltonian_lswt_nambu}
    \hat{H}_\text{LSWT} = {\swscale \over 2} \sum_{\km} \nambu^\dag_{\km} \left[ \pauli_0 + \gamma_{\km} \pauli_1 \right] \nambu_{\km} + \text{constant}.
\end{equation}
Here, $\pauli_j$ denote the Pauli matrices, and we defined the spin-wave energy scale $\swscale = \holepf z J S$. The above is straightforward to diagonalize via a Bogoliubov transformation,
\begin{equation} \label{eq:bogoliubov_transform_undoped}
    \nambu_{\km} = \left[  \sqrt{1 + \e_{\km} \over 2 \e_{\km}} \; \pauli_0 - \rm{sgn}(\gamma_{\km}) \sqrt{1 - \e_{\km} \over 2 \e_{\km}} \; \pauli_1 \right] \begin{pmatrix}
        \mgn_{\km} \\ \mgn^\dag_{-\km}
    \end{pmatrix},
\end{equation}
where the dimensionless dispersion is given by $\e_\km = \sqrt{1 - \gamma_{\km}^2}$. We find
\begin{equation}
    \hat{H}_\text{LSWT} = \swscale  \sum_{\km} \e_\km \; \mgn^\dag_{\km} \mgn_{\km} + \text{constant},
\end{equation}
with the usual excitation spectrum $E_{\km} = \swscale \e_{\km}$. However, this approach will no longer work at finite doping. Indeed, the dynamical interaction between holes and magnons will introduce nontrivial frequency dependence in the propagators of the latter, such that a single Bogoliubov transformation will not be able to simultaneously diagonalize the magnon Green's function at all frequencies.\\

We therefore work with a $2 \times 2$ matrix propagator for the magnons, in the original $\hat{b}$ basis. Employing the zero-temperature, real-time formalism, we will need to calculate the time-ordered Green's functions
\begin{subequations} \label{eq:greens_function_definitions}
\begin{align}
    \hpr_{\km} (t - t') &= -i \left\langle \mathcal{T} \hat{h}_{\km}(t) \hat{h}_{\km}^\dag(t') \right\rangle, \\
    \mpr_\km (t - t') &= -i \left\langle \mathcal{T} \nambu_{\km}(t) \nambu^\dag_{\km}(t') \right\rangle = -i \; \begin{pmatrix}
        \langle \mathcal{T}  \hat{b}_{\km}(t) \hat{b}_{\km}^\dag(t') \rangle & \langle \mathcal{T}  \hat{b}_{\km}(t) \hat{b}_{-\km}(t') \rangle \\
        \langle \mathcal{T}  \hat{b}^\dag_{-\km}(t) \hat{b}_{\km}^\dag(t') \rangle & \langle \mathcal{T}  \hat{b}^\dag_{-\km}(t) \hat{b}_{-\km}(t') \rangle
    \end{pmatrix},
\end{align}
\end{subequations}
or, in the frequency domain,
\begin{equation}
    \mpr_{\km}(\omega) = \int dt \; e^{i \omega t} \; \mpr_{\km} (t) ,
\end{equation}
and analogously for the holes. Note that nontrivial (anti--)commutation relations of $\hat{h}$ and respectively $\hat{b}$ operators will yield discontinuities of \eqref{eq:greens_function_definitions} at $t-t' = 0$. Specifically, we have the instantaneous correlations at $t-t' = 0^+$,
\begin{subequations} \label{eq:propagators_time_zero_plus}
    \begin{align}
        \hpr_{\km} (0^+) &= -i \left\langle \hat{h}_{\km} \hat{h}_{\km}^\dag \right\rangle, \\ 
        \mpr_\km (0^+) &= -i \; \begin{pmatrix}
        \langle \hat{b}_{\km} \hat{b}_{\km}^\dag \rangle & \langle \hat{b}_{\km} \hat{b}_{-\km} \rangle \\
        \langle \hat{b}^\dag_{-\km} \hat{b}_{\km}^\dag \rangle & \langle \hat{b}^\dag_{-\km} \hat{b}_{-\km} \rangle
        \end{pmatrix},
    \end{align}
\end{subequations}
while in the opposite case $t-t' = 0^-$, the ordering is reversed:
\begin{subequations} \label{eq:propagators_time_zero_minus}
    \begin{align}
        \hpr_{\km} (0^-) &= +i \left\langle \hat{h}_{\km}^\dag \hat{h}_{\km} \right\rangle, \\
        \mpr_\km (0^-) &= -i \; \begin{pmatrix}
        \langle \hat{b}_{\km}^\dag \hat{b}_{\km}  \rangle & \langle \hat{b}_{-\km} \hat{b}_{\km}  \rangle \\
        \langle  \hat{b}_{\km}^\dag \hat{b}^\dag_{-\km} \rangle & \langle \hat{b}_{-\km} \hat{b}^\dag_{-\km} \rangle
    \end{pmatrix}.
    \end{align}
\end{subequations}
We adopt a symmetrized convention for the zero-time value of the magnon propagator,
\begin{equation} \label{eq:symmetrization_zero_time}
    \mpr_\km (0) = {1 \over 2} \left[ \mpr_\km (0^-) + \mpr_\km (0^+) \right],
\end{equation}
which will be important when computing the Oguchi correction down the line. In order to analyze the single-particle response, it is also useful to define the spectral functions
\begin{subequations}
\begin{align}
    \hsf_{\km} (\omega) &= -{\sign (\omega - \mu) \over \pi} \imag \hpr_{\km} (\omega - \mu), \\ 
    \msf_{\km} (\omega) &= -{\sign (\omega) \over \pi} \imag \mpr_{\km} (\omega), \label{eq:magnon_spectral_function_matrix}
\end{align} 
\end{subequations}
as well as a scalar version for the magnons,
\begin{equation} \label{eq:magnon_spectral_function_scalar}
    \msfsc_{\km} (\omega) = {1 \over 2} \trace \left[ \pauli_3 \; \msf_{\km} (\omega) \right].
\end{equation}
For each momentum $\km$, both $\msfsc_{\km}(\omega)$ and $\hsf_{\km}(\omega)$ are normalized such that their frequency integral yields 1.

\subsection{Symmetry properties}
Throughout the following derivations, symmetry properties of the propagators will be helpful. We list them here, and discuss along the way why they hold true; the spirit of the proof is inductive, and for simplicity we will do it directly in the limit. Namely, we will first show that the properties hold true for the bare propagators. Afterwards, we will self-consistently prove them for the dressed ones.\\

Starting with the point group of the lattice, let $\qrot$ represent a 90-degree rotation acting on momentum,
\begin{equation} \label{eq:momentum_rotation_definition}
    \mathbf R (k_x, k_y) = (k_y, -k_x),
\end{equation}
which squares to the inversion operation: $\qrot^2 \km = -\km$. Both the hole and magnon Green's functions will be invariant under $\qrot$,
\begin{subequations} \label{eq:symm_lattice_both}
    \begin{align}
        \hpr_{\qrot \km}(\omega) &= \hpr_{\km}(\omega), \label{eq:symm_lattice_hole} \\
        \mpr_{\qrot \km}(\omega) &= \mpr_{\km}(\omega). \label{eq:symm_lattice_magnon}
    \end{align}
\end{subequations}
The other important operation is shifting the momentum by $\qafm = (\pi, \pi)$. The hole propagator, as well as the diagonal components of the magnon one, are even under this:
\begin{subequations} \label{eq:symm_hole_mag_diag_qafm}
    \begin{align}
        \hpr_{\km + \qafm}(\omega) &= \hpr_{\km}(\omega), \label{eq:symm_hole_qafm} \\
        \mpr^{\rm ab}_{\km + \qafm}(\omega) &= \mpr^{\rm ab}_{\km}(\omega) \qquad \forall a=b, \label{eq:symm_mag_diag_qafm}
    \end{align}
\end{subequations}
while the off-diagonal magnon components are odd,
\begin{equation} \label{eq:symm_mag_offdiag_qafm}
    \mpr^{\rm ab}_{\km + \qafm}(\omega) = -\mpr^{\rm ab}_{\km}(\omega) \qquad \forall a \neq b.
\end{equation}
Previously, we have defined the square lattice form factor $\gamma_\km = (\cos k_x + \cos k_y) / 2$, which can be seen to obey $\gamma_{\qrot \km} = \gamma_\km$ as well as $\gamma_{\km + \qafm} = -\gamma_\km$. It is also useful to consider the modified version $\tilde{\gamma}_\km = (\cos k_x - \cos k_y) / 2$, which instead will have $\tilde{\gamma}_{\qrot \km} = -\tilde{\gamma}_\km$ but still $\tilde{\gamma}_{\km + \qafm} = -\tilde{\gamma}_\km$. Note the convenient identity, following from the definitions of $\gamma, \tilde{\gamma}$:
\begin{equation} \label{eq:magic_identity}
    \gamma_{\km + \q} + \gamma_{\km - \q} = 2 \left( \gamma_{\km} \gamma_{\q} + \tilde{\gamma}_{\km} \tilde{\gamma}_{\q} \right).
\end{equation}
Moreover, it is straightforward to see that, for any function $f$ which is even under the $\qafm$ shift, $f_{ \km + \qafm} = f_{\km}$, and for any other momentum $\q$, we have
\begin{equation} \label{eq:symm_generic_integral_afm_even}
    \int_{\km} \gamma_{\km + \q} f_{\km} = \int_{\km} \gamma_{(\km + \qafm) + \q} f_{\km + \qafm} = \int_{\km} \left( -\gamma_{\km + \q} \right) f_{\km} = 0.
\end{equation}
Similarly, for an odd function under this shift, $g_{ \km + \qafm} = -g_{\km}$, we find
\begin{equation} \label{eq:symm_generic_integral_afm_odd}
    \int_{\km} g_{\km} = \int_{\km} g_{\km + \qafm} = \int_{\km} \left( - g_{\km} \right) = 0.
\end{equation}
Returning to the point group, for any function $h$ obeying the lattice symmetry $h_{\qrot \km} = h_{\km}$, its integral against $\tilde{\gamma}_{\km}$ vanishes, since the latter is odd under $\qrot$:
\begin{equation} \label{eq:symm_lattice_generic_integral}
    \int_{\km} \tilde{\gamma}_{\km} h_{\km} = 0.
\end{equation}
Finally, note that the dimensionless SW dispersion $\e_{\km} = \sqrt{1 - \gamma_{\km}^2}$ is invariant under both $\qrot$ and $\qafm$ shifts.

\subsection{Diagram conventions}

\begin{figure}[ht]
	\centering	
    \scalebox{\fwfigsize}{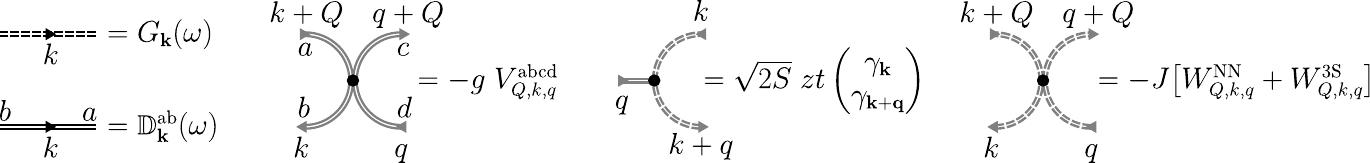}
	\caption{Propagator and vertex conventions for diagrammatic computations.}
	\label{fig:diagrammatic_conventions}
\end{figure}

Since we will employ diagrammatic perturbation theory for the calculations, we now introduce all the relevant conventions. When labeling diagrams, we use four-momentum notation $k = (\mathbf k, \omega)$. As depicted in Figure~\ref{fig:diagrammatic_conventions}, we represent hole propagators by dashed lines. Magnons are in turn denoted by solid lines, and the propagators have one Nambu index at each end. The magnon interaction \eqref{eq:hamiltonian_magnon_int} gives rise to a vertex $V^{\rm abcd}_{Q,k,q}$ with four Nambu indices, whose structure will be discussed in more detail later; to keep $V$ dimensionless, the overall energy scale is extracted as $g = {\holepf z J / 4}$, and we explicitly keep a minus sign in front to emphasize the attractive nature of the interaction. The hole-magnon vertex arising from \eqref{eq:simp_ham_coupling} has a single Nambu index, and can be directly written in vector form. Lastly, we have the two hole-hole interaction terms, eqns. \eqref{eq:simp_ham_holes} and \eqref{eq:hamiltonian_holes_corr_hop}, which we combine into the last vertex. Again keeping a minus sign to emphasize the attractive character, and taking out the dimensionful scale $J$, we have:
\begin{subequations}
\begin{align}
    W_{Q,k,q}^{\rm NN} &= z S^2 \lambda (\gamma_{\Qm} + \gamma_{\km - \q}), \\ 
    W_{Q,k,q}^{\rm 3S} &= 2 S^2 \left( \xi_{\q, \km - \q} + \xi_{\km + \Qm, \q - \km} + \xi_{\q, \Qm} + \xi_{\km + \Qm, -\Qm} \right). \label{eq:diagrammatic_vertex_holes_3s}
\end{align}
\end{subequations}
Note that writing the magnon-magnon and hole-hole interaction vertices with a single dot is shorthand for including all possible decouplings, e.g. Hartree as well as Fock in the case of the holes. This is reflected in the structure of $W_{Q,k,q}^{\rm 3S}$, eq.~\eqref{eq:diagrammatic_vertex_holes_3s}, as well as $V^{\rm abcd}_{Q,k,q}$, see Table~\ref{tab:magnon_vertex_entries}. In terms of other Feynman rules, we have one prefactor of $i$ for each appearance of the magnon-magnon and respectively hole-hole vertices, as well as for every two hole-magnon interactions. Each fermion loop gets a $-1$ as usual.

\section{Conserving calculation}

From the Hamiltonian \eqref{eq:simp_ham_total}, one can extract the bare propagators for magnons and holes. The linear spin-wave term \eqref{eq:hamiltonian_lswt_nambu} yields
\begin{equation} \label{eq:bare_magnon_propagator}
    \mpr^0_{\km}(\omega) = \Big[ [\omega + i \eta_B \sign (\omega)] \pauli_3 - \swscale [\pauli_0 + \gamma_{\km} \pauli_1] \Big]^{-1},
\end{equation}
where $\eta_B = 0^+$ is a positive infinitesimal giving the time-ordering prescription, $\pauli_0$ is the $2 \times 2$ identity matrix, and $\pauli_{1 \dots 3}$ are the Pauli matrices. The only momentum dependence comes through the $\gamma_{\km}$ factor, which is invariant under the action of $\qrot$ defined in \eqref{eq:momentum_rotation_definition}; then, the property \eqref{eq:symm_lattice_magnon} clearly follows at the level of $\mpr^0$. As for the behavior under $\qafm$ shifts, note that direct inversion of \eqref{eq:bare_magnon_propagator} yields
\begin{equation} \label{eq:bare_magnon_propagator_explicit_inverted}
    \mpr^0_{\km}(\omega) =  -{[\omega + i \eta_B \sign (\omega)] \pauli_3 + \swscale [\pauli_0 - \gamma_{\km} \pauli_1] \over \left[\swscale \e_{\km}\right]^2 - \left[\omega + i \eta_B \sign (\omega) \right]^2}.
\end{equation}
The denominator is invariant under shifting by $\qafm$ since $\e_{\km}$ is, while in the numerator we have momentum-independent coefficients for $\pauli_0$ and $\pauli_3$. So the diagonal components of $\mpr^0$ indeed obey \eqref{eq:symm_mag_diag_qafm}. On the other hand, the coefficient of $\pauli_1$ contains a single factor of $\gamma_{\km}$, which changes sign upon the $\qafm$ shift; it follows that \eqref{eq:symm_mag_offdiag_qafm} also holds true at this level.\\

Meanwhile, the holes only have a bare dispersion as a result of the three-site terms, eq.~\eqref{eq:hamiltonian_holes_corr_hop}; besides this, they can hop via \eqref{eq:simp_ham_coupling}, which however couples them to the magnons. On the other hand, the quadratic term in \eqref{eq:simp_ham_holes} only provides a correction of $-2 z J S^2 \lambda$ to the chemical potential. Defining $\tilde \mu = \mu - 2 z J S^2 \lambda$ and $\eta_F = 0^+$, we have
\begin{equation} \label{eq:bare_hole_propagator}
\hpr^0_{\km}(\omega) = \Big[ \omega - 4 J S^2 \chi_{\km} + i \eta_F \sign (\omega) + \tilde{\mu} \Big]^{-1}.
\end{equation}
Since the momentum dependence of the above arises purely from $\chi_{\km}$, which is invariant under both $\qrot$ and $\qafm$ translations, as can be seen directly from eq.~\eqref{eq:three_site_dispersion_def}, we find that \eqref{eq:symm_lattice_hole} and \eqref{eq:symm_hole_qafm} hold true at this point.\\

\begin{figure}[ht]
	\centering	
    \scalebox{\fwfigsize}{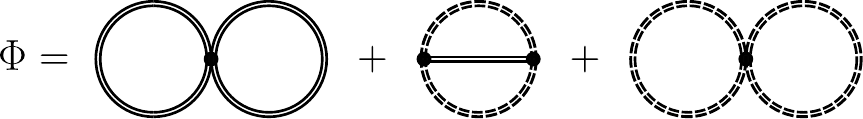}
	\caption{Luttinger-Ward functional used in the calculations.}
	\label{fig:luttinger_ward_functional}
\end{figure}

We employ the Luttinger-Ward functional to self-consistently determine dressed single-particle propagators for the magnons and holes, as well as the bimagnon Raman response. As shown in Figure \ref{fig:luttinger_ward_functional}, included are the lowest-order diagrams that can be constructed using each type of vertex introduced in the previous section.\\

\subsection{Magnon self-energy}

Begin by computing the magnon self-energy, which corresponds to cutting a single solid line from the diagrams depicted in Figure \ref{fig:luttinger_ward_functional}. The resulting contributions are presented in Figure \ref{fig:magnon_self_energy}, and can be interpreted as follows:

\begin{figure}[ht]
	\centering	
    \scalebox{\fwfigsize}{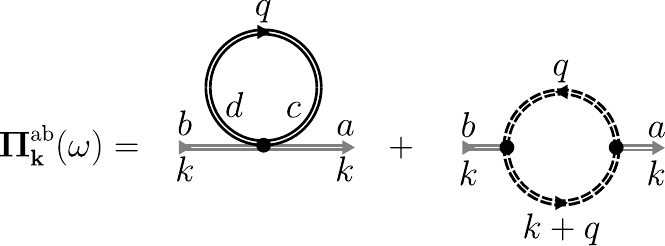}
	\caption{Diagrams for magnon self-energy.}
	\label{fig:magnon_self_energy}
\end{figure}

\begin{itemize}
    \item The effect of magnon-magnon interactions on the single-particle propagator. In the absence of doping, this is known as the Oguchi correction, and is particularly well-behaved since it only scales the overall energy of the magnon response, without modifying its shape. It can therefore be understood as a renormalization of $J = 4 t^2/U$ to $J_0 = (1 + \oco_0) J$, where in 2D we know $\oco_0 \approx 0.157$ \cite{Oguchi.1960}. We will argue that this convenient property continues to hold even in the presence of holes, although the correction will become doping-dependent, $\oco(\delta)$.
    
    \item An RPA-type contribution due to a hole polarization bubble. It encodes the kinetic effect of carrier delocalization trying to frustrate AFM order, and will be the main driver of magnon softening upon increasing the hole concentration.
\end{itemize}
Writing the self-energy as $\mse_{\km}(\omega) = \mse^\og_{\km}(\omega) + \mse^\rpa_{\km}(\omega)$, we have
\begin{subequations}
\begin{align}
    \left[ \mse^\og_{\km}(\omega) \right]^{\rm ab} &= -i g \int_{\q, \e} V^{\rm badc}_{0,k,q} \; \mpr^{\rm cd}_{\q}(\e), \\ 
    \mse^\rpa_{\km}(\omega) &= -2 i S z^2 t^2 \int_{\q, \e} \begin{pmatrix}
        \gamma_{\q} & \gamma_{\km + \q}
    \end{pmatrix} \begin{pmatrix}
        \gamma_{\q} \\ \gamma_{\km + \q}
    \end{pmatrix} \; \hpr_{\q} (\e) \; \hpr_{\km + \q} (\omega + \e), \label{eq:rpa_term_first_definition}
\end{align}
\end{subequations}
where the Einstein summation convention was used in the first line.\\

\begin{center}
\begin{table}[h]
    \centering
    \begin{tabular}{|c|c|c|c||c|}
    \hline
        $\rm a$ & $\rm b$ & $\rm c$ & $\rm d$ & $V^{\rm abcd}_{0, \km, \q}$  \\ \hline \hline
        0 & 0 & 0 & 0 &  $2 \gamma_{\km - \q} + 2 $  \\\hline
        0 & 0 & 0 & 1 &  $\gamma_{\km} + 2 \gamma_{\q}$  \\\hline
        0 & 0 & 1 & 0 &  $\gamma_{\km} + 2 \gamma_{\q}$  \\\hline
        0 & 0 & 1 & 1 &  $2 \gamma_{\km + \q } + 2 $ \\\hline
        0 & 1 & 0 & 0 &  $2\gamma_{\km} + \gamma_{\q}$  \\\hline
        0 & 1 & 0 & 1 &  $2 \gamma_{\km - \q} + 2 \gamma_{\km + \q }$  \\\hline
        0 & 1 & 1 & 0 &  $0$  \\\hline
        0 & 1 & 1 & 1 &  $2 \gamma_{\km} + \gamma_{\q }$  \\\hline
        1 & 0 & 0 & 0 &  $2 \gamma_{\km} + \gamma_{\q }$  \\\hline
        1 & 0 & 0 & 1 &  $0$  \\\hline
        1 & 0 & 1 & 0 &  $2 \gamma_{\km - \q} + 2 \gamma_{\km + \q }$  \\\hline
        1 & 0 & 1 & 1 &  $2 \gamma_{\km} + \gamma_{\q} $  \\\hline
        1 & 1 & 0 & 0 &  $2 \gamma_{\km + \q} + 2$  \\\hline
        1 & 1 & 0 & 1 &  $\gamma_{\km } + 2 \gamma_{\q} $  \\\hline
        1 & 1 & 1 & 0 &  $\gamma_{\km} + 2 \gamma_{\q} $  \\\hline
        1 & 1 & 1 & 1 &  $2 \gamma_{\km - \q} + 2 $  \\\hline
    \end{tabular}
    \caption{Magnon-magnon vertex entries, for all possible Nambu index combinations, at $\Qm = 0$. Note the symmetries $V^{\rm abcd} = V^{\rm badc}$ and $V^{\rm abcd} = V^{\rm \overline{abcd}}$, where we denote $\overline{j} = 1-j$. }
    \label{tab:magnon_vertex_entries}
\end{table}
\end{center}

Start with the Oguchi correction; to find it, we need the values of $V_{Q, k, q}$ at $\Qm = 0$, which are listed in Table \ref{tab:magnon_vertex_entries}. Since the magnon-magnon interaction is instantaneous, the vertex in principle depends only on momenta, but not on frequencies. However, extra care must be taken when two legs from the same vertex are contracted with each other, as is the case in Oguchi-type diagrams. The frequency integral over a single propagator corresponds to its value at zero time difference, 
\begin{equation}
    \int \dbar \e \; \mpr_{\km}(\epsilon) = \mpr_{\km} (0),
\end{equation}
which, as we have seen, is a point of discontinuity in \eqref{eq:greens_function_definitions} due to the nontrivial commutation relation $[\hat{b}, \hat{b}^\dag] = 1$. Since \eqref{eq:hamiltonian_magnon_int} is in fact normal-ordered, we should respect this by evaluating $\mpr^{00}_{\km} (t = 0^-)$ but $\mpr^{11}_{\km} (t = 0^+)$. The off-diagonal components do not suffer from this issue, as they are continuous at $t=0$. Therefore, we obtain
\begin{equation}
    \int \dbar \e \; V^{\rm badc}_{0,k,q} \; \mpr^{\rm cd}_{\q}(\e) = V^{\rm badc}_{0, \km, \q} \int \dbar \e \; \exp \left[ i \pauli_3^{\rm cd} \e 0^+ \right] \; \mpr^{\rm cd}_{\q}(\e).
\end{equation}
Under the symmetrized convention \eqref{eq:symmetrization_zero_time}, we have $\mpr^{00}_{\km} (0^-) = \mpr^{00}_{\km} (0) + i/2$, as well as $\mpr^{11}_{\km} (0^+) = \mpr^{11}_{\km} (0) + i/2$, so that the above simplifies to
\begin{equation} \label{eq:oguchi_simplification_1}
    V^{\rm badc}_{0, \km, \q} \int \dbar \e \; \exp \left[ i \pauli_3^{\rm cd} \e 0^+ \right] \; \mpr^{\rm cd}_{\q}(\e) = V^{\rm badc}_{0, \km, \q} \left[ {i \pauli_0^{\rm cd} \over 2} + \int \dbar \e \; \mpr^{\rm cd}_{\q}(\e) \right] = i V^{\rm abcd}_{0, \km, \q} \left[ {\pauli_0^{\rm cd} \over 2} + \imag \int \dbar \e \; \mpr^{\rm cd}_{\q}(\e) \right],
\end{equation}
where in the second step we have used the symmetry $V^{\rm badc} = V^{\rm abcd}$, as well as the fact that $\mpr_{\q}(0)$ is purely imaginary. One may now write the individual components of the Oguchi self-energy, by integrating \eqref{eq:oguchi_simplification_1} over $\q$ and multiplying by $-ig$; we will only deal with the $00$ and $01$ ones explicitly, since the remaining two follow in a completely analogous manner, owing to the symmetry of $V$:
\begin{subequations}
\begin{align}
    \left[ \mse^\og_{\km}(\omega) \right]^{\rm 00} &= g \int_{\q} \bigg[ \left( 2 + \gamma_{\km + \q} + \gamma_{\km - \q} \right) + \imag \int \dbar \e \Big[ \left( 2 \gamma_{\km - \q} + 2 \right) \mpr^{\rm 00}_{\q}(\e) \nonumber \\
    & \qquad \qquad + \left( \gamma_{\km} + 2 \gamma_{\q} \right) \left[ \mpr^{\rm 01}_{\q}(\e) + \mpr^{\rm 10}_{\q}(\e) \right] + \left( 2 \gamma_{\km + \q } + 2 \right) \mpr^{\rm 11}_{\q}(\e) \Big] \bigg], \\
    \left[ \mse^\og_{\km}(\omega) \right]^{\rm 01} &= g \int_{\q} \bigg[ \left( 2 \gamma_{\km} + \gamma_{\q} \right) + \imag \int \dbar \e \Big[ \left( 2 \gamma_{\km} + \gamma_{\q} \right) \mpr^{00}_{\q}(\e) \nonumber \\
    & \qquad \qquad  + \left( 2 \gamma_{\km - \q} + 2 \gamma_{\km + \q } \right) \mpr^{\rm 01}_{\q}(\e) + \left( 2 \gamma_{\km} + \gamma_{\q } \right) \mpr^{\rm 11}_{\q}(\e) \Big] \bigg].
\end{align}  
\end{subequations}
Momentum integrals of the form $\int_{\q} \gamma_{\km \pm \q} \; \mpr^{\rm 00 / 11}_{\q}(\e)$ vanish due to the integrand's behavior under $\qafm$ shifts, as evidenced by Eqns.~\eqref{eq:symm_mag_diag_qafm} and \eqref{eq:symm_generic_integral_afm_even}. Similarly, we have $\gamma_{\km} \int_{\q} \mpr^{\rm 01 / 10}_{\q}(\e) = 0$, via \eqref{eq:symm_mag_offdiag_qafm} and \eqref{eq:symm_generic_integral_afm_odd}. Momentum integrals over a single factor of $\gamma$ vanish by \eqref{eq:symm_generic_integral_afm_odd} as well, so we arrive at
\begin{subequations}
\begin{align}
    \left[ \mse^\og_{\km}(\omega) \right]^{\rm 00} &= 2 g \bigg[ 1 + \imag \int_{\q, \e} \Big[ \mpr^{\rm 00}_{\q}(\e) + \gamma_{\q} \left[ \mpr^{\rm 01}_{\q}(\e) + \mpr^{\rm 10}_{\q}(\e) \right] + \mpr^{\rm 11}_{\q}(\e) \Big] \bigg], \\
    \left[ \mse^\og_{\km}(\omega) \right]^{\rm 01} &= 2 g \bigg[ \gamma_{\km} + \imag \int_{\q, \e} \Big[ \gamma_{\km} \left[ \mpr^{00}_{\q}(\e) + \mpr^{\rm 11}_{\q}(\e)\right] + \left( \gamma_{\km - \q} + \gamma_{\km + \q } \right) \mpr^{\rm 01}_{\q}(\e) \Big] \bigg].
\end{align}  
\end{subequations}
In the second line, use the identity \eqref{eq:magic_identity} to rewrite the prefactor of $\mpr^{\rm 01}_{\q}(\e)$, and the lattice symmetry \eqref{eq:symm_lattice_generic_integral} to eliminate the $\tilde{\gamma}$ terms. This leaves us with $2 \gamma_{\km}  \gamma_{\q} \mpr^{\rm 01}_{\q}(\e)$, and since the frequency $\e$ is integrated over to yield instantaneous correlation functions, we can make the replacement $2 \mpr^{\rm 01}_{\q}(\e) \to \mpr^{\rm 01}_{\q}(\e) + \mpr^{\rm 10}_{\q}(\e)$. Indeed, $\ex{\hat{b} \hat{b}}$ and $\ex{\hat{b}^\dag \hat{b}^\dag}$ are complex conjugates, but can in fact be gauged to purely real, equal values; we will always do so throughout this work.\\

It then becomes clear that $\left[ \mse^\og_{\km}(\omega) \right]^{\rm 01} = \gamma_{\km} \left[ \mse^\og_{\km}(\omega) \right]^{\rm 00}$, and an identical derivation reveals $\left[ \mse^\og_{\km}(\omega) \right]^{\rm 10} = \left[ \mse^\og_{\km}(\omega) \right]^{\rm 01}$, as well as $\left[ \mse^\og_{\km}(\omega) \right]^{\rm 11} = \left[ \mse^\og_{\km}(\omega) \right]^{\rm 00}$. The conclusion is compactly expressed as
\begin{equation} \label{eq:oguchi_self_energy_result}
    \mse^\og_{\km}(\omega) = 2 g \left[ \pauli_0 + \gamma_{\km} \pauli_1 \right] \bigg[ 1 + \imag \int_{\q, \e} \trace \Big[ \left[ \pauli_0 + \gamma_{\q} \pauli_1 \right] \mpr_{\q}(\e) \Big] \bigg],
\end{equation}
and by comparison with \eqref{eq:bare_magnon_propagator}, we extract the doping-dependent Oguchi correction to $J$:
\begin{equation} \label{eq:oguchi_result_general}
    \oco(\delta) = {1 \over 2S} \bigg[ 1 + \imag \int_{\q, \e} \trace \Big[ \left[ \pauli_0 + \gamma_{\q} \pauli_1 \right] \mpr_{\q}(\e) \Big] \bigg].
\end{equation}
An alternative form can be found by relating the frequency integral of the magnon propagator to the instantaneous correlation functions via \eqref{eq:propagators_time_zero_plus}--\eqref{eq:symmetrization_zero_time},
\begin{equation} \label{eq:oguchi_result_corr}
    \oco(\delta) = {1 \over 2S} \left[1 - \int_{\q} \left[ \ex{\hat{b}^\dag_{\q} \hat{b}_{\q}} + \ex{\hat{b}_{-\q} \hat{b}^\dag_{-\q}} + \gamma_{\q} \left( {\ex{\hat{b}_{ - \q} \hat{b}_{\q}} + \ex{\hat{b}_{\q}^\dag \hat{b}_{-\q}^\dag}} \right) \right] \right].
\end{equation}
As a check, we can consider the zero-doping case. From the Bogoliubov approach \eqref{eq:bogoliubov_transform_undoped} we know
\begin{subequations}
    \begin{align}
        \ex{\hat{b}^\dag_{\q} \hat{b}_{\q}}_0 + \ex{\hat{b}_{-\q} \hat{b}^\dag_{-\q}}_0 &= 2 v_{\q}^2 = {1 \over \e_{\q}}, \\
        \ex{\hat{b}_{-\q} \hat{b}_{\q}}_0 + \ex{\hat{b}_{\q}^\dag \hat{b}_{-\q}^\dag}_0 &= 2 u_{\q} v_{\q} = -{\gamma_{\q} \over \e_{\q}},
    \end{align}
\end{subequations}
which upon plugging back into \eqref{eq:oguchi_result_corr} and using $1 - \gamma_{\q}^2 = \e_{\q}^2$, yields the familiar expression
\begin{equation} \label{eq:oguchi_result_undoped}
    \oco(0) = {1 \over 2S} \left[1 - \int_{\q} \e_{\q} \right].
\end{equation}
Now consider the RPA term \eqref{eq:rpa_term_first_definition}. Since the $\gamma$ factors have no frequency dependence, we can rearrange it as
\begin{equation} \label{eq:magnon_self_energy_rpa}
    \mse^\rpa_{\km}(\omega) = -2 i S z^2 t^2 \int_{\q} \begin{pmatrix}
        \gamma_{\q}^2 & \gamma_{\q} \gamma_{\km + \q} \\
        \gamma_{\q} \gamma_{\km + \q} & \gamma_{\km + \q}^2
    \end{pmatrix} \int \dbar \e \; \hpr_{\q} (\e) \; \hpr_{\km + \q} (\omega + \e).
\end{equation}
Its matrix structure can be highlighted by writing 
\begin{equation} \label{eq:magnon_rpa_pauli_separation}
    \mse^\rpa_{\km} (\omega) = \Pi^0_{\km}(\omega) \; \pauli_0 + \Pi^1_{\km}(\omega) \; \pauli_1 + \Pi^3_{\km}(\omega) \; \pauli_3,
\end{equation}
with the corresponding coefficients
\begin{subequations} \label{eq:magnon_rpa_pauli_coefficients}
\begin{align}
    \Pi^0_{\km}(\omega) &= - i S z^2 t^2 \int_{\q} \left( \gamma_{\q}^2 + \gamma_{\km + \q}^2 \right) \int \dbar \e \; \hpr_{\q} (\e) \; \hpr_{\km + \q} (\omega + \e), \\
    \Pi^1_{\km}(\omega) &= - i S z^2 t^2 \int_{\q} 2 \gamma_{\q} \gamma_{\km + \q}  \int \dbar \e \; \hpr_{\q} (\e) \; \hpr_{\km + \q} (\omega + \e), \\
    \Pi^3_{\km}(\omega) &= - i S z^2 t^2 \int_{\q} \left( \gamma_{\q}^2 - \gamma_{\km + \q}^2 \right) \int \dbar \e \; \hpr_{\q} (\e) \; \hpr_{\km + \q} (\omega + \e).
\end{align}
\end{subequations}

\subsection{Hole self-energy}

We turn to the corresponding calculation for the holes. Cutting one dashed line from the Luttinger-Ward functional of Fig. \ref{fig:luttinger_ward_functional}, we arrive at the two diagrams depicted in Fig. \ref{fig:hole_self_energy}, with the following physical interpretations:
\begin{figure}[ht]
	\centering	
    \scalebox{\fwfigsize}{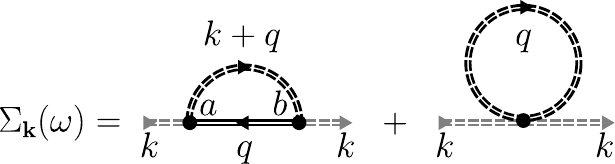}
	\caption{Diagrams for hole self-energy.}
	\label{fig:hole_self_energy}
\end{figure}
\begin{itemize}
    \item Scattering of a hole to different momentum states, via the emission / absorption of a magnon, through the coupling given by the kinetic term \eqref{eq:simp_ham_coupling} in the Hamiltonian. This describes the tendency of the hole to delocalize in the lattice, working against the energy cost of frustrating the AFM background. This term captures the essential polaronic physics which arises from the competition of kinetic and exchange terms in the $t-J$ model.
    \item The effect of hole-hole interactions, as described in eqns. \eqref{eq:simp_ham_holes} and \eqref{eq:hamiltonian_holes_corr_hop}. We will see that this only leads to a renormalization of the bare hole dispersion by $(1-\delta)$, as well as some doping-dependent chemical potential corrections. The latter are completely irrelevant for this work, as we always consider fixed doping.
\end{itemize}
We separate the contributions as $\hse_{\km}(\omega) = \hse^\scba_{\km}(\omega) + \hse^\hnn_{\km}(\omega) + \hse^{\rm 3S}_{\km}(\omega)$, and write for the kinetic term
\begin{equation} \label{eq:hole_self_energy_scba}
    \hse^\scba_{\km}(\omega) = 2 iS z^2 t^2 \int_{\q, \e} \hpr_{\km + \q}(\omega + \e) \; \begin{pmatrix}
        \gamma_{\km} & \gamma_{\km + \q}
    \end{pmatrix} \mpr_{\q}(\e) \begin{pmatrix}
        \gamma_{\km} \\ \gamma_{\km + \q}
    \end{pmatrix}. 
\end{equation}
For the hole-hole interactions, one must take into account the factor of two arising from the possible ways of contracting a pair of legs from the vertex. The nearest-neighbor interaction $- J \; W_{Q,k,q}^{\rm NN}$ yields a self-energy
\begin{align}
    \hse^\hnn_{\km}(\omega) &= 2 i z J S^2 \lambda \int_{\q, \e} (\gamma_0 + \gamma_{\km - \q}) \; \hpr_{\q}(\e),
\end{align}
which can in turn be understood to consist of a Hartree contribution, arising from the $\gamma_0 \equiv 1$ factor, and a Fock one, due to the $\gamma_{\km - \q}$ term. The latter will vanish due to the $\qafm$ shift properties \eqref{eq:symm_hole_qafm} and \eqref{eq:symm_generic_integral_afm_even}, while the former has no $\km$ or $\omega$ dependence and therefore renormalizes the chemical potential:
\begin{align} \label{eq:hole_self_energy_hartree_mu}
    \hse^\hnn &= 2 i z J S^2 \lambda \int_{\q, \e} \hpr_{\q}(\e) = -2 z J S^2 \lambda \delta.
\end{align}
Meanwhile, the three-site interaction $- J \; W_{Q,k,q}^{\rm 3S}$ gives 
\begin{align}
    \hse^{\rm 3S}_{\km}(\omega) &= 4 i J S^2 \int_{\q, \e} \hpr_{\q}(\e) \left( \xi_{\q, \km - \q} + \xi_{\km, \q - \km} + \xi_{\q, 0} + \xi_{\km, 0} \right) ,
\end{align}
and we also aim to simplify this using the $\qafm$ shift properties. From the definition \eqref{eq:three_site_interaction_form_def}, it follows that $\xi$ is invariant under $\qafm$ shifts of its first argument, but changes sign under shifting the second. Therefore, upon $\q \to \q + \qafm$, the terms involving $\xi_{\q, \km - \q}$ and $\xi_{\km, \q - \km}$ will flip sign, making their integrals vanish. Meanwhile, the contribution involving $\xi_{\q, 0}$ has no $\km$ or $\omega$ dependence, so it again renormalizes the chemical potential. By comparing \eqref{eq:three_site_dispersion_def} and \eqref{eq:three_site_interaction_form_def}, we see that $\xi_{\q, 0} \equiv \chi_\q$, so we find
\begin{equation} \label{eq:hole_self_energy_3s_hartree_mu}
    \hse^{\rm 3S}_{\mu} = 4 i J S^2 \int_{\q, \e} \chi_{\q} \; \hpr_{\q}(\e).
\end{equation}
Finally, the $\xi_{\km, 0} \equiv \chi_{\km}$ term is independent of $\q$ and $\e$, so it can be taken out of the integral. It yields a simple doping-dependent renormalization to the bare hole dispersion,
\begin{equation}
    \hse^{\rm 3S}_{\chi; \km} = 4 i J S^2 \chi_{\km} \int_{\q, \e} \hpr_{\q}(\e) = - 4 J S^2 \chi_{\km} \; \delta.
\end{equation}

\subsection{Single-particle propagators}
We may now collect all the results above, and write self-consistent expressions for the dressed propagators of holes and magnons. Incorporating \eqref{eq:hole_self_energy_hartree_mu} and \eqref{eq:hole_self_energy_3s_hartree_mu} into a doping-dependent effective chemical potential
\begin{equation}
    \tilde{\mu}(\delta) = \mu - 2 z J S^2 \lambda (1 - \delta) - \hse^{\rm 3S}_{\mu},
\end{equation}
and recalling that the Oguchi correction \eqref{eq:oguchi_self_energy_result} can be understood as an overall renormalization \eqref{eq:oguchi_result_general} of $J$, we find
\begin{subequations} \label{eq:full_single_particle_propagators}
    \begin{align}
    \label{eq:full_hole_propagator}
    \hpr_{\km}(\omega) &= \Big[ \omega - 4 J S^2 (1 - \delta) \chi_{\km} + i \eta_F \sign (\omega)  + \tilde{\mu}(\delta) - \hse^\scba_{\km}(\omega) \Big]^{-1},  \\ \label{eq:full_magnon_propagator}
    \mpr_{\km}(\omega) &= \Big[ [\omega + i \eta_B \sign (\omega)] \pauli_3 - \swscale \left[ 1 + \oco(\delta) \right] \left[ \pauli_0 + \gamma_{\km} \pauli_1 \right] - \mse^\rpa_{\km}(\omega) \Big]^{-1}.
    \end{align}
\end{subequations}
Starting from the noninteracting initial guesses \eqref{eq:bare_magnon_propagator} and \eqref{eq:bare_hole_propagator}, we employ an iterative approach to self-consistently solve the system formed by Eqns.~\eqref{eq:oguchi_result_general}, \eqref{eq:magnon_self_energy_rpa}, \eqref{eq:hole_self_energy_scba}, and \eqref{eq:full_single_particle_propagators}. Namely, at the $n^{\rm th}$ step, we start with a set of guesses $\hpr^{(n)}, \mpr^{(n)}$ for the propagators. Based on these, we find the self-energies $\hse^{(n)}, \oco^{(n)}, \mse^{(n)}$. From the new self-energies, we compute new propagators $\tilde{\hpr}^{(n)}, \tilde{\mpr}^{(n)}$ using Eqns.~\eqref{eq:full_single_particle_propagators}. At this point, we adjust the effective chemical potential $\tilde{\mu}$ in order to fix the doping $\delta$ to the desired value. Then, we pick the new propagator guesses to be fed into the next step as linear combinations of the old guess and the new result: $\hpr^{(n + 1)} = \theta \; \tilde{\hpr}^{(n)} + (1 - \theta) \; \hpr^{(n)}$, and respectively $\mpr^{(n + 1)} = \theta \; \tilde{\mpr}^{(n)} + (1 - \theta) \; \mpr^{(n)}$. The weight $\theta$ is dynamically chosen in order to accelerate convergence, based on the difference between $\tilde{\hpr}^{(n)}$ and $\hpr^{(n)}$, and respectively between $\tilde{\mpr}^{(n)}$ and $\mpr^{(n)}$. These differences, together with changes in $\tilde{\mu}$, are also used to track solver convergence.\\

\begin{figure}[ht]
	\centering	
    \scalebox{0.4}{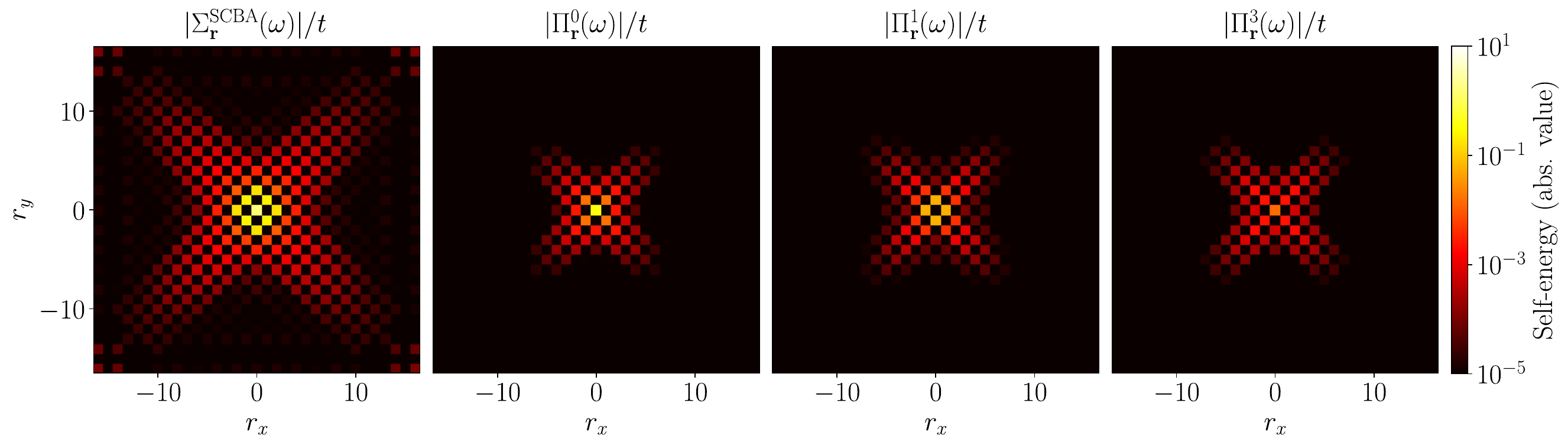}
	\caption{Absolute values of self-energies $|\hse_{\mathbf r}^{\rm SCBA} (\omega)|$ and $|\mse_{\mathbf r}^{\rm RPA} (\omega)|$ versus spatial separation $\mathbf r$, at frequency $\omega=t/2$, doping $\delta = 7\%$, interaction strength $U/t = 8$, and broadenings $\eta_F / t = 0.2$ respectively $\eta_B / t = 0.025$. Note that the color scale is logarithmic, highlighting the exponential decay of the self-energies with the separation $\mathbf r$.}
	\label{fig:sm_self_energies}
\end{figure}

All propagators and self-energies are sampled on a frequency grid of width $30t$ and step size $\Delta \omega / t = 5 \times 10^{-3}$. Momentum space is discretized using a $32 \times 32$ or $64 \times 64$ grid, and the converged results are later extended to larger system sizes, if necessary for detailed plots. This extension works as follows: from the converged solution of the self-consistent equations, we take the self-energies $\hse^{\rm SCBA}, \mse^{\rm RPA}$ and Fourier transform them in the spatial directions to obtain $\hse_{\mathbf r}^{\rm SCBA} (\omega), \mse_{\mathbf r}^{\rm RPA} (\omega)$. With respect to $\mathbf r$, the self-energies are finite-range; namely, they exponentially decay over characteristic length scales on the order of 10 sites (see the example in Fig.~\ref{fig:sm_self_energies}). With square lattice sizes of $32$ or $64$, the self-energies at the real-space grid edge are negligible. We may then further pad $\hse^{\rm SCBA}$ and $\mse^{\rm RPA}$ with zeros up to any desired size for plotting (e.g. 128 or 256), Fourier transform back to momentum space, and calculate the corresponding propagators.\\

\begin{figure}[ht]
	\centering	
    \scalebox{0.42}{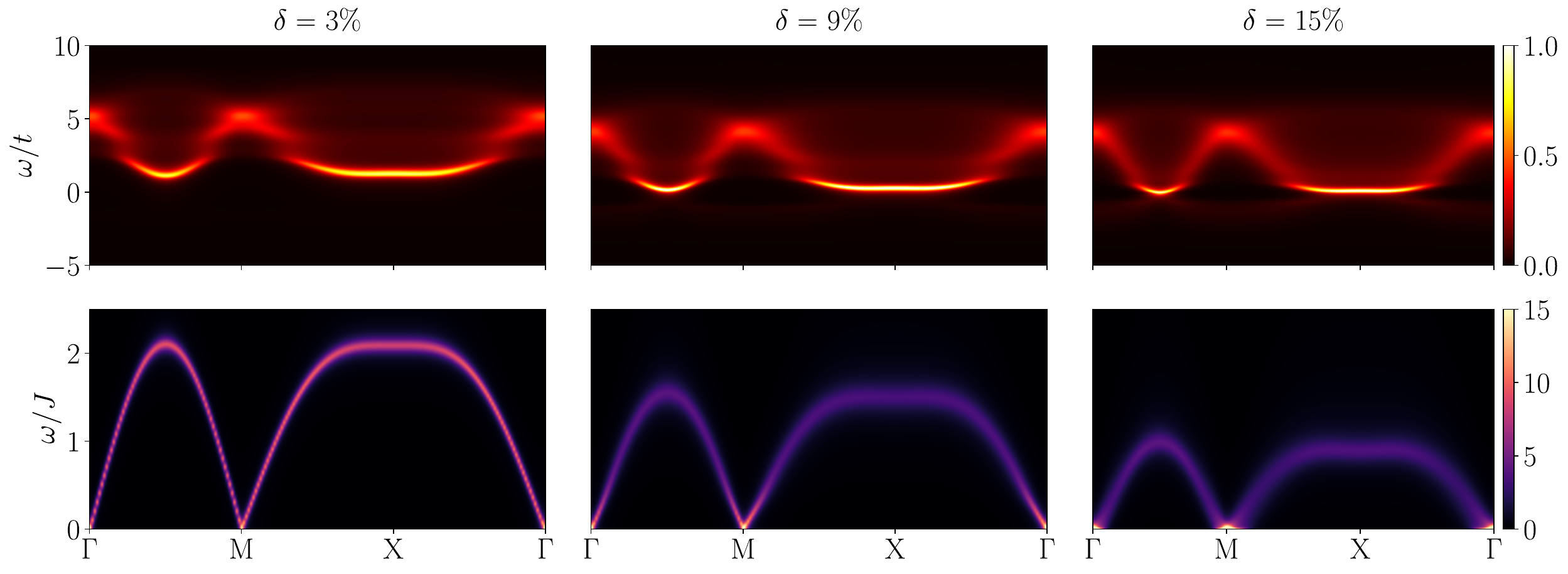}
	\caption{Typical single-particle spectral functions for holes (top row) and magnons (bottom row), at $U/t = 7$ and three different doping values $\delta \in \{3\%, 9\%, 15\%\}$. Calculations performed for a $64 \times 64$ square lattice, on a frequency grid with spacing $\Delta \omega / t = 5 \times 10^{-3}$. Broadenings are $\eta_B/t = 0.025$ and $\eta_F / t = 0.25$, the latter being necessary for convergence at high dopings. We observe gradual closing of the gap in the polaron dispersion upon doping, as the magnons soften. Note also the appearance at high doping of hole spectral weight below the conventional polaron band bottom, as well as the broadening of magnons.}
	\label{fig:sm_spectral_functions}
\end{figure}

We note that, as seen in the main text, our self-consistent approach leads to nonperturbative results which capture the essential polaronic physics at low doping, as well as its modifications manifesting at higher dopant densities. Typical results for hole and magnon spectral functions, at $U/t = 7$ and along a $\Gamma \to \rm M \to X \to \Gamma$ cut in the BZ, are presented in Fig.~\ref{fig:sm_spectral_functions}.\\

It remains to argue that the symmetry properties \eqref{eq:symm_lattice_both}~--~\eqref{eq:symm_mag_offdiag_qafm} indeed hold for the dressed propagators \eqref{eq:full_single_particle_propagators}. With regard to holes, the momentum dependence arising from $\chi_{\km}$ has already been discussed. The only new feature is the dependence due to $\hse^\scba_{\km}(\omega)$, which obeys
\begin{align}
     \hse^\scba_{\qrot \km}(\omega) &= 2 iS z^2 t^2 \int_{\q, \e} \hpr_{\qrot \km + \q}(\omega + \e) \; \begin{pmatrix}
        \gamma_{\qrot \km} & \gamma_{\qrot \km + \q}
    \end{pmatrix} \mpr_{\q}(\e) \begin{pmatrix}
        \gamma_{\qrot \km} \\ \gamma_{\qrot \km + \q}
    \end{pmatrix} \nonumber \\
    &= 2 iS z^2 t^2 \int_{\q, \e} \hpr_{\qrot (\km + \q)}(\omega + \e) \; \begin{pmatrix}
        \gamma_{\qrot \km} & \gamma_{\qrot (\km + \q)}
    \end{pmatrix} \mpr_{\qrot \q}(\e) \begin{pmatrix}
        \gamma_{\qrot \km} \\ \gamma_{\qrot (\km + \q)}
    \end{pmatrix}, \label{eq:scba_lattice_symmetry_self_consistent_argument}
\end{align}
where in going from the first to the second line we have changed the integration variable $\q \to \qrot \q$. All $\gamma$ factors and propagators are assumed to be invariant under $\qrot$, so the RHS indeed reduces to $\hse^\scba_{\km}(\omega)$, which self-consistently results in \eqref{eq:symm_lattice_hole}. Under shifting $\km \to \km + \qafm$, we use the change in sign of $\gamma$ to write
\begin{equation}
     \hse^\scba_{\km + \qafm}(\omega) = 2 iS z^2 t^2 \int_{\q, \e} \hpr_{\km + \qafm + \q}(\omega + \e) \; \begin{pmatrix}
        -\gamma_{\km} & -\gamma_{\km + \q}
    \end{pmatrix} \mpr_{\q}(\e) \begin{pmatrix}
        -\gamma_{\km} \\ -\gamma_{\km + \q}
    \end{pmatrix},
\end{equation}
so the sign flips due to $\gamma$ factors cancel out, and under the assumption that $\hpr$ is invariant, so is $\hse^\scba$; the property \eqref{eq:symm_hole_qafm} follows.\\

Moving on to the magnons, we note that the scalar description of the Oguchi correction, via $\oco(\delta)$, does not affect the symmetry properties at all. We consider instead $\mse^\rpa$, which is clearly invariant under the action of $\qrot$ by the same argument as \eqref{eq:scba_lattice_symmetry_self_consistent_argument}: all of the propagators and form factors entering its expression are not affected by $\qrot$. With \eqref{eq:symm_lattice_magnon} settled, it remains to focus on the shift by $\qafm$, which is slightly complicated by the matrix structure of $\mpr$. Investigation of \eqref{eq:magnon_rpa_pauli_coefficients} shows that $\Pi^0_{\km}(\omega)$ and $\Pi^3_{\km}(\omega)$ are even, while $\Pi^1_{\km}(\omega)$ is odd under this shift. Indeed, all the hole propagators are assumed to be even, and $\gamma_{\q}^2 \pm \gamma_{\km + \q}^2$ is also invariant due to the squaring of individual $\gamma$ factors. On the other hand, $\gamma_{\q} \gamma_{\km + \q}$ is odd since the shift only applies to $\km$. Without inverting \eqref{eq:full_magnon_propagator} explicitly, we can still see that the diagonal entries in $[\mpr_{\km}(\omega)]^{-1}$ are even under the $\qafm$ shift, while the off-diagonal ones are odd. By the same argument as for the bare propagator, then, we arrive at the properties \eqref{eq:symm_mag_diag_qafm} and \eqref{eq:symm_mag_offdiag_qafm}.

\section{Bimagnon spectrum and Raman-type probing}

In this section, we compute the bimagnon Raman response at finite doping, consistent with the single-particle behavior found in the previous part. We also describe a simple estimate of the charge contribution to Raman-type spectra, which features suppressed response below a threshold set by $J^*$. \\

We apply the d-wave perturbation
\begin{equation} \label{eq:lf_vertex_real_space}
    \hat{H}_\text{LF} \propto \sum_{n} \vec S_{n} \cdot (\vec S_{n + \hat y} + \vec S_{n - \hat y} - \vec S_{n + \hat x} - \vec S_{n - \hat x}),
\end{equation}
and also measure the same correlation function. This corresponds to the B$_{1g}$ channel in optical Raman measurements, which for e.g. cuprates displays the usual bimagnon peak. With the sublattice rotation and auxiliary-boson transformation, we found in eq.~\eqref{eq:raman_vetex_bimagnon} the vertex
\begin{equation} \label{eq:lf_vertex_k_space}
    \hat{V}_{\rm 2M} = 4 \alpha_h z S \; {t  \over U} \; \Delta t \sum_{\mathbf k} \tilde{\gamma}_{\mathbf k} \left( \hat{b}_{\mathbf k} \hat{b}_{-\mathbf k} + \hat{b}^\dag_{\mathbf k} \hat{b}^\dag_{-\mathbf k} \right).
\end{equation}
Compared to the usual spin-wave Hamiltonian \eqref{eq:hamiltonian_lswt}, which had $s-$wave symmetry, the $d-$wave nature of \eqref{eq:lf_vertex_real_space} leads to the cancellation of diagonal $\hat{b}^\dag_{\km} \hat{b}_{\km}$ terms, as well as to the replacement $\gamma_{\km} \to \tilde \gamma_{\km}$. Diagrammatically, this gives rise to a Raman vertex $\bsv_0$, creating or annihilating a pair of magnons at zero total momentum. Making for brevity the notation $\rampr = 4 \alpha_h z S t \Delta t / U$, we have the Nambu structure
\begin{equation} \label{eq:raman_vertex_bare_explicit}
    [\bsv_0]_{\km}^{\rm ab} = \rampr  \begin{pmatrix}
        0 & \tilde \gamma_{\km} \\
        \tilde \gamma_{\km} & 0
    \end{pmatrix}^{\rm ab} = \rampr \tilde \gamma_{\km} \tau_1^{\rm ab}.
\end{equation}

\subsection{Ladder Bethe-Salpeter equation}

\begin{figure}[ht]
	\centering	
    \scalebox{\fwfigsize}{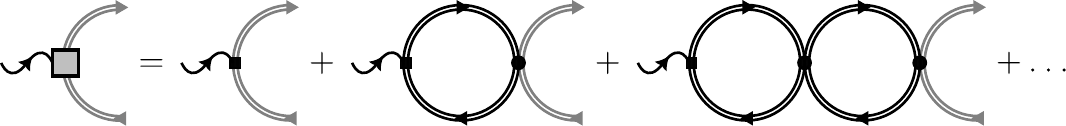}
	\caption{Definition of the dressed Raman vertex for bimagnons.}
	\label{fig:raman_dressed_vertex}
\end{figure}

\begin{figure}[ht]
	\centering	
    \scalebox{\fwfigsize}{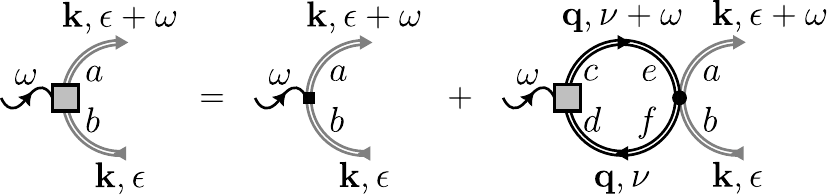}
	\caption{Bethe-Salpeter equation for the dressed bimagnon Raman vertex.}
	\label{fig:bethe_salpeter_vertex}
\end{figure}

It is well-known that magnon-magnon interactions are crucial for recovering the correct position of the bimagnon peak in Raman response. In consequence, we also define a dressed vertex $\bsv$, shown in Figure \ref{fig:raman_dressed_vertex}, which contains the corrections resulting from such interactions. Using the ladder approximation for the Bethe-Salpeter equation governing this vertex, depicted in Figure \ref{fig:bethe_salpeter_vertex}, we arrive at
\begin{equation} \label{eq:bethe_salpeter_0}
    \bsv_{\km}^{\rm ab} (\omega, \e) = [\bsv_0]_{\km}^{\rm ab}(\omega, \e) - i g \int_{\q, \nu} \bsv_{\q}^{\rm cd} (\omega, \nu) \; \mpr_{\q}^{\rm df}(\nu) \; \mpr_{\q}^{\rm ec}(\nu + \omega) \; V^{\rm efab}_{0, \q, \km}.
\end{equation}
Here, we have already used the fact that magnon-magnon interactions $V^{\rm efab}_{0, \q, \km}$ are instantaneous, and therefore do not depend on any of the frequencies associated with their legs. The bare vertex $\bsv_0$, given in \eqref{eq:raman_vertex_bare_explicit}, is also frequency-independent. It follows that the RHS of \eqref{eq:bethe_salpeter_0} does not contain any $\e$, and with regard to the LHS, this means that the dressed vertex $\bsv$ only depends on the external frequency $\omega$. We arrive at
\begin{equation} \label{eq:bethe_salpeter_reference}
    \bsv_{\km}^{\rm ab} (\omega) = [\bsv_0]_{\km}^{\rm ab}(\omega) - i g \int_{\q} V^{\rm efab}_{0, \q, \km} \; \bsv_{\q}^{\rm cd} (\omega) \int \dbar \nu \; \mpr_{\q}^{\rm df}(\nu) \; \mpr_{\q}^{\rm ec}(\nu + \omega),  
\end{equation}
and for simplicity make the following notation of the magnon bubble frequency integral:
\begin{equation}
    \mb_{\q}^{\rm cdef}(\omega) = \swscale \int \dbar \nu \; \mpr_{\q}^{\rm df}(\nu) \; \mpr_{\q}^{\rm ec}(\nu + \omega).
\end{equation}
One factor of the spin-wave energy scale $\swscale$ has been included in the definition, to keep $\mb_{\q}$ dimensionless. In turn, we define the dimensionless magnon-magnon coupling constant,
\begin{equation}
    \tilde{g} = {g \over \swscale} = {1 \over 4S}.
\end{equation}
We must now deal with the momentum integral in \eqref{eq:bethe_salpeter_reference}. The important observation is that the bare vertex $[\bsv_0]_\km$ has $d-$wave symmetry with respect to $\km$, and the interaction $V_{0, \q, \km}$ does not mix different symmetry channels. The magnon bubble $\mb_{\q}^{\rm cdef}$ is invariant under the action of $\qrot$ because each magnon propagator is, and it is also even under the $\qafm$ shift since it contains two magnon propagtors. As $\mb_{\q}$ does not affect symmetry properties of $\bsv$, it follows that we can separate the momentum dependence of the dressed Raman vertex as
\begin{equation} \label{eq:dressed_raman_vertex_k_dependence}
    \bsv_{\km}^{\rm ab}(\omega) = \rampr \tilde \gamma_{\km} \bsvf^{\rm ab}(\omega).
\end{equation}
The rigorous proof is, just as for the symmetry properties of single-particle propagators, achieved by induction. The base case is the bare vertex \eqref{eq:raman_vertex_bare_explicit}, which of course obeys the desired property. One may then define a sequence $\bsv_n$ of vertices dressed with $n$ magnon interactions, using an equation akin to \eqref{eq:bethe_salpeter_reference} to relate $\bsv_{n+1}$ to $\bsv_n$. To avoid such steps which bring no additional enlightenment, we again work directly in the limit \eqref{eq:bethe_salpeter_reference}. Plugging \eqref{eq:raman_vertex_bare_explicit} and \eqref{eq:dressed_raman_vertex_k_dependence} into the RHS, we find
\begin{equation} \label{eq:bethe_salpeter_1}
    \bsv_{\km}^{\rm ab} (\omega) = \rampr \tilde \gamma_{\km} \tau_1^{\rm ab} - i \tilde{g} \rampr \bsvf^{\rm cd}(\omega) \int_{\q} V^{\rm efab}_{0, \q, \km} \; \tilde \gamma_{\q} \; \mb_{\q}^{\rm cdef}(\omega),
\end{equation}
and we need to plug in the vertex $V^{\rm efab}_{0, \q, \km}$. Individual entries are already presented in Table \ref{tab:magnon_vertex_entries}, and the information can be reformulated as
\begin{equation} \label{eq:magnon_interaction_symmetry_decomposition}
    V^{\rm efab}_{0, \q, \km} = \mmvc^{\rm efab} + \gamma_{\km} \mmvk^{\rm efab} + \gamma_{\q} \mmvq^{\rm efab} + {1 \over 2} (\gamma_{\km + \q} - \gamma_{\km - \q}) \mmva^{\rm efab} + {1 \over 2} (\gamma_{\km + \q} + \gamma_{\km - \q}) \mmvs^{\rm efab},
\end{equation}
where each of $\{\mmvc, \mmvk, \mmvq, \mmva, \mmvs \}$ is a constant tensor with four Nambu indices. Upon integrating over $\q$ in \eqref{eq:bethe_salpeter_1}, all but the last of these terms will give vanishing contributions. Indeed, recalling that $\mb_{\q}$ is invariant under all relevant symmetry operations, 
\begin{itemize}
    \item the $\mmvc$ and $\gamma_{\km} \mmvk$ terms have no $\q$ dependence and can be taken out of the integral, leaving the integrand as $\tilde \gamma_{\q} \; \mb_{\q}$. Since this is odd under the action of $\qrot$, the result is zero;
    \item from the term containing $\mmvq$, an additional $\gamma_\q$ remains in the integrand. However, this is even under $\qrot$ and the rest is still odd, so it will vanish;
    \item for the contribution with $\mmva$, note that $(\gamma_{\km + \q} - \gamma_{\km - \q})$ is odd under inversion of $\q$, i.e. two actions of $\qrot$. But the rest of the integrand is even under this, so we again get zero.
\end{itemize}
\begin{center}
\begin{table}[h]
    \centering
    \begin{tabular}{|c|c|c|c||c|}
    \hline
        $\rm a$ & $\rm b$ & $\rm c$ & $\rm d$ & $\mmvs^{\rm abcd}$  \\ \hline \hline
        0 & 0 & 0 & 0 &  2  \\\hline
        0 & 0 & 1 & 1 &  2  \\\hline
        0 & 1 & 0 & 1 &  4  \\\hline
        1 & 0 & 1 & 0 &  4  \\\hline
        1 & 1 & 0 & 0 &  2  \\\hline
        1 & 1 & 1 & 1 &  2  \\\hline
    \end{tabular}
    \caption{Individual nonzero components of the constant tensor $\mmvs$, which governs the d-wave bimagnon states.}
    \label{tab:s_tensor_explicit}
\end{table}
\end{center}
The $\mmvs$ tensor is explicitly presented in Table \ref{tab:s_tensor_explicit}. We indeed see that only its contribution in \eqref{eq:magnon_interaction_symmetry_decomposition} survives integration, and furthermore can be re-expressed using the identity \eqref{eq:magic_identity} as
\begin{equation}
    {1 \over 2} (\gamma_{\km + \q} + \gamma_{\km - \q}) \mmvs^{\rm efab} = \left( \gamma_{\km} \gamma_{\q} + \tilde{\gamma}_{\km} \tilde{\gamma}_{\q} \right) \mmvs^{\rm efab},
\end{equation}
and the contribution of $\gamma_{\km} \gamma_{\q}$ will also vanish by the same arguments as above. We obtain
\begin{equation} \label{eq:bethe_salpeter_2}
    \bsv_{\km}^{\rm ab} (\omega) = \rampr \tilde \gamma_{\km} \left[ \tau_1^{\rm ab} - i \tilde{g} \bsvf^{\rm cd}(\omega) \; \mmvs^{\rm efab} \int_{\q} \tilde{\gamma}_{\q}^2 \; \mb_{\q}^{\rm cdef}(\omega) \right],
\end{equation}
which not only completes the proof of \eqref{eq:dressed_raman_vertex_k_dependence}, but also yields the self-consistency equation for $\bsvf$. Making the notation
\begin{equation}
    \lfn^{\rm cdef}(\omega) = \int_{\q} \tilde{\gamma}_{\q}^2 \; \mb_{\q}^{\rm cdef}(\omega),
\end{equation}
we obtain
\begin{equation} \label{eq:bethe_salpeter_3}
    \bsvf^{\rm ab} (\omega) = \tau_1^{\rm ab} - i \tilde{g} \bsvf^{\rm cd}(\omega) \lfn^{\rm cdef}(\omega) \mmvs^{\rm efab} .
\end{equation}
The remaining linear algebra problem is straightforward to solve if we define the flattened indices
\begin{subequations}
\begin{align}
    x &\to (a,b), \\
    y &\to (c,d), \\
    z &\to (e,f),
\end{align}
\end{subequations}
in terms of which $\bsvf$ and $\pauli_1$ become vectors, while $\lfn$ and $\mmvs$ matrices. We have
\begin{equation} \label{eq:bethe_salpeter_4}
    \bsvf^{\rm x} (\omega) = \tau_1^{\rm x} - i \tilde{g} \bsvf^{\rm y}(\omega) \lfn^{\rm yz}(\omega) \mmvs^{\rm zx},
\end{equation}
with the solution
\begin{equation} \label{eq:dressed_raman_vertex_solution}
    \bsvf(\omega) = \tau_1 \left[ \id + i \tilde{g} \lfn(\omega) \mmvs \right]^{-1}.
\end{equation}

\subsection{Bimagnon response}

Knowing the dressed vertex $\bsv_{\km} (\omega)$, we may now extract the two-magnon spectrum as $-\imag \ramanbm(\omega)$, where $\ramanbm(\omega)$ is the correlation function depicted in Fig. \ref{fig:bimagnon_spectrum}:
\begin{figure}[ht]
	\centering	
    \scalebox{\fwfigsize}{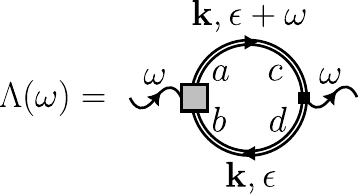}
	\caption{Bimagnon Raman spectrum, as a function of the dressed vertex.}
	\label{fig:bimagnon_spectrum}
\end{figure}
\begin{equation}
    \ramanbm(\omega) = i \int_{\km, \e} \bsv_{\km}^{\rm ab} (\omega) \; [\bsv_0^*]_{\km}^{\rm cd} \; \mpr_{\km}^{\rm bd} (\e) \; \mpr_{\km}^{\rm ca} (\e + \omega).
\end{equation}
Using the notation from before, this simplifies to
\begin{equation}
    \ramanbm(\omega) = i \; {\rampr^2 \over \Omega} \; \bsvf^{\rm ab}(\omega) \left[ \int_{\km} \tilde \gamma_{\km}^2 \; \mb_{\km}^{\rm abcd}(\omega) \right] \tau_1^{\rm cd} = i \; {\rampr^2 \over \Omega} \; \bsvf(\omega) \lfn(\omega) \; \tau_1,
\end{equation}
and together with \eqref{eq:dressed_raman_vertex_solution}, we find the final answer
\begin{equation} \label{eq:bimagnon_spectrum_full_expression}
    \ramanbm(\omega) = i \; {\rampr^2 \over \Omega} \; \; \tau_1 \left[ \id + i \tilde{g} \lfn(\omega) \mmvs \right]^{-1} \lfn(\omega) \; \tau_1.
\end{equation}

\subsection{Estimation of Raman-type charge contribution}

In order to illustrate the interplay between magnetic and charge responses in doped systems -- in particular, how the energy scale $J^*$ controls pseudogap-like behavior in observables primarily related to charge DOF -- we estimate the charge contribution to Raman-type spectra at finite doping. Recall that a hopping modulation \eqref{eq:perturbation_definition} in the Hubbard model maps, under the Schrieffer-Wolff and auxiliary-boson transformations, to three effective Raman vertices: the Loudon-Fleury one \eqref{eq:raman_vetex_bimagnon} governing spin response, a kinetic one \eqref{eq:raman_vetex_kinetic} describing hole scattering together with magnon creation / annihilation, as well as a 3-site one \eqref{eq:raman_vetex_3_site} which concerns only the charge DOF at leading order. Previously, we have considered in detail the Loudon-Fleury contribution to the Raman spectrum, and found an RPA-type solution \eqref{eq:bimagnon_spectrum_full_expression} to the two-magnon Bethe-Salpeter equation, yielding the well-known bimagnon Raman peak. Here, we turn to the contributions from the remaining two vertices.\\

In solid-state systems such as certain underdoped cuprates, a rather flat and featureless continuum appears in the Raman spectrum at finite doping, with total weight proportional to the doping level. Its intensity is negligible at very low energy transfers, it typically picks up weight around the frequency where the bimagnon peak is located, and then extends up to much higher energies (see for example Fig.~1 of Ref.~\cite{Devereaux.2007}). The continuum is attributed to charge response, and the reduced weight at low frequencies is a typical signature of the pseudogap regime. Owing to the practically featureless nature of this continuum, we are motivated to estimate it even in the absence of vertex corrections. We therefore consider the direct contraction of two kinetic Raman vertices $\hat{V}_{\rm kin}$ with the self-consistent single-particle propagators for holes and magnons. The resulting charge-charge contribution to the signal has the expression
\begin{equation} \label{eq:raman_charge_estimation}
    \ramanbm_{\rm kin}(\omega) = 2S \; (z \Delta t)^2 \int_{\q, \nu} \trace \left[ \mathbb{H}_{\q} (\omega + \nu) \; \mpr_{\q}(\nu) \right],
\end{equation}
where we have defined the hole bubble contracted with $d-$wave vertices,
\begin{equation} \label{eq:raman_charge_hole_bubble}
    \mathbb{H}_{\q} (\omega + \nu) \equiv i \int_{\km, \e} \begin{pmatrix}
        \tilde{\gamma}_{\km} \\ \tilde{\gamma}_{\km + \q}
    \end{pmatrix} \hpr_{\km + \q}(\e + \omega + \nu) \; \hpr_{\km}(\e) \begin{pmatrix}
        \tilde{\gamma}_{\km} & \tilde{\gamma}_{\km + \q}
    \end{pmatrix}.
\end{equation}
In practice, the appearance of the magnon propagator $\mpr_{\q}(\nu)$ in \eqref{eq:raman_charge_estimation} will favor $\q$ lying on the diamond of the BZ, as well as the frequency condition $\nu \sim 2 J^*$, since this is where the single-magnon DOS is peaked. The hole bubble $\mathbb{H}$, on the other hand, yields a rather flat response with characteristic width $8t$, due to the markedly incoherent nature of $\hpr$. Bringing the two together, we obtain a flat continuum starting at a frequency comparable to that of the bimagnon peak (and thus dictated by $J^*$), extending to much higher frequencies, and enclosing total weight proportional to $\delta$ (See Fig.~\ref{fig:assembled_3} in the main text). It is thus in good qualitative agreement with the typical experimental features discussed above, and motivates the understanding of low-frequency response suppression -- the signature of the pseudogap regime -- as being intimately connected to the \emph{magnetic} energy scale $J^*$.\\

Finally, let us comment on other possible combinations of vertices. If we take mixed terms between $\hat{V}_{\rm kin}$ and $\hat{V}_{\rm 2M}$, we should use the dressed vertex $\bsv_{\km}^{\rm ab} (\omega)$ for the Loudon-Fleury part. As this has a strong resonance at $\omega = \omega_{\rm 2M}(\delta)$, the result will have most of its weight concentrated around the bimagnon peak. This amounts to a doping-dependent renormalization of the bimagnon peak weight, but should not yield important qualitative changes. The contraction of $\hat{V}_{3-\rm{site}}$ with itself (again in the absence of vertex corrections) yields a contribution which we have numerically found to be negligible, while the pairing of $\hat{V}_{3-\rm{site}}$ with $\hat{V}_{\rm kin}$ or $\hat{V}_{\rm 2M}$ is also expected to be unimportant, owing to the very small amplitude of $\hat{V}_{3-\rm{site}}$ itself. We are led to the conclusion that Eqns.~\eqref{eq:bimagnon_spectrum_full_expression} and \eqref{eq:raman_charge_estimation} capture the essential features of Raman-type spectra in doped systems, and highlight further connections between $J^*$ and pseudogap behavior.

\section{Emergent magnetic energy scales in the RPA approach}

In this section, we closely investigate the magnon self-energy $\mse^\rpa_{\km}(\omega)$, Eq. \eqref{eq:magnon_self_energy_rpa}, which is the essential driver of the physics explored in our work. We argue why the magnon modes naturally split into two categories, namely the ones in the immediate vicinity of the gapless points $\Gamma$ and M, and respectively everything else; and why they are governed by two different energy scales $J_\rho$ and respectively $J^*$. We provide further insight into how the hole spectral function determines the distinct doping behavior of these two magnetic energy scales.\\

In the main text, we take $t>0$ for simplicity; here, we relax that assumption in order to provide a more general argument. This simply amounts to making the replacement $t \to |t|$ in most of the final results.

\subsection{Kinetic renormalization of exchange at high energies}

We begin by noting that all three components of the RPA self-energy $\mse^\rpa_{\km} (\omega)$, described in Eqns. \eqref{eq:magnon_rpa_pauli_coefficients}, contain the same frequency integral over the hole bubble,
\begin{equation} \label{eq:hole_bubble_rpa_full}
    i \int \dbar \e \; \hpr_{\q} (\e) \; \hpr_{\km + \q} (\omega + \e).
\end{equation}
Our self-consistent solution for the hole spectral function shows that the polaron is coherent only within a narrow range of momenta around its dispersion minimum (see Fig.~\ref{fig:sm_spectral_functions}). Therefore, in order to obtain an overlap of two such sharp features in the integral \eqref{eq:hole_bubble_rpa_full}, the frequency transfer $\omega$ must be small, on the order of the coherent hole bandwidth. Further restrictions arise from the condition that the momentum transfer $\km$ connects two low-energy pockets in the polaron dispersion, i.e. it must be close enough to one of $(0, 0)$, $(\pi, 0)$, $(0, \pi)$, or $(\pi, \pi)$. For the purpose of finding the magnon dispersion, we are interested in $\mse^\rpa_{\km} (\omega)$ at the new pole location, i.e. $(\km, \omega)$ must be on-shell with the (renormalized) magnons. Since the $(\pi, 0)$ and $(0, \pi)$ magnons are very high-energy, values of $\km$ around these points are immediately disqualified.\\

The conclusion is that, in order to have any relevant overlap of coherent features from hole spectral functions, we are restricted to low $\omega$, as well as momenta $\km$ close to the gapless points of the magnon dispersion ($\Gamma$ or M). This region of momentum space is extremely narrow, as the hole excitation must be nearly on-shell with a magnon. That condition is difficult to satisfy because the polaron is very heavy, whereas the magnon is massless. In the next subsection, we will take a close look at this narrow region of low-energy magnons, and indeed find that their behavior upon doping is drastically different from the universal scaling \eqref{Eq:Universal_Scaling}. Here, however, we focus on the vast majority of magnon modes, which are not immediately near $\Gamma$ or M. For them, it is an excellent approximation to ignore overlap of coherent hole features in \eqref{eq:hole_bubble_rpa_full}.\\

When the sharp peaks in hole spectral functions do not overlap, even the existence of such features is completely irrelevant: a narrow peak integrated against a featureless, almost-flat background will only yield the product of the background height and the peak weight. We may therefore vastly simplify our approach by retaining the hole bandwidth, but otherwise ignoring the particular dispersion details. In fact, we can entirely replace the propagator $\hpr_{\q} (\e)$ by a \emph{dispersionless}, \emph{fully incoherent} approximation (e.g. box or semicircle spectral function), with characteristic width set by $|t|$; we find that qualitative conclusions are the same regardless of the shape of this incoherent substitute.\\

As the spectral function is normalized, $\int d\e \; B(\e) = 1$, typical values of $B(\e)$ will be on the order of $|t|^{-1}$. The frequency integral of $\hpr (\e) \; \hpr (\omega + \e)$, for $|\omega| \ll |t|$, will then scale with $|t|^{-1}$. Moreover, it should be proportional to $\delta$, at least for small dopings. We arrive at the approximation
\begin{equation}
    i \int \dbar \e \; \hpr_{\q} (\e) \; \hpr_{\km + \q} (\omega + \e) \to {\delta \over |t|}\; g(\omega),
\end{equation}
which has no dependence on the momentum $\km$. Here, the dimensionless function $g(\omega)$ obeys $g(|\omega| \ll |t|) \sim \mathcal{O}(1)$ real and positive, as well as $g(|\omega| \gg |t|) \to 0$. Note that in the intermediate regime $|\omega| \sim |t|$, the function $g(\omega)$ can acquire a significant imaginary part, as the scattering of charge excitations can damp magnon modes. The momentum integrals in \eqref{eq:magnon_rpa_pauli_coefficients} can be performed analytically, giving
\begin{subequations} \label{eq:magnon_rpa_pauli_coefficient_q_integrals}
\begin{align}
    \int_{\q} \left( \gamma_{\q}^2 + \gamma_{\km + \q}^2 \right)  &= {1 \over 2}, \\
    2 \int_{\q} \gamma_{\q} \gamma_{\km + \q}  &= {\gamma_{\km} \over 2}, \\
    \int_{\q} \left( \gamma_{\q}^2 - \gamma_{\km + \q}^2 \right) &= 0.
\end{align}
\end{subequations}
Since $\Pi^3_{\km}(\omega) = 0$ and $\Pi^1_{\km}(\omega) = \gamma_{\km} \Pi^0_{\km}(\omega)$, we again have a self-energy contribution whose Nambu and momentum structure are summarized as $[\pauli_0 + \gamma_{\km} \pauli_1]$. We have seen that this can be understood as a (scalar) correction to $J$. By including the finite-doping Oguchi correction \eqref{eq:oguchi_result_general} and comparing with \eqref{eq:bare_magnon_propagator}, we find the effective exchange interaction,
\begin{equation} \label{eq:anderson_toy_model}
    J^*(\omega) = J_{\delta} - \left[ {z g(\omega) \over 2} \right] |t| \; \delta,
\end{equation}
where $J_{\delta} \equiv \alpha_h [1 + r(\delta)] J$. The above recovers the Anderson-type competition between the superexchange $J$ on the one hand, and the delocalization tendency of charge carriers, quantified by $|t| \delta$, on the other. As discussed in the main text, the middle of the magnon spectrum is typically low enough in energy to make the further approximation $g(\omega) \approx g(0) > 0$, which directly yields a frequency-independent effective exchange. Dividing \eqref{eq:anderson_toy_model} by $J_0 = (1 + r_0) J$, and plugging in $J = 4t^2 / U$, we find in this case
\begin{equation} \label{eq:anderson_toy_model_divided}
    {J^* \over J_0} \approx \alpha_h \; {1 + r(\delta) \over 1 + r_0} - \left[ {z g(0) \over 8 (1 + r_0)} \right] {U \over |t|} \; \delta,
\end{equation}
which is identical to the universal scaling \eqref{Eq:Universal_Scaling} if we ignore the doping dependence of $\alpha_h$ and $r(\delta)$. In practice, these effects are subleading, while most of the reduction in $J$ upon doping indeed comes from the kinetic term. As discussed in a footnote to the main text, including these subleading effects will only correct the slope $a$ in Eq.~\eqref{Eq:Universal_Scaling} at orders $t/U$ and above. We also note that towards the top of the magnon spectrum, the approximation $g(\omega) \approx g(0)$ begins to lose some of its accuracy. The main deviation consists in $g(\omega)$ acquiring an imaginary part, which in turn leads to visible damping of high-energy magnons. Although those peaks in the magnon spectral function are broadened, their position still retains the same linear behavior with doping as described by the universal scaling \eqref{Eq:Universal_Scaling}.\\

The simple picture described in this section remarkably reproduces the linear scaling of $J^*$ with doping $\delta$, in agreement with the full numerical calculation as well as experiments. It further reveals that the origin of this magnetic energy scale renormalization -- and so perhaps of the pseudogap regime boundary -- is fundamentally kinetic in nature. The foundations of this reasoning -- the shape of polaron dispersion, its incoherence away from the band bottom, and the dispersion mismatch between polaron and magnon -- are general, and transcend the specific assumptions of our current approach. We therefore expect this line of argumentation to be generally applicable to a wide range of systems.

\subsection{Low-energy magnons}

Having established the doping-induced renormalization of magnons at intermediate frequencies, we now consider the modes around the gapless points $\Gamma$ and M. For simplicity, we focus on low $(\km, \omega)$ in the vicinity of $\Gamma$, while the results around M are identical due to the symmetry properties \eqref{eq:symm_hole_mag_diag_qafm} and \eqref{eq:symm_mag_offdiag_qafm}. The full inverse propagator for the magnons, eq. \eqref{eq:full_magnon_propagator}, with the RPA self-energy separated as in \eqref{eq:magnon_rpa_pauli_separation}, gives
\begin{equation}
    \mpr_{\km}^{-1}(\omega) = [\omega + i \eta_B \operatorname{sgn} (\omega) - \Pi^3_{\km}(\omega)] \tau_3 - \left[ z J_{\delta} S + \Pi^0_{\km}(\omega) \right] \tau_0 - \left[ z J_{\delta} S \; \gamma_{\km} + \Pi^1_{\km}(\omega)\right] \tau_1.
\end{equation}
Note that the magnon dispersion $\omega_{\km}$ is set by the solution of $\det \mpr_{\km}^{-1}(\omega_{\km}) = 0$. This translates to
\begin{align}
    0 = \left[ z J_{\delta} S (1 - \gamma_{\km}) + \Pi^0_{\km}(\omega_{\km}) - \Pi^1_{\km}(\omega_{\km}) \right] \left[ z J_{\delta} S (1 + \gamma_{\km}) + \Pi^0_{\km}(\omega_{\km}) + \Pi^1_{\km}(\omega_{\km}) \right] - \left[ \omega_{\km} + i \eta_B - \Pi^3_{\km}(\omega_{\km}) \right]^2. \label{eq:dispersion_condition}
\end{align}
In general, the solution $\omega_\km$ cannot be extracted directly. However, by expanding at $|\km| \ll 1$ and $0 \le \omega_{\km} \ll J$, we can determine the spin stiffness. Based on symmetry considerations, we can write the lowest terms in the series:
\begin{subequations} \label{eq:self_energy_expansions}
\begin{align}
    \Pi^0_{\km}(\omega) &\approx \tilde{\Pi}^0_{0} + \omega \; \tilde{\Pi}^0_{\omega, 1} + \omega^2 \; \tilde{\Pi}^0_{\omega, 2} +  |\km|^2 \; \tilde{\Pi}^0_{\km} + \dots, \\
    \Pi^1_{\km}(\omega) &\approx \tilde{\Pi}^1_{0} + \omega \; \tilde{\Pi}^1_{\omega, 1} + \omega^2 \; \tilde{\Pi}^1_{\omega, 2} +  |\km|^2 \; \tilde{\Pi}^1_{\km} + \dots, \\
    \Pi^3_{\km}(\omega) &\approx \omega \; |\km|^2 \; \tilde{\Pi}^3 + \dots.
\end{align}   
\end{subequations}
The reasoning for these forms, based on the properties of the definitions \eqref{eq:magnon_rpa_pauli_coefficients}, is as follows: 
\begin{itemize}
    \item All components of $\mse_{\km}(\omega)$ must have the symmetries of the lattice with respect to $\km$, by the definition of the self-energy. So with respect to momentum they can only contain even-order terms. Furthermore, due to the symmetry between the $\hat x$ and $\hat y$ axes, the only allowed term at second order is $k_x^2 + k_y^2 = |\km|^2$.
    \item $\Pi^3$ starts linear in $\omega$ since it is an odd function in frequency. However, since it identically vanishes at zero momentum, $\Pi^3_{0}(\omega) = 0$, the lowest term in its expansion has to be $\omega \; |\km|^2$.
\end{itemize}
Furthermore, it is clear from their definition that the first two components of $\Pi$ are identical for zero momentum, $\Pi^0_{0}(\omega) = \Pi^1_{0}(\omega)$. In particular, this means that in the expansion we have the following equalities between coefficients: $\tilde{\Pi}^0_{0} = \tilde{\Pi}^1_{0} \equiv \tilde{\Pi}_{0}$, together with $\tilde{\Pi}^0_{\omega, 1} = \tilde{\Pi}^1_{\omega, 1} \equiv \tilde{\Pi}_{\omega, 1}$, as well as $\tilde{\Pi}^0_{\omega, 2} = \tilde{\Pi}^1_{\omega, 2} \equiv \tilde{\Pi}_{\omega, 2}$.\\

We now plug the expansion \eqref{eq:self_energy_expansions} into equation \eqref{eq:dispersion_condition}, and find the spin-wave velocity. The goal is to find an expression for the dispersion relation $\omega_{\km}$ related to the low-energy exchange $J_\rho$ as follows: 
\begin{equation} \label{eq:j_rho_definition}
    \omega_{\km} = {z S \over \sqrt{2}} \; J_{\rho} \; |\km|.
\end{equation}
This confirms that $\omega_\km$ and $|\km|$ appear at the same order, and the simultaneous expansion in both variables is well-defined. To second order in the expansion, then, the result is
\begin{align}
    0 = \det \mpr_{\km}^{-1}(\omega_{\km}) &\approx {1 \over 2} |\km|^2 \left( z J_\delta S + \tilde{\Pi}_{0} \right) \left[ z J_\delta S + 4 \left( \tilde{\Pi}^0_{\km} - \tilde{\Pi}^1_{\km} \right) \right] - \omega_\km^2,
    \label{eq:dispersion_condition_expanded}
\end{align}
whose positive-frequency solution is straightforwardly extracted, and yields the linear dispersion \eqref{eq:j_rho_definition} with
\begin{equation} \label{eq:j_rho_solution_coefficients}
    J_{\rho} = \sqrt{ \left( J_\delta +  {\tilde{\Pi}_{0} \over z S} \right) \left[ J_\delta + {4 \over z S} \left( \tilde{\Pi}^0_{\km} - \tilde{\Pi}^1_{\km} \right) \right]}.
\end{equation}
It remains to find the coefficients $\tilde{\Pi}_{0}$ and respectively $\tilde{\Pi}^0_{\km} - \tilde{\Pi}^1_{\km}$. Recall from Eq.~\eqref{eq:self_energy_expansions} and the following paragraph that $\tilde{\Pi}_{0}$ is just the $(\km = 0$, $\omega = 0)$ value of $\Pi^0_{\km}(\omega)$ as well as $\Pi^1_{\km}(\omega)$, i.e. by comparing with \eqref{eq:magnon_rpa_pauli_coefficients} we find
\begin{equation}
    \tilde{\Pi}_{0} = - 2 i S z^2 t^2 \int_{\q} \gamma_{\q}^2 \int \dbar \e \; \hpr_{\q}^2 (\e).
\end{equation}
It is advantageous to define the frequency integral
\begin{equation}
    \mathcal{B}_{\q}(\delta) \equiv i |t| \int \dbar \e \; \hpr_{\q}^2 (\e) \in \mathbb{R},
\end{equation}
where we emphasized the doping dependence by writing $\delta$ as an argument. In terms of this, we then find
\begin{equation}
    {\tilde{\Pi}_{0} \over z S} = - 2 z |t| \int_{\q} \gamma_{\q}^2 \; \mathcal{B}_{\q}(\delta).
\end{equation}
On the other hand, to find $\tilde{\Pi}^0_{\km} - \tilde{\Pi}^1_{\km}$, we must investigate the momentum dependence of $\Pi^0_{\km}(\omega) - \Pi^1_{\km}(\omega)$ for vanishing frequency, $\omega = 0$. Using the definitions \eqref{eq:magnon_rpa_pauli_coefficients}, we write this difference as
\begin{equation} \label{eq:pi_components_difference}
    \Pi^0_{\km}(0) - \Pi^1_{\km}(0) = - i S z^2 t^2 \int_{\q} \left( \gamma_{\km + \q} - \gamma_{\q} \right)^2 \int \dbar \e \; \hpr_{\q} (\e) \; \hpr_{\km + \q} (\e).
\end{equation}
By writing $\gamma_{\km + \q} - \gamma_{\q} \approx \km \cdot (\nabla_{\q} \gamma_\q) + \dots$, we see that the momentum prefactor in \eqref{eq:pi_components_difference} is already at order $|\km|^2$. This means that for the second-order momentum contribution we can already take $\km = 0$ for hole propagators inside the frequency integral. After basic manipulations, and using the symmetry property \eqref{eq:symm_lattice_hole}, we arrive at
\begin{equation}
    {4 \over z S} \left( \tilde{\Pi}^0_{\km} - \tilde{\Pi}^1_{\km} \right) = - z |t| \int_{\q} {\sin^2 q_x + \sin^2 q_y \over 2} \; \mathcal{B}_{\q}(\delta).
\end{equation}
Combining the above, we have the expression for $J_\rho$ corresponding to eq. \eqref{eq:j_rho_main} in the main text:
\begin{equation} \label{eq:j_rho_full_expr}
    J_{\rho} = \sqrt{ \left[ J_\delta  - 2 z |t| \int_{\q} \gamma_{\q}^2 \; \mathcal{B}_{\q}(\delta) \right] \left[ J_\delta - z |t| \int_{\q} {\sin^2 q_x + \sin^2 q_y \over 2} \; \mathcal{B}_{\q}(\delta) \right]}.
\end{equation}
When both terms inside the square root have the same sign, we find a real and positive $J_\rho$. However, if they have opposite signs, $J_\rho$ becomes imaginary, signaling a simultaneous instability of all low-energy magnon modes, and therefore a collapse of AFM order. In practice, the second term inside the square root of \eqref{eq:j_rho_full_expr} will dominate the doping dependence, and thus determine the collapse. The reasoning is as follows:
\begin{itemize}
    \item The polaron dispersion has minima at $(\pm \pi/2, \pm \pi/2)$, and is rather dispersionless on the entire diamond of the Brillouin zone. These low-energy regions have strong quasiparticle peaks, which will get doped first, and give rise to strong contributions in $\mathcal{B}_{\q}$. Points such as $(0, 0)$ and $(\pi, \pi)$ will have much broader features lying at higher energies, so their corresponding $\mathcal{B}_{\q}$ will be less relevant.
    \item The form factor $\gamma_{\q}^2$, appearing in the first integral, vanishes on the entire diamond and peaks at $(0, 0)$ as well as $(\pi, \pi)$. Therefore, it has weak overlap with $\mathcal{B}_{\q}$, and yields a small integral.
    \item In contrast, $\sin^2 q_x + \sin^2 q_y$ peaks at $(\pm \pi/2, \pm \pi/2)$ and vanishes at $(0, 0)$ as well as $(\pi, \pi)$, having very strong overlap with $\mathcal{B}_{\q}$. Thus, the corresponding momentum integral will be much larger.
\end{itemize}
So, in the regime of low dopings, it is a reasonable approximation to take
\begin{equation} \label{eq:j_rho_full_expr_simplified}
    J_{\rho} \approx \sqrt{J_\delta \left[ J_\delta - z |t| \int_{\q} {\sin^2 q_x + \sin^2 q_y \over 2} \; \mathcal{B}_{\q}(\delta) \right]},
\end{equation}
and the stability condition for AFM order can be summarized as
\begin{equation}
    J_\delta - z |t| \int_{\q} {\sin^2 q_x + \sin^2 q_y \over 2} \; \mathcal{B}_{\q}(\delta) > 0.
\end{equation}
As discussed above, for very small dopings, the bubble $\mathcal{B}_{\q}(\delta)$ only acquires nontrivial values near the polaron band minima $\q = (\pm \pi/2, \pm \pi/2)$, where the momentum form factor $(\sin^2 q_x + \sin^2 q_y) / 2$ is equal to 1; we can therefore define $\mathcal{B}(\delta) \equiv \int_{\q} \mathcal{B}_{\q}(\delta)$ and simplify the above condition to
\begin{equation} \label{eq:simplified_stability_condition}
    J_\delta - z |t| \mathcal{B}(\delta) > 0.
\end{equation}
This expression allows us to reexamine the fate of long-range AFM order at zero temperature and small $\delta$. If the polaron was a perfectly coherent quasiparticle, $\mathcal{B}(\delta)$ would be given by the Lindhard function at zero momentum and frequency. Since the latter is related to the density of states at the Fermi level, Eq. \eqref{eq:simplified_stability_condition} translates to a Stoner-type stability condition. In practice, due to the high polaron mass, we tend to obtain a large DOS which violates the inequality. This has led to the theoretical suggestion that AFM order is unstable upon any infinitesimal doping.\\

On the other hand, if the polaron spectral function is broadened by any means, the low-doping behavior of our bubble will instead tend to be $\mathcal{B}(\delta) \propto \delta$. The connection between this regime and the Stoner one described above, which is an issue closely related to the order of limits when we consider $\eta_F \to 0$, is beyond the scope of this work. In general, however, the $\mathcal{B}(\delta) \propto \delta$ behavior leads to a stability condition of the form $J_\delta - b \; \delta > 0$, where we defined the constant $b \equiv z |t| \mathcal{B}(\delta) / \delta$. For very small dopings the difference between $J_\delta$ and $J_0$ is negligible, and we can therefore estimate the critical doping $\delta_{\rm AFM} \approx J_0 / b$. This not only yields a finite range of dopings supporting AFM order, but also gives rise to the following approximate behavior for $J_\rho$:
\begin{equation} \label{eq:j_rho_scaling_generic}
    J_{\rho} \approx \sqrt{ J_0  \left( J_0 - b \; \delta \right)} = J_0 \sqrt{1 - {\delta \over \delta_{\rm AFM}}},
\end{equation}
in agreement with previous theoretical approaches. In our calculations, a straightforward way to implement such broadening is to take $\eta_F>0$, in which case we obtain $\delta_{\rm AFM} \propto \sqrt{\eta_F}$ (see Fig.~\ref{fig:sm_afm_doping}). As mentioned in the main text, when $U/t = 8$, typical values for the AFM critical doping $\delta_{\rm AFM} \sim 4-5\%$ are obtained from relatively small broadening $\eta_F/t \sim 0.05$. Since even the existence of perfectly sharp, $\delta-$like quasiparticle peaks for polarons in the $t-J$ model is contested (see Refs.~\cite{Bohrdt2020, monte_carlo_single_hole_z, mischchenko_delta_function_peak} versus \cite{phase_string_1, phase_string_2}), and real-world experimental systems will always include sources of decoherence and broadening, we may indeed expect to find finite $\delta_{\rm AFM}$. Moreover, the conclusion of our approach is that by tuning $\eta_F$ experimentally, one may indeed control the range of dopings over which long-range AFM order survives.

\subsection{Numerical results of self-consistent approach}

\begin{figure}[ht]
	\centering	
    \scalebox{0.4}{\includegraphics{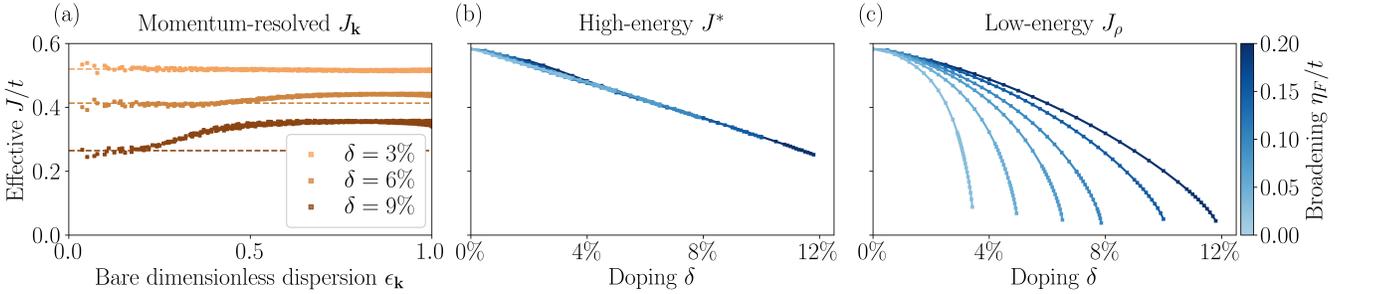}}
	\caption{Extracting magnetic energy scales from the self-consistent numerical solution for the magnon and hole propagators, at $U/t = 8$ and $\eta_B/t = 0.025$. (a)~Momentum-resolved effective exchange, defined by $J_{\km} = \omega_{\km} / 2 \epsilon_{\km}$, at three representative dopings, for $\eta_F/t = 0.2$. Dashed lines denote the $J_\rho$ calculated using the hole propagator, cf.~Eq.~\eqref{eq:j_rho_full_expr}. For the higher dopings, we see a clear distinction between $J^*$ governing the top of the magnon spectrum, and $J_\rho$ determining the bottom. (b)~Doping dependence of $J^*$ for various fermion broadenings $\eta_F / t \in \{ 0.03, 0.05, 0.075, 0.1, 0.15, 0.2 \}$. Note the linear behavior and complete independence on $\eta_F$. Here, $J^* (\delta)$ is numerically extracted by the best fit of $\omega_\km(\delta) = 2 J^*(\delta) \; \e_{\km}$ to the high-energy magnon modes $\e_{\km} \in [0.2, 0.8]$. (c)~Doping dependence of $J_\rho$, for the same parameters as above. Note the $\eta_F$ dependence of the critical doping $\delta_{\rm AFM}$ for AFM collapse.}
	\label{fig:sm_j_measures}
\end{figure}

Having gathered analytical intuition about the two magnetic energy scales which emerge upon doping, we compare our approximate results with the full numerical solution of the self-consistent problem, Eq. \eqref{eq:full_single_particle_propagators}. Using Eqns. \eqref{eq:magnon_spectral_function_matrix} and \eqref{eq:magnon_spectral_function_scalar}, we can obtain for every doping $\delta$ the magnon spectral function $A_\km(\omega)$. Then, at every momentum $\km$, we extract the (positive) frequency $\omega_\km$ of the peak in $A_\km(\omega)$, which gives the magnon dispersion. Recalling that spin-wave theory for the undoped case predicts $\omega_\km = 2 J \e_\km$, where $\e_\km = \sqrt{1 - \gamma_{\km}^2}$, we are motivated to define
\begin{equation}
    J_\km(\delta) = {\omega_\km(\delta) \over 2 \e_\km}.
\end{equation}
If a single energy scale indeed governs an entire region of the Brillouin zone, all the corresponding momentum points should yield the same value of $J_\km$. Figure ~\ref{fig:sm_j_measures}(a) presents numerical results for $J_\km(\delta)$ at $U/t = 8$ and three different dopings. In order to meaningfully include data from the entire 2D Brillouin zone within this plot, the $\hat{x}-$axis represents the $\e_\km$ value of the 2D momentum $\km$, i.e. each magnon energy contour of the undoped system gets mapped to a single horizontal coordinate. We can see that, for low dopings, the entire spectrum is governed by one energy scale. On the other hand, upon increasing $\delta$, a clear distinction appears between the low-energy modes, which are governed by a smaller $J_\rho$, and the higher ones, whose dispersion is set by the larger $J^*$.\\

We independently calculate $J_\rho$ based on the solution for the hole propagator, from Eq.~\eqref{eq:j_rho_full_expr}, and depict it by the dashed horizontal lines in Fig.~\ref{fig:sm_j_measures}(a). The excellent agreement with $J_\km$ in the low-energy region of the magnon spectrum confirms the validity of the analytic expansion, and motivates us to use Eq.~\eqref{eq:j_rho_full_expr} for extracting $J_\rho$ from now on. This avoids the numerical issue of the alternative approach -- fitting $J_\rho$ to the very small number of momentum points near $\Gamma$ and M. To extract $J^*$, on the other hand, we fit the expression $\omega_\km(\delta) = 2 J^*(\delta) \; \e_\km$ to the upper part of the magnon spectrum, which corresponds to the vast majority of momentum points in the BZ (recall that the linear spin-wave theory DOS vanishes at zero energy and strongly peaks at the top of the magnon band). In practice, the significant broadening at the top of the magnon band negatively impacts the extraction of $J^*$; for this reason, we fit it using $\omega_\km(\delta) = 2 J^*(\delta) \; \e_\km$ in the range $\e_{\km} \in [0.2, 0.8]$.\\

Having extracted the two scales $J^*$ and $J_\rho$ from single-particle propagators, we turn to investigating their dependence on doping, as well as on system parameters. Of particular interest is the polaron broadening $\eta_F$, which by the arguments of the previous section should play a crucial role in the stability of AFM order. 
On the other hand, we have numerically verified that varying $\eta_B / t$ within the range $[0.025, 0.2]$ does not lead to quantitative changes in the aforementioned results, and therefore we fix it to the lowest value compatible with our frequency grid, $\eta_B / t = 0.025$. In Figure \ref{fig:sm_j_measures} (b) and (c) we show the doping dependence of $J^*$ and respectively $J_\rho$, for six distinct values of $\eta_F$. We find for $J^*$ a clear linear dependence on doping, completely independent of $\eta_F$, in excellent qualitative agreement with \eqref{eq:anderson_toy_model_divided} and the arguments preceding it. On the other hand, $J_\rho$ curves down and collapses much earlier than $J^*$, with typical behavior $J_\rho \propto \sqrt{\delta_{\rm AFM} - \delta}$, and exhibiting a dependence of $\delta_{\rm AFM}$ on $\eta_F$. We remark that our self-consistent solver never converges for $J_\rho^2 < 0$, and therefore we require higher $\eta_F$ to access higher doping values.\\

\begin{figure}[ht]
	\centering	
    \scalebox{0.38}{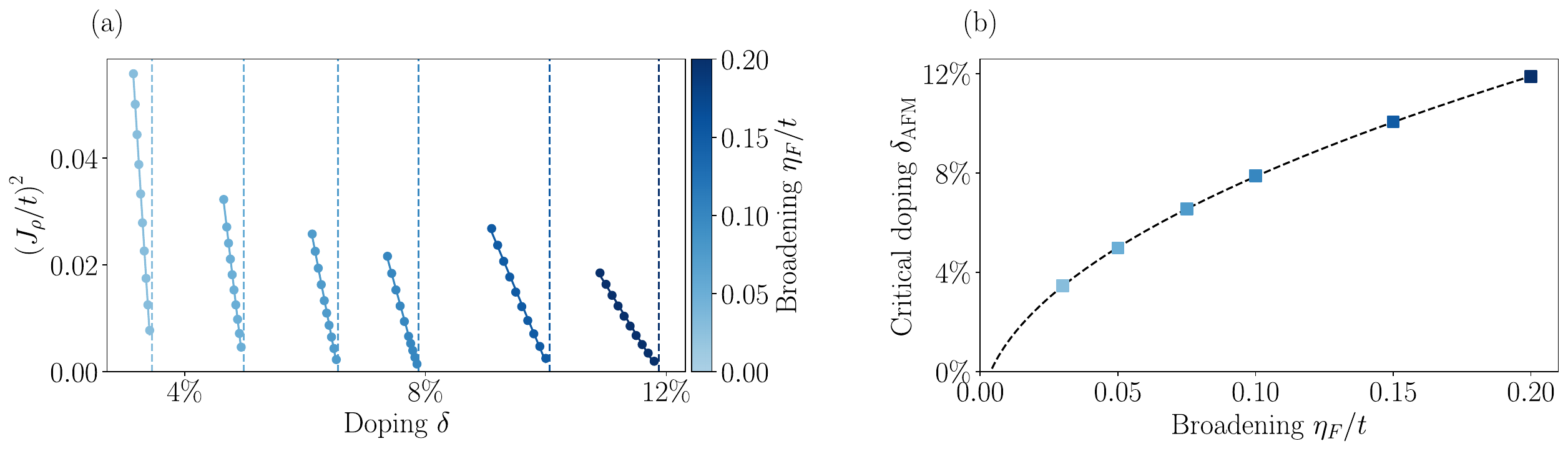}
	\caption{AFM critical doping $\delta_{\rm AFM}$ and its relation to the polaron broadening $\eta_F$. (a)~Doping dependence of $J_\rho^2$ near the AFM collapse point. The observed linear behavior confirms the scaling \eqref{eq:j_rho_scaling_generic} and facilitates the extraction of the collapse point $\delta_{\rm AFM}$, highlighted by vertical dashed lines. (b)~Values of the critical doping $\delta_{\rm AFM}$ for different broadenings $\eta_F / t$, together with a square root fit (dashed line), suggesting the scaling $\delta_{\rm AFM} \propto \sqrt{\eta_F}$.}
	\label{fig:sm_afm_doping}
\end{figure}

In order to confirm the square-root scaling $J_\rho \propto \sqrt{\delta_{\rm AFM} - \delta}$ and also pinpoint $\delta_{\rm AFM}$, we plot in Fig.~\ref{fig:sm_afm_doping}~(a) the doping dependence of $J_\rho^2$ near the collapse point. The visible linear trend is consistent with Eq.~\eqref{eq:j_rho_scaling_generic}, and allows for the extraction of $\delta_{\rm AFM}$. We then plot the latter versus $\eta_F$ in Fig.~\ref{fig:sm_afm_doping}~(b), together with a square-root curve, which we have obtained by performing a linear fit of $\delta_{\rm AFM}$ against $\sqrt{\eta_F}$ \footnote{The linear fit returns a small, nonzero intercept - and so the dahsed line in Fig.~\ref{fig:sm_afm_doping}~(b) does not go exactly through the origin. This is likely an artifact of our frequency grid discretization.}. The significant tunability of $\delta_{\rm AFM}$ via changes in $\eta_F$ highlights the powerful role of the latter as an experimental tuning knob of the antiferromagnetic phase.\\

\begin{figure}[ht]
	\centering	
    \scalebox{0.38}{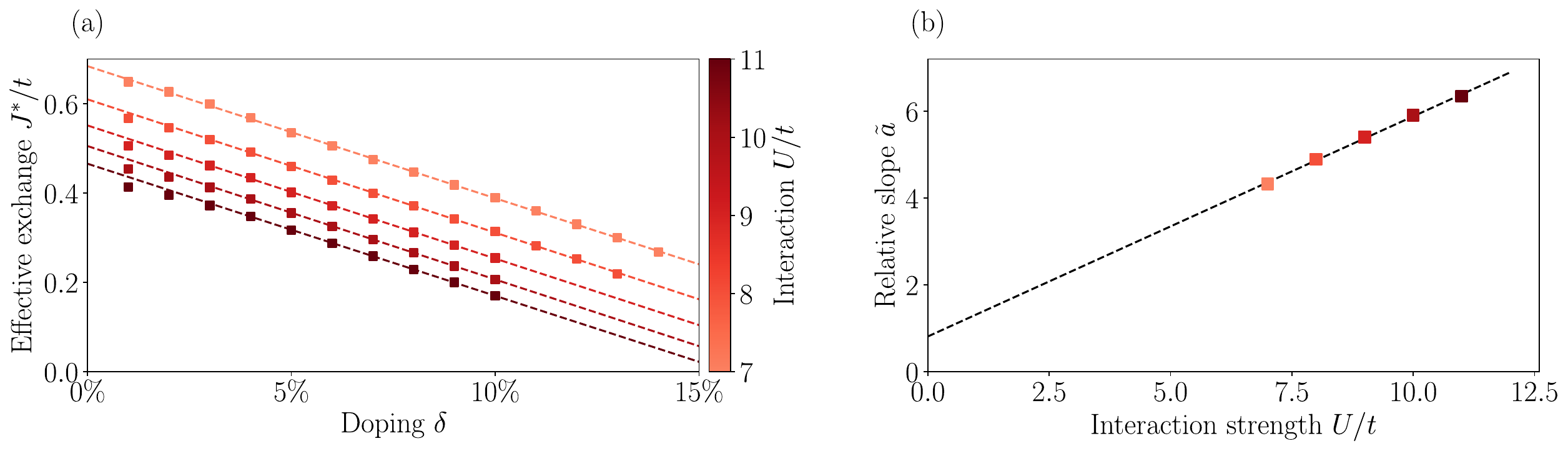}
	\caption{Behavior of magnetic energy scale $J^*$ as a function of Hubbard interaction strength $U/t$. (a)~Doping dependence of $J^*(\delta)$, for $U/t \in \{7,8,9,10,11\}$, together with linear fits (dashed lines). The trends are almost parallel, in accordance with the kinetic picture, where the slope of $J^*(\delta)$ is set by $t$ but not by $U$. (b)~Relative slope $\tilde a$, defined by $J^*(\delta) / J_0 = 1 - \tilde{a} \; \delta$, as a function of $U/t$. Note the linear behavior $\tilde{a} \approx a \times U/t$, which immediately leads to the universal scaling \eqref{Eq:Universal_Scaling}. The presence of a small intercept, on the other hand, signifies the presence of $\mathcal{O}(t/U)$ corrections to the prefactor $a$, which is thus not a perfect constant. Calculations for both panels performed with $\eta_F/t = 0.25$ in order to access higher dopings.}
	\label{fig:j_star_u_t}
\end{figure}

Turning our attention back to the high-energy scale $J^*$, it remains to investigate its doping dependence at different interaction strengths $U/t$, in order to fully obtain the universal scaling, Eq.~\eqref{Eq:Universal_Scaling}. We plot the behavior of $J^*$ versus $\delta$ in Figure~\ref{fig:j_star_u_t}~(a), for five different values of $U/t$, together with linear fits. Recall that the kinetic picture of $J^*$ renormalization predicts $J^* \approx J_{\delta} - 2 g(0) \times t \delta$. Excluding subleading effects from $J_\delta$, the slope should then be set exclusively by $t$, and thus independent of $U$. Visually, the linear trends in Figure~\ref{fig:j_star_u_t}~(a) are almost parallel, in agreement with our expectation. To analyze this in more detail, we divide out the intercept from the linear fits, to extract a single scaling parameter $\tilde{a}$, defined by
\begin{equation} \label{eq:almost_universal_scaling}
    {J^*(\delta) \over J_0} = 1 - \tilde{a} \; \delta.
\end{equation}
Upon comparison with Eq.~\eqref{eq:anderson_toy_model_divided}, we expect that the relative slope $\tilde{a}$ consists of a strong kinetic contribution, given by ${2 g(0) \over 1 + r_0} \; {t \over J} \propto {U \over t}$, as well as a subleading effect obtained from expanding $\alpha_h \; {1 + r(\delta) \over 1 + r_0}$ to linear order in $\delta$. The latter need not be proportional to $U/t$, and can indeed give an offset to the dependence of $\tilde{a}$ on $U/t$. Figure~\ref{fig:j_star_u_t}~(b) shows the extracted relative slopes $\tilde{a}$ for our five values of $U/t$, which indeed display a clear linear behavior. Upon writing $\tilde{a} \approx a \times U/t$, Eq.~\eqref{eq:almost_universal_scaling} immediately yields the universal scaling \eqref{Eq:Universal_Scaling}. On the other hand, the linear fit in Fig.~\ref{fig:j_star_u_t}~(b) does yield a small intercept, which leads to the conclusion that $a \equiv \tilde{a} \times t/U$ is not fully independent of the Hubbard interaction strength, but rather contains corrections at $\mathcal{O}(t/U)$. In practice, however, in the strongly-interacting regime $U/t \gg 1$, this is a subleading effect. We also remark that the universal scaling \eqref{Eq:Universal_Scaling} smoothly connects to Nagaoka's result: for $U/t \to \infty$, any infinitesimal doping $\delta = 0^+$ will effectively turn $J^*$ ferromagnetic.

\section{Instability of AFM state towards incommensurate magnetic ordering}

In the previous section, we have shown that including a finite $\eta_F$ will stabilize AFM ordering, at wavevector $\mathbf Q = (\pi, \pi)$, up to a finite critical doping $\delta_{\rm AFM}$. Nevertheless, for larger doping values $\delta > \delta_{\rm AFM}$, we find our self-consistent approach to be unstable, signaling that magnetic order should instead appear at a different, incommensurate wavevector $\tilde{\mathbf Q}$. In principle, an extension of our self-consistent diagrammatic expansion, allowing for the underlying magnetic order to emerge at arbitrary wavevectors, is expected to capture the correct behavior; however, this approach is left as a subject of future studies. Here, we aim to learn about the incommensurate ordered state by studying the instability towards it from our initial AFM ansatz.\\

Recall that the result of expanding the magnon dispersion equation \eqref{eq:dispersion_condition} to second order in $\mathbf k$ and $\omega$ was
\begin{equation} \label{eq:gl_instability_order_2}
    \omega_{\km}^2 = {(z S)^2 \over 2} \; J_{\rho}^2(\delta) \; |\km|^2,
\end{equation}
with the doping dependence of $J_{\rho}$ near the critical value $\delta_{\rm AFM}$ obeying 
\begin{equation}
    {(z S)^2 \over 2} \; J_{\rho}^2(\delta) \approx \mathcal{A} \; (\delta_{\rm AFM} - \delta).
\end{equation}
Simply plugging $\delta > \delta_{\rm AFM}$ into the above will yield $J_{\rho}^2 < 0$, signaling a simultaneous instability of all low-lying magnon modes. The solution of \eqref{eq:gl_instability_order_2} will be
\begin{equation}
    \omega_{\km} = \pm i \; \sqrt{\mathcal{A} \; (\delta - \delta_{\rm AFM})} \; |\km|,
\end{equation}
meaning that the instability rate of a given mode is directly proportional to its momentum $|\mathbf k|$. However, this cannot continue to arbitrarily high momenta, as the high-energy magnons should be well described by $J^*(\delta)$, thus surviving well beyond $\delta_{\rm AFM}$. Of course, the resolution lies in expanding the self-energies to higher order in $\mathbf k$, which will cut off the instability at some finite momentum. By lattice symmetry, the next allowed terms are of the form $k_x^4$, $k_y^4$, and respectively $k_x^2 k_y^2$. Focusing for simplicity on momenta parallel to a coordinate axis, e.g. $\mathbf k = (k, 0)$, one can envision extending \eqref{eq:gl_instability_order_2} to the Ginzburg-Landau-type expression
\begin{equation} \label{eq:gl_instability_order_4}
    \omega_{k}^2 = - \mathcal{A} \; (\delta - \delta_{\rm AFM}) \; k^2 + \mathcal{K} \; k^4,
\end{equation}
where $\mathcal{K}>0$, and in the vicinity of $\delta_{\rm AFM}$ we may neglect the doping dependence of $\mathcal{A}$ and $\mathcal{K}$. In this case, we find a finite range of unstable modes with $\omega_k^2 < 0$, extending from $k=0$ to a finite momentum $k_{\rm S} = \sqrt{(\delta - \delta_{\rm AFM}) \mathcal{A} / \mathcal{K}}$. Crucially, since $\omega_k^2$ now has a minimum, there exists a momentum $\km_{\rm IC}$ whose instability rate is highest. It is then natural to conclude that the strongest instability wins, and incommensurate magnetic order will emerge at the wavevector $\tilde{\mathbf Q} = \mathbf Q \pm \km_{\rm IC}$. We can find the incommensuration $k_{\rm IC}$ by straightforward differentiation of \eqref{eq:gl_instability_order_4}, yielding
\begin{equation}
    k_{\rm IC} = \sqrt{(\delta - \delta_{\rm AFM}) \; {\mathcal{A} \over 2 \mathcal{K}}},
\end{equation}
while the corresponding maximal instability rate is
\begin{equation}
    \Gamma_{\rm IC} = |\omega_{k_{\rm IC}}| = {\mathcal{A} \over 2 \sqrt{\mathcal{K}}}  \; (\delta - \delta_{\rm AFM}).
\end{equation}
While the qualitative conclusions of the above reasoning are correct, one should do the self-energy expansion more carefully in order to be consistent. Namely, reducing \eqref{eq:dispersion_condition} to a condition of the simple form $\omega_{k}^2 = f(k)$ relied on the cancellation of frequency derivatives in $\Pi^0_{\km}(\omega)$ and $\Pi^1_{\km}(\omega)$, see Eq.~\eqref{eq:self_energy_expansions}. This cancellation, in turn, was only possible because we simultaneously expanded to second order in $\mathbf k$ and $\omega$, together with the equality $\Pi^0_{\km}(\omega) = \Pi^1_{\km}(\omega)$ at zero momentum. When performing the momentum expansions to higher order, as necessary for \eqref{eq:gl_instability_order_4}, these simplifying cancellations no longer occur. We turn to a detailed extraction of $\omega_k$ in this case.\\

For simplicity, we will fix the doping $\delta > \delta_{\rm AFM}$ and one momentum of interest $\mathbf k$, and omit the corresponding indices on $\Pi$ and $\omega$ in the following derivation. With $\eta_B \to 0$, the dispersion condition to be solved for $\omega$ is
\begin{align} \label{eq:dispersion_condition_single_k_specialized}
    0 = \left[ z J_{\delta} S (1 - \gamma_{\km}) + \Pi^0(\omega) - \Pi^1(\omega) \right] \left[ z J_{\delta} S (1 + \gamma_{\km}) + \Pi^0(\omega) + \Pi^1(\omega) \right] - \left[ \omega - \Pi^3(\omega) \right]^2. 
\end{align}
Since the last term generates an $\omega^2$ contribution, we should expand the entire equation \eqref{eq:dispersion_condition_single_k_specialized} to second order in $\omega$ at least. On the other hand, in the vicinity of $\delta_{\rm AFM}$, all the relevant instability rates are expected to be small, so we can indeed stop at order $\omega^2$. We write the frequency expansions for $\omega \ge 0$ as 
\begin{subequations} \label{eq:second_order_frequency_expansions_pi}
    \begin{align}
        \Pi^0(\omega) &\approx \tilde{\Pi}^0_0 + i \omega \; \tilde{\Pi}^0_1 + \omega^2 \; \tilde{\Pi}^0_2 + \dots, \\
        \Pi^1(\omega) &\approx \tilde{\Pi}^1_0 + i \omega \; \tilde{\Pi}^1_1 + \omega^2 \; \tilde{\Pi}^1_2 + \dots, \\
        \Pi^3(\omega) &\approx \omega \; \tilde{\Pi}^3_1 + \dots,
    \end{align}
\end{subequations}
where we no longer have any particular relation between the coefficients of $\Pi^0(\omega)$ and $\Pi^1(\omega)$, since we are working at an arbitrary $\mathbf k$. The factor of $i$ in the linear coefficients of $\Pi^0(\omega)$ and $\Pi^1(\omega)$ was added because the real parts of these two functions only contain even powers of $\omega$, while the imaginary parts only have odd powers; thus, with the expansion defined as above, all the $\tilde{\Pi}$ coefficients on the RHS are real \footnote{From the definition \eqref{eq:magnon_rpa_pauli_coefficients}, the time-ordered self-energies $\Pi^0_{\mathbf k}(\omega)$ and $\Pi^1_{\mathbf k}(\omega)$ are even functions of frequency. However, they are not analytic: the imaginary parts start as $|\omega|$. We can make the functions analytic near the origin by working with the retarded versions, i.e. flipping the sign of the imaginary part at negative frequencies. Then the real parts are even in $\omega$, while the imaginary ones are odd, yielding the expansion from the text. Of course, this mapping does not change the $\omega>0$ values of the self-energy.}. To second order in frequency, Eq.~\eqref{eq:dispersion_condition_single_k_specialized} yields the quadratic equation
\begin{equation}
    - C_2 \; \omega^2 + 2 i C_1 \; \omega + C_0 = 0,
\end{equation}
which has the solutions
\begin{equation} \label{eq:omega_solution_2nd_order}
    \omega_{\pm} = {i C_1 \pm \sqrt{C_0 C_2 - C_1^2} \over C_2}.
\end{equation}
The coefficients $C_{0 \dots 2}$, implicitly depending on $\delta$ and $\mathbf k$, are purely real and given by:
\begin{subequations} \label{eq:quadratic_omega_coefficients}
    \begin{align}
        C_0 &= \left[ z J_{\delta} S (1 - \gamma_{\km}) + \tilde{\Pi}^0_0 - \tilde{\Pi}^1_0 \right] \left[ z J_{\delta} S (1 + \gamma_{\km}) + \tilde{\Pi}^0_0 + \tilde{\Pi}^1_0 \right], \\
        C_1 &= \tilde{\Pi}^0_0 \tilde{\Pi}^0_1 - \tilde{\Pi}^1_0 \tilde{\Pi}^1_1 + z J_{\delta} S \left[ \tilde{\Pi}^0_1 - \gamma_{\km} \tilde{\Pi}^1_1 \right], \\
        C_2 &= \left[ 1 - \tilde{\Pi}^3_1 \right]^2 - 2 z J_{\delta} S \left[ \tilde{\Pi}^0_2 - \gamma_{\km} \tilde{\Pi}^1_2 \right] - 2 \left[ \tilde{\Pi}^0_0 \tilde{\Pi}^0_2 - \tilde{\Pi}^1_0 \tilde{\Pi}^1_2 \right] + \left[ \tilde{\Pi}^0_1 \right]^2 - \left[ \tilde{\Pi}^1_1 \right]^2.
    \end{align}
\end{subequations}
Note that, if we were to consider the momentum expansion of each such coefficient, we would find $C_0, C_1 \sim \mathcal{O}(|\mathbf k|^2)$, while $C_2 \sim 1 + \mathcal{O}(|\mathbf k|^2)$. Therefore, to leading order in $|\mathbf k|$, the solution \eqref{eq:omega_solution_2nd_order} is just given by the leading term in $\sqrt{C_0}$, which precisely recovers Eq.~\eqref{eq:j_rho_solution_coefficients}. Moreover, the proposed extension \eqref{eq:gl_instability_order_4} would arise from continuing to assume $C_2 \approx 1$ and $C_1 \approx 0$, but expanding $C_0$ to order $k^4$. As one can see from the previous arguments, this would not be a consistent expansion in $k$, and we should use the solution \eqref{eq:omega_solution_2nd_order} instead.\\ 

Since $C_{0 \dots 2}$ are real, we can distinguish three regimes in \eqref{eq:omega_solution_2nd_order}:
\begin{enumerate}
    \item Stable magnon regime $C_0 C_2 > C_1^2$. The square-root is real, and we find a pair of solutions with opposite-sign real parts $\pm \sqrt{C_0 C_2 - C_1^2} / C_2$. Since our expansion focused on $\omega \ge 0$, we take the positive solution in this regime. The imaginary part is $C_1 / C_2$, which should be negative, and in practice we indeed find $C_1<0 < C_2$.
    \item Overdamped regime $C_1^2 > C_0 C_2 > 0$. The square-root is imaginary, but its magnitude is less than $|C_1|$. Thus, both solutions in \eqref{eq:omega_solution_2nd_order} are negative imaginary, and no instability arises yet.
    \item Unstable regime $C_0 C_2 < 0$. Both solutions are imaginary, with one of them positive, yielding the instability rate of that mode:
    \begin{equation} \label{eq:instability_rate_2nd_order}
        \Gamma = { C_1 + \sqrt{C_1^2 - C_0 C_2} \over C_2}.
    \end{equation}
\end{enumerate}
The boundary of the unstable regime is given by the condition $C_0 C_2 = 0$, which in practice occurs when $C_0$ vanishes. Note that setting $C_0 = 0$ is how we found $k_{\rm S}$ in our previous simplified approach. We indeed expect the answers to be identical, since $k_{\rm S}$ is defined as the momentum where $\omega = 0$ is a solution to our original determinant equation. In that case, no frequency expansion of the self energies is needed, and plugging $\omega = 0$ into \eqref{eq:dispersion_condition_single_k_specialized} directly yields the exact condition $C_0 = 0$.\\

\begin{figure}[ht]
	\centering	
    \scalebox{0.4}{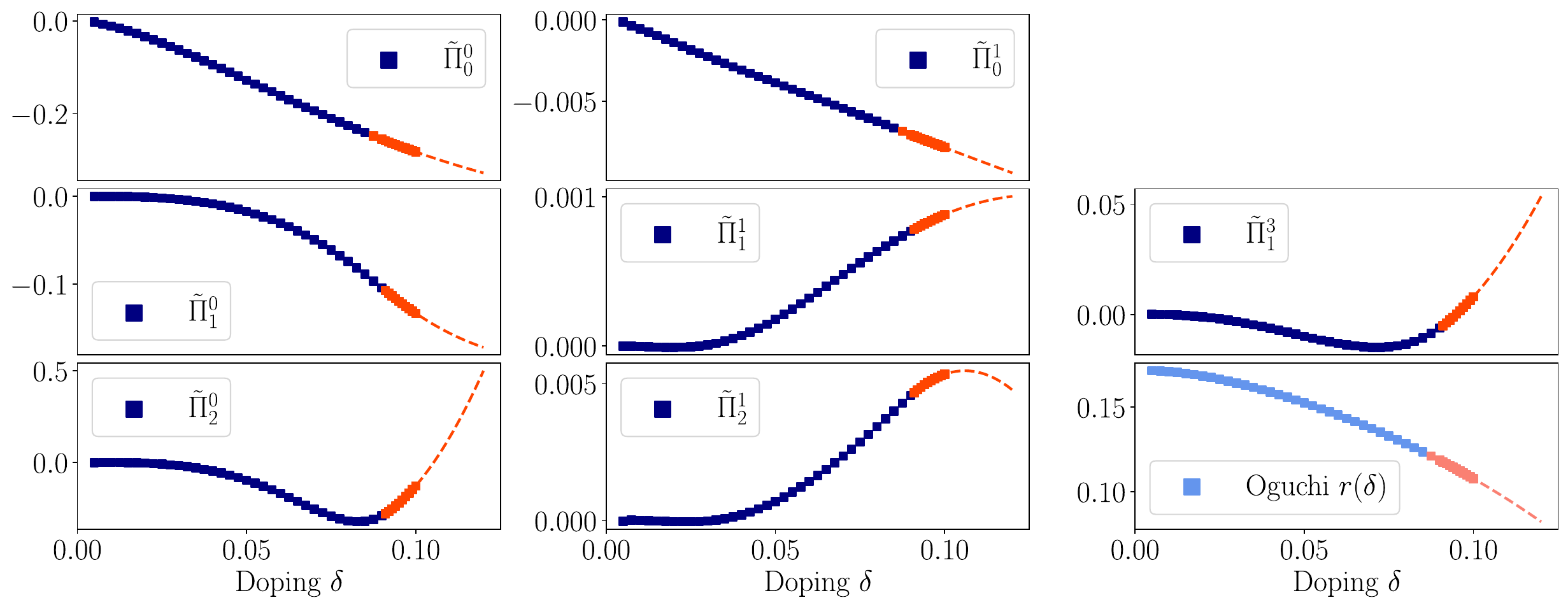}
	\caption{Doping extrapolation of RPA self-energy coefficients $\tilde{\Pi}$ and Oguchi self-energy $r$, into the regime $\delta \gtrsim \delta_{\rm AFM}$. Squares denote results of self-consistent calculation. Orange squares are used for fitting a second-degree polynomial in $\delta$, and the dashed orange lines represent the resulting extrapolation beyond $\delta_{\rm AFM}$.}
	\label{fig:rpa_doping_extrapolation}
\end{figure}

While the range of unstable modes is captured correctly by Eq.~\eqref{eq:gl_instability_order_4}, the momentum dependence of the instability rate $\Gamma_{\km}$ will be slightly renormalized in the improved expansion \eqref{eq:omega_solution_2nd_order}. Thus, to improve the precision of our numerical estimations, we directly work with Eqs.~\eqref{eq:omega_solution_2nd_order} and \eqref{eq:instability_rate_2nd_order}. The procedure for numerically estimating $\Gamma_{\km}(\delta)$ at arbitrary momenta $\km$ and dopings $\delta \gtrsim \delta_{\rm AFM}$ beyond the AFM instability is the following:
\begin{itemize}
    \item From self-consistent calculation results at dopings just below the instability, $\delta \lesssim \delta_{\rm AFM}$, we extract the magnon self-energy components $\Pi^0_{\km}(\omega), \Pi^1_{\km}(\omega), \Pi^3_{\km}(\omega)$ as well as the Oguchi self-energy $r$.
    \item For every available doping $\delta$ and momentum $\km$, we perform the frequency expansions to second order for $\Pi^0_{\km}(\omega)$ and $\Pi^1_{\km}(\omega)$, respectively to first order for $\Pi^3_{\km}(\omega)$, see Eq.~\eqref{eq:second_order_frequency_expansions_pi}. This gives us the $\tilde{\Pi}$ coefficients at dopings $\delta \lesssim \delta_{\rm AFM}$ and momenta $\km$ on the original calculation grid.
    \item We rely on the smooth behavior of the RPA self-energy with doping (Fig.~\ref{fig:rpa_doping_extrapolation}) to extrapolate the $\tilde{\Pi}$ coefficients and Oguchi self-energy $r$, at every $\km$ on the original calculation grid, to dopings $\delta \gtrsim \delta_{\rm AFM}$.
    \item For a given doping $\delta \gtrsim \delta_{\rm AFM}$, we interpolate $\tilde{\Pi}$ to arbitrary momenta $\km$ by relying on the exponential decay of self-energies with real-space separation (see Fig.~\ref{fig:sm_self_energies} and surrounding discussion).
    \item Once we have the $\tilde{\Pi}$ coefficients and Oguchi self-energy $r$ for arbitrary $\km$ and dopings $\delta \gtrsim \delta_{\rm AFM}$, we calculate $J_\delta$, the $C_{0 \dots 2}$ coefficients defined in Eq.~\eqref{eq:quadratic_omega_coefficients}, and extract the instability rate $\Gamma_{\km}(\delta)$ from Eq.~\eqref{eq:instability_rate_2nd_order}.
\end{itemize}

Once momentum-dependent instability rates are calculated, we can look for the maximum value $\Gamma_{\rm IC}$, as well as its location $\km_{\rm IC}$. Since $\Gamma_{\km}$ should have the lattice symmetries with respect to $\km$, its maxima must lie along high-symmetry directions, e.g. $\Gamma \to \rm M$ or $\Gamma \to \rm X$. In practice, the latter case occurs, and the four maxima are found at $\km_{\rm IC} = (\pm k_{\rm IC}, 0)$ and $(0, \pm k_{\rm IC})$. In the main text, we directly refer to the scalar value $k_{\rm IC}$.


\begin{thebibliography}{98}%
	\makeatletter
	\providecommand \@ifxundefined [1]{%
		\@ifx{#1\undefined}
	}%
	\providecommand \@ifnum [1]{%
		\ifnum #1\expandafter \@firstoftwo
		\else \expandafter \@secondoftwo
		\fi
	}%
	\providecommand \@ifx [1]{%
		\ifx #1\expandafter \@firstoftwo
		\else \expandafter \@secondoftwo
		\fi
	}%
	\providecommand \natexlab [1]{#1}%
	\providecommand \enquote  [1]{``#1''}%
	\providecommand \bibnamefont  [1]{#1}%
	\providecommand \bibfnamefont [1]{#1}%
	\providecommand \citenamefont [1]{#1}%
	\providecommand \href@noop [0]{\@secondoftwo}%
	\providecommand \href [0]{\begingroup \@sanitize@url \@href}%
	\providecommand \@href[1]{\@@startlink{#1}\@@href}%
	\providecommand \@@href[1]{\endgroup#1\@@endlink}%
	\providecommand \@sanitize@url [0]{\catcode `\\12\catcode `\$12\catcode `\&12\catcode `\#12\catcode `\^12\catcode `\_12\catcode `\%12\relax}%
	\providecommand \@@startlink[1]{}%
	\providecommand \@@endlink[0]{}%
	\providecommand \url  [0]{\begingroup\@sanitize@url \@url }%
	\providecommand \@url [1]{\endgroup\@href {#1}{\urlprefix }}%
	\providecommand \urlprefix  [0]{URL }%
	\providecommand \Eprint [0]{\href }%
	\providecommand \doibase [0]{https://doi.org/}%
	\providecommand \selectlanguage [0]{\@gobble}%
	\providecommand \bibinfo  [0]{\@secondoftwo}%
	\providecommand \bibfield  [0]{\@secondoftwo}%
	\providecommand \translation [1]{[#1]}%
	\providecommand \BibitemOpen [0]{}%
	\providecommand \bibitemStop [0]{}%
	\providecommand \bibitemNoStop [0]{.\EOS\space}%
	\providecommand \EOS [0]{\spacefactor3000\relax}%
	\providecommand \BibitemShut  [1]{\csname bibitem#1\endcsname}%
	\let\auto@bib@innerbib\@empty
	\bibitem [{\citenamefont {Auerbach}(1998)}]{Auerbach1998}%
	\BibitemOpen
	\bibfield  {author} {\bibinfo {author} {\bibfnamefont {A.}~\bibnamefont {Auerbach}},\ }\href@noop {} {\emph {\bibinfo {title} {Interacting electrons and quantum magnetism}}},\ \bibinfo {edition} {1st}\ ed.,\ Graduate Texts in Contemporary Physics\ (\bibinfo  {publisher} {Springer},\ \bibinfo {address} {New York, NY},\ \bibinfo {year} {1998})\BibitemShut {NoStop}%
	\bibitem [{\citenamefont {Scalapino}(2012)}]{ScalapinoReviewHighTc}%
	\BibitemOpen
	\bibfield  {author} {\bibinfo {author} {\bibfnamefont {D.~J.}\ \bibnamefont {Scalapino}},\ }\bibfield  {title} {\bibinfo {title} {A common thread: The pairing interaction for unconventional superconductors},\ }\href {https://doi.org/10.1103/RevModPhys.84.1383} {\bibfield  {journal} {\bibinfo  {journal} {Rev. Mod. Phys.}\ }\textbf {\bibinfo {volume} {84}},\ \bibinfo {pages} {1383} (\bibinfo {year} {2012})}\BibitemShut {NoStop}%
	\bibitem [{\citenamefont {Keimer}\ \emph {et~al.}(2015)\citenamefont {Keimer}, \citenamefont {Kivelson}, \citenamefont {Norman}, \citenamefont {Uchida},\ and\ \citenamefont {Zaanen}}]{KeimerReviewHighTc}%
	\BibitemOpen
	\bibfield  {author} {\bibinfo {author} {\bibfnamefont {B.}~\bibnamefont {Keimer}}, \bibinfo {author} {\bibfnamefont {S.~A.}\ \bibnamefont {Kivelson}}, \bibinfo {author} {\bibfnamefont {M.~R.}\ \bibnamefont {Norman}}, \bibinfo {author} {\bibfnamefont {S.}~\bibnamefont {Uchida}},\ and\ \bibinfo {author} {\bibfnamefont {J.}~\bibnamefont {Zaanen}},\ }\bibfield  {title} {\bibinfo {title} {From quantum matter to high-temperature superconductivity in copper oxides},\ }\href {https://doi.org/10.1038/nature14165} {\bibfield  {journal} {\bibinfo  {journal} {Nature}\ }\textbf {\bibinfo {volume} {518}},\ \bibinfo {pages} {179} (\bibinfo {year} {2015})}\BibitemShut {NoStop}%
	\bibitem [{\citenamefont {Lee}\ \emph {et~al.}(2006)\citenamefont {Lee}, \citenamefont {Nagaosa},\ and\ \citenamefont {Wen}}]{Lee_review_hightc}%
	\BibitemOpen
	\bibfield  {author} {\bibinfo {author} {\bibfnamefont {P.~A.}\ \bibnamefont {Lee}}, \bibinfo {author} {\bibfnamefont {N.}~\bibnamefont {Nagaosa}},\ and\ \bibinfo {author} {\bibfnamefont {X.-G.}\ \bibnamefont {Wen}},\ }\bibfield  {title} {\bibinfo {title} {Doping a mott insulator: Physics of high-temperature superconductivity},\ }\href {https://doi.org/10.1103/RevModPhys.78.17} {\bibfield  {journal} {\bibinfo  {journal} {Rev. Mod. Phys.}\ }\textbf {\bibinfo {volume} {78}},\ \bibinfo {pages} {17} (\bibinfo {year} {2006})}\BibitemShut {NoStop}%
	\bibitem [{\citenamefont {Timusk}\ and\ \citenamefont {Statt}(1999)}]{Tom_review_pseudogap}%
	\BibitemOpen
	\bibfield  {author} {\bibinfo {author} {\bibfnamefont {T.}~\bibnamefont {Timusk}}\ and\ \bibinfo {author} {\bibfnamefont {B.}~\bibnamefont {Statt}},\ }\bibfield  {title} {\bibinfo {title} {The pseudogap in high-temperature superconductors: an experimental survey},\ }\href {https://doi.org/10.1088/0034-4885/62/1/002} {\bibfield  {journal} {\bibinfo  {journal} {Reports on Progress in Physics}\ }\textbf {\bibinfo {volume} {62}},\ \bibinfo {pages} {61} (\bibinfo {year} {1999})}\BibitemShut {NoStop}%
	\bibitem [{\citenamefont {Chubukov}\ and\ \citenamefont {Morr}(1997)}]{Chubukov.1997}%
	\BibitemOpen
	\bibfield  {author} {\bibinfo {author} {\bibfnamefont {A.~V.}\ \bibnamefont {Chubukov}}\ and\ \bibinfo {author} {\bibfnamefont {D.~K.}\ \bibnamefont {Morr}},\ }\bibfield  {title} {\bibinfo {title} {{Electronic structure of underdoped cuprates}},\ }\href {https://doi.org/10.1016/s0370-1573(97)00033-1} {\bibfield  {journal} {\bibinfo  {journal} {Physics Reports}\ }\textbf {\bibinfo {volume} {288}},\ \bibinfo {pages} {355} (\bibinfo {year} {1997})},\ \Eprint {https://arxiv.org/abs/cond-mat/9701196} {cond-mat/9701196} \BibitemShut {NoStop}%
	\bibitem [{\citenamefont {Lee}\ and\ \citenamefont {Nagaosa}(1992)}]{Lee.1992}%
	\BibitemOpen
	\bibfield  {author} {\bibinfo {author} {\bibfnamefont {P.~A.}\ \bibnamefont {Lee}}\ and\ \bibinfo {author} {\bibfnamefont {N.}~\bibnamefont {Nagaosa}},\ }\bibfield  {title} {\bibinfo {title} {{Gauge theory of the normal state of high-Tc superconductors}},\ }\href {https://doi.org/10.1103/physrevb.46.5621} {\bibfield  {journal} {\bibinfo  {journal} {Physical Review B}\ }\textbf {\bibinfo {volume} {46}},\ \bibinfo {pages} {5621} (\bibinfo {year} {1992})}\BibitemShut {NoStop}%
	\bibitem [{\citenamefont {Wen}\ and\ \citenamefont {Lee}(1996)}]{Wen.1996}%
	\BibitemOpen
	\bibfield  {author} {\bibinfo {author} {\bibfnamefont {X.-G.}\ \bibnamefont {Wen}}\ and\ \bibinfo {author} {\bibfnamefont {P.~A.}\ \bibnamefont {Lee}},\ }\bibfield  {title} {\bibinfo {title} {{Theory of Underdoped Cuprates}},\ }\href {https://doi.org/10.1103/physrevlett.76.503} {\bibfield  {journal} {\bibinfo  {journal} {Physical Review Letters}\ }\textbf {\bibinfo {volume} {76}},\ \bibinfo {pages} {503} (\bibinfo {year} {1996})},\ \Eprint {https://arxiv.org/abs/cond-mat/9506065} {cond-mat/9506065} \BibitemShut {NoStop}%
	\bibitem [{\citenamefont {Gegenwart}\ \emph {et~al.}(2008)\citenamefont {Gegenwart}, \citenamefont {Si},\ and\ \citenamefont {Steglich}}]{Gegenwart_review_heavy_fermion}%
	\BibitemOpen
	\bibfield  {author} {\bibinfo {author} {\bibfnamefont {P.}~\bibnamefont {Gegenwart}}, \bibinfo {author} {\bibfnamefont {Q.}~\bibnamefont {Si}},\ and\ \bibinfo {author} {\bibfnamefont {F.}~\bibnamefont {Steglich}},\ }\bibfield  {title} {\bibinfo {title} {Quantum criticality in heavy-fermion metals},\ }\href {https://doi.org/10.1038/nphys892} {\bibfield  {journal} {\bibinfo  {journal} {Nature Physics}\ }\textbf {\bibinfo {volume} {4}},\ \bibinfo {pages} {186} (\bibinfo {year} {2008})}\BibitemShut {NoStop}%
	\bibitem [{\citenamefont {Si}\ and\ \citenamefont {Steglich}(2010)}]{Qimiao_review_heavy_fermion}%
	\BibitemOpen
	\bibfield  {author} {\bibinfo {author} {\bibfnamefont {Q.}~\bibnamefont {Si}}\ and\ \bibinfo {author} {\bibfnamefont {F.}~\bibnamefont {Steglich}},\ }\bibfield  {title} {\bibinfo {title} {Heavy fermions and quantum phase transitions},\ }\href {https://doi.org/10.1126/science.1191195} {\bibfield  {journal} {\bibinfo  {journal} {Science}\ }\textbf {\bibinfo {volume} {329}},\ \bibinfo {pages} {1161} (\bibinfo {year} {2010})},\ \Eprint {https://arxiv.org/abs/https://www.science.org/doi/pdf/10.1126/science.1191195} {https://www.science.org/doi/pdf/10.1126/science.1191195} \BibitemShut {NoStop}%
	\bibitem [{\citenamefont {Powell}\ and\ \citenamefont {McKenzie}(2011)}]{Powell_review_organic}%
	\BibitemOpen
	\bibfield  {author} {\bibinfo {author} {\bibfnamefont {B.~J.}\ \bibnamefont {Powell}}\ and\ \bibinfo {author} {\bibfnamefont {R.~H.}\ \bibnamefont {McKenzie}},\ }\bibfield  {title} {\bibinfo {title} {Quantum frustration in organic mott insulators: from spin liquids to unconventional superconductors},\ }\href {https://doi.org/10.1088/0034-4885/74/5/056501} {\bibfield  {journal} {\bibinfo  {journal} {Reports on Progress in Physics}\ }\textbf {\bibinfo {volume} {74}},\ \bibinfo {pages} {056501} (\bibinfo {year} {2011})}\BibitemShut {NoStop}%
	\bibitem [{\citenamefont {Balents}\ \emph {et~al.}(2020)\citenamefont {Balents}, \citenamefont {Dean}, \citenamefont {Efetov},\ and\ \citenamefont {Young}}]{Young_review_moire}%
	\BibitemOpen
	\bibfield  {author} {\bibinfo {author} {\bibfnamefont {L.}~\bibnamefont {Balents}}, \bibinfo {author} {\bibfnamefont {C.~R.}\ \bibnamefont {Dean}}, \bibinfo {author} {\bibfnamefont {D.~K.}\ \bibnamefont {Efetov}},\ and\ \bibinfo {author} {\bibfnamefont {A.~F.}\ \bibnamefont {Young}},\ }\bibfield  {title} {\bibinfo {title} {Superconductivity and strong correlations in moir{\'e} flat bands},\ }\href {https://doi.org/10.1038/s41567-020-0906-9} {\bibfield  {journal} {\bibinfo  {journal} {Nature Physics}\ }\textbf {\bibinfo {volume} {16}},\ \bibinfo {pages} {725} (\bibinfo {year} {2020})}\BibitemShut {NoStop}%
	\bibitem [{\citenamefont {Andrei}\ and\ \citenamefont {MacDonald}(2020)}]{MacDonald_review_moire}%
	\BibitemOpen
	\bibfield  {author} {\bibinfo {author} {\bibfnamefont {E.~Y.}\ \bibnamefont {Andrei}}\ and\ \bibinfo {author} {\bibfnamefont {A.~H.}\ \bibnamefont {MacDonald}},\ }\bibfield  {title} {\bibinfo {title} {Graphene bilayers with a twist},\ }\href {https://doi.org/10.1038/s41563-020-00840-0} {\bibfield  {journal} {\bibinfo  {journal} {Nature Materials}\ }\textbf {\bibinfo {volume} {19}},\ \bibinfo {pages} {1265} (\bibinfo {year} {2020})}\BibitemShut {NoStop}%
	\bibitem [{\citenamefont {L.~N.~Bulaevskii}\ and\ \citenamefont {Khomskii}(1968)}]{Bulaevskii1968}%
	\BibitemOpen
	\bibfield  {author} {\bibinfo {author} {\bibfnamefont {E.~L.~N.}\ \bibnamefont {L.~N.~Bulaevskii}}\ and\ \bibinfo {author} {\bibfnamefont {D.~I.}\ \bibnamefont {Khomskii}},\ }\bibfield  {title} {\bibinfo {title} {A new type of auto-localized state of a conduction electron in an antiferromagnetic semiconductor},\ }\href@noop {} {\bibfield  {journal} {\bibinfo  {journal} {Zh. Eksp. Teor. Fiz.}\ }\textbf {\bibinfo {volume} {54}},\ \bibinfo {pages} {1562} (\bibinfo {year} {1968})}\BibitemShut {NoStop}%
	\bibitem [{\citenamefont {Brinkman}\ and\ \citenamefont {Rice}(1970)}]{Brinkman1970}%
	\BibitemOpen
	\bibfield  {author} {\bibinfo {author} {\bibfnamefont {W.~F.}\ \bibnamefont {Brinkman}}\ and\ \bibinfo {author} {\bibfnamefont {T.~M.}\ \bibnamefont {Rice}},\ }\bibfield  {title} {\bibinfo {title} {Single-particle excitations in magnetic insulators},\ }\href {https://doi.org/10.1103/PhysRevB.2.1324} {\bibfield  {journal} {\bibinfo  {journal} {Phys. Rev. B}\ }\textbf {\bibinfo {volume} {2}},\ \bibinfo {pages} {1324} (\bibinfo {year} {1970})}\BibitemShut {NoStop}%
	\bibitem [{\citenamefont {Trugman}(1988)}]{Trugman1988}%
	\BibitemOpen
	\bibfield  {author} {\bibinfo {author} {\bibfnamefont {S.~A.}\ \bibnamefont {Trugman}},\ }\bibfield  {title} {\bibinfo {title} {Interaction of holes in a hubbard antiferromagnet and high-temperature superconductivity},\ }\href {https://doi.org/10.1103/PhysRevB.37.1597} {\bibfield  {journal} {\bibinfo  {journal} {Phys. Rev. B}\ }\textbf {\bibinfo {volume} {37}},\ \bibinfo {pages} {1597} (\bibinfo {year} {1988})}\BibitemShut {NoStop}%
	\bibitem [{\citenamefont {Schmitt-Rink}\ \emph {et~al.}(1988)\citenamefont {Schmitt-Rink}, \citenamefont {Varma},\ and\ \citenamefont {Ruckenstein}}]{Schmitt1988}%
	\BibitemOpen
	\bibfield  {author} {\bibinfo {author} {\bibfnamefont {S.}~\bibnamefont {Schmitt-Rink}}, \bibinfo {author} {\bibfnamefont {C.~M.}\ \bibnamefont {Varma}},\ and\ \bibinfo {author} {\bibfnamefont {A.~E.}\ \bibnamefont {Ruckenstein}},\ }\bibfield  {title} {\bibinfo {title} {Spectral function of holes in a quantum antiferromagnet},\ }\href {https://doi.org/10.1103/PhysRevLett.60.2793} {\bibfield  {journal} {\bibinfo  {journal} {Phys. Rev. Lett.}\ }\textbf {\bibinfo {volume} {60}},\ \bibinfo {pages} {2793} (\bibinfo {year} {1988})}\BibitemShut {NoStop}%
	\bibitem [{\citenamefont {Shraiman}\ and\ \citenamefont {Siggia}(1988)}]{Shraiman1988}%
	\BibitemOpen
	\bibfield  {author} {\bibinfo {author} {\bibfnamefont {B.~I.}\ \bibnamefont {Shraiman}}\ and\ \bibinfo {author} {\bibfnamefont {E.~D.}\ \bibnamefont {Siggia}},\ }\bibfield  {title} {\bibinfo {title} {Mobile vacancies in a quantum heisenberg antiferromagnet},\ }\href {https://doi.org/10.1103/PhysRevLett.61.467} {\bibfield  {journal} {\bibinfo  {journal} {Phys. Rev. Lett.}\ }\textbf {\bibinfo {volume} {61}},\ \bibinfo {pages} {467} (\bibinfo {year} {1988})}\BibitemShut {NoStop}%
	\bibitem [{\citenamefont {Sachdev}(1989)}]{Sachdev1989}%
	\BibitemOpen
	\bibfield  {author} {\bibinfo {author} {\bibfnamefont {S.}~\bibnamefont {Sachdev}},\ }\bibfield  {title} {\bibinfo {title} {Hole motion in a quantum n\'eel state},\ }\href {https://doi.org/10.1103/PhysRevB.39.12232} {\bibfield  {journal} {\bibinfo  {journal} {Phys. Rev. B}\ }\textbf {\bibinfo {volume} {39}},\ \bibinfo {pages} {12232} (\bibinfo {year} {1989})}\BibitemShut {NoStop}%
	\bibitem [{\citenamefont {Kane}\ \emph {et~al.}(1989)\citenamefont {Kane}, \citenamefont {Lee},\ and\ \citenamefont {Read}}]{Kane1989}%
	\BibitemOpen
	\bibfield  {author} {\bibinfo {author} {\bibfnamefont {C.~L.}\ \bibnamefont {Kane}}, \bibinfo {author} {\bibfnamefont {P.~A.}\ \bibnamefont {Lee}},\ and\ \bibinfo {author} {\bibfnamefont {N.}~\bibnamefont {Read}},\ }\bibfield  {title} {\bibinfo {title} {Motion of a single hole in a quantum antiferromagnet},\ }\href {https://doi.org/10.1103/PhysRevB.39.6880} {\bibfield  {journal} {\bibinfo  {journal} {Phys. Rev. B}\ }\textbf {\bibinfo {volume} {39}},\ \bibinfo {pages} {6880} (\bibinfo {year} {1989})}\BibitemShut {NoStop}%
	\bibitem [{\citenamefont {Dagotto}\ \emph {et~al.}(1989)\citenamefont {Dagotto}, \citenamefont {Moreo},\ and\ \citenamefont {Barnes}}]{Dagotto1989}%
	\BibitemOpen
	\bibfield  {author} {\bibinfo {author} {\bibfnamefont {E.}~\bibnamefont {Dagotto}}, \bibinfo {author} {\bibfnamefont {A.}~\bibnamefont {Moreo}},\ and\ \bibinfo {author} {\bibfnamefont {T.}~\bibnamefont {Barnes}},\ }\bibfield  {title} {\bibinfo {title} {Hubbard model with one hole: Ground-state properties},\ }\href {https://doi.org/10.1103/PhysRevB.40.6721} {\bibfield  {journal} {\bibinfo  {journal} {Phys. Rev. B}\ }\textbf {\bibinfo {volume} {40}},\ \bibinfo {pages} {6721} (\bibinfo {year} {1989})}\BibitemShut {NoStop}%
	\bibitem [{\citenamefont {Trugman}(1990)}]{Trugman1990}%
	\BibitemOpen
	\bibfield  {author} {\bibinfo {author} {\bibfnamefont {S.~A.}\ \bibnamefont {Trugman}},\ }\bibfield  {title} {\bibinfo {title} {Spectral function of a hole in a hubbard antiferromagnet},\ }\href {https://doi.org/10.1103/PhysRevB.41.892} {\bibfield  {journal} {\bibinfo  {journal} {Phys. Rev. B}\ }\textbf {\bibinfo {volume} {41}},\ \bibinfo {pages} {892} (\bibinfo {year} {1990})}\BibitemShut {NoStop}%
	\bibitem [{\citenamefont {Auerbach}\ and\ \citenamefont {Larson}(1991{\natexlab{a}})}]{Auerbach1991}%
	\BibitemOpen
	\bibfield  {author} {\bibinfo {author} {\bibfnamefont {A.}~\bibnamefont {Auerbach}}\ and\ \bibinfo {author} {\bibfnamefont {B.~E.}\ \bibnamefont {Larson}},\ }\bibfield  {title} {\bibinfo {title} {Small-polaron theory of doped antiferromagnets},\ }\href {https://doi.org/10.1103/PhysRevLett.66.2262} {\bibfield  {journal} {\bibinfo  {journal} {Phys. Rev. Lett.}\ }\textbf {\bibinfo {volume} {66}},\ \bibinfo {pages} {2262} (\bibinfo {year} {1991}{\natexlab{a}})}\BibitemShut {NoStop}%
	\bibitem [{\citenamefont {Martinez}\ and\ \citenamefont {Horsch}(1991{\natexlab{a}})}]{Martinez.1991}%
	\BibitemOpen
	\bibfield  {author} {\bibinfo {author} {\bibfnamefont {G.}~\bibnamefont {Martinez}}\ and\ \bibinfo {author} {\bibfnamefont {P.}~\bibnamefont {Horsch}},\ }\bibfield  {title} {\bibinfo {title} {{Spin polarons in the t-J model}},\ }\href {https://doi.org/10.1103/physrevb.44.317} {\bibfield  {journal} {\bibinfo  {journal} {Physical Review B}\ }\textbf {\bibinfo {volume} {44}},\ \bibinfo {pages} {317} (\bibinfo {year} {1991}{\natexlab{a}})}\BibitemShut {NoStop}%
	\bibitem [{\citenamefont {Liu}\ and\ \citenamefont {Manousakis}(1991)}]{Manousakis1991}%
	\BibitemOpen
	\bibfield  {author} {\bibinfo {author} {\bibfnamefont {Z.}~\bibnamefont {Liu}}\ and\ \bibinfo {author} {\bibfnamefont {E.}~\bibnamefont {Manousakis}},\ }\bibfield  {title} {\bibinfo {title} {Spectral function of a hole in the t-j model},\ }\href {https://doi.org/10.1103/PhysRevB.44.2414} {\bibfield  {journal} {\bibinfo  {journal} {Phys. Rev. B}\ }\textbf {\bibinfo {volume} {44}},\ \bibinfo {pages} {2414} (\bibinfo {year} {1991})}\BibitemShut {NoStop}%
	\bibitem [{\citenamefont {Martinez}\ and\ \citenamefont {Horsch}(1991{\natexlab{b}})}]{Horsch1991}%
	\BibitemOpen
	\bibfield  {author} {\bibinfo {author} {\bibfnamefont {G.}~\bibnamefont {Martinez}}\ and\ \bibinfo {author} {\bibfnamefont {P.}~\bibnamefont {Horsch}},\ }\bibfield  {title} {\bibinfo {title} {Spin polarons in the t-j model},\ }\href {https://doi.org/10.1103/PhysRevB.44.317} {\bibfield  {journal} {\bibinfo  {journal} {Phys. Rev. B}\ }\textbf {\bibinfo {volume} {44}},\ \bibinfo {pages} {317} (\bibinfo {year} {1991}{\natexlab{b}})}\BibitemShut {NoStop}%
	\bibitem [{\citenamefont {Khaliullin}\ and\ \citenamefont {Horsch}(1993)}]{Khaliullin.1993}%
	\BibitemOpen
	\bibfield  {author} {\bibinfo {author} {\bibfnamefont {G.}~\bibnamefont {Khaliullin}}\ and\ \bibinfo {author} {\bibfnamefont {P.}~\bibnamefont {Horsch}},\ }\bibfield  {title} {\bibinfo {title} {{Doping dependence of long-range magnetic order in the t-J model}},\ }\href {https://doi.org/10.1103/physrevb.47.463} {\bibfield  {journal} {\bibinfo  {journal} {Physical Review B}\ }\textbf {\bibinfo {volume} {47}},\ \bibinfo {pages} {463} (\bibinfo {year} {1993})}\BibitemShut {NoStop}%
	\bibitem [{\citenamefont {Sherman}\ and\ \citenamefont {Schreiber}(1994)}]{Sherman.1994}%
	\BibitemOpen
	\bibfield  {author} {\bibinfo {author} {\bibfnamefont {A.}~\bibnamefont {Sherman}}\ and\ \bibinfo {author} {\bibfnamefont {M.}~\bibnamefont {Schreiber}},\ }\bibfield  {title} {\bibinfo {title} {{Evolution of hole and magnon spectra of the two-dimensional t-J model with doping}},\ }\href {https://doi.org/10.1103/physrevb.50.12887} {\bibfield  {journal} {\bibinfo  {journal} {Physical Review B}\ }\textbf {\bibinfo {volume} {50}},\ \bibinfo {pages} {12887} (\bibinfo {year} {1994})}\BibitemShut {NoStop}%
	\bibitem [{\citenamefont {Vojta}\ and\ \citenamefont {Becker}(1996)}]{Vojta.1996}%
	\BibitemOpen
	\bibfield  {author} {\bibinfo {author} {\bibfnamefont {M.}~\bibnamefont {Vojta}}\ and\ \bibinfo {author} {\bibfnamefont {K.~W.}\ \bibnamefont {Becker}},\ }\bibfield  {title} {\bibinfo {title} {{Cumulant approach to weakly doped antiferromagnets}},\ }\href {https://doi.org/10.1103/physrevb.54.15483} {\bibfield  {journal} {\bibinfo  {journal} {Physical Review B}\ }\textbf {\bibinfo {volume} {54}},\ \bibinfo {pages} {15483} (\bibinfo {year} {1996})},\ \Eprint {https://arxiv.org/abs/cond-mat/9609098} {cond-mat/9609098} \BibitemShut {NoStop}%
	\bibitem [{\citenamefont {Grusdt}\ \emph {et~al.}(2018)\citenamefont {Grusdt}, \citenamefont {K\'anasz-Nagy}, \citenamefont {Bohrdt}, \citenamefont {Chiu}, \citenamefont {Ji}, \citenamefont {Greiner}, \citenamefont {Greif},\ and\ \citenamefont {Demler}}]{Grusdt2018}%
	\BibitemOpen
	\bibfield  {author} {\bibinfo {author} {\bibfnamefont {F.}~\bibnamefont {Grusdt}}, \bibinfo {author} {\bibfnamefont {M.}~\bibnamefont {K\'anasz-Nagy}}, \bibinfo {author} {\bibfnamefont {A.}~\bibnamefont {Bohrdt}}, \bibinfo {author} {\bibfnamefont {C.~S.}\ \bibnamefont {Chiu}}, \bibinfo {author} {\bibfnamefont {G.}~\bibnamefont {Ji}}, \bibinfo {author} {\bibfnamefont {M.}~\bibnamefont {Greiner}}, \bibinfo {author} {\bibfnamefont {D.}~\bibnamefont {Greif}},\ and\ \bibinfo {author} {\bibfnamefont {E.}~\bibnamefont {Demler}},\ }\bibfield  {title} {\bibinfo {title} {Parton theory of magnetic polarons: Mesonic resonances and signatures in dynamics},\ }\href {https://doi.org/10.1103/PhysRevX.8.011046} {\bibfield  {journal} {\bibinfo  {journal} {Phys. Rev. X}\ }\textbf {\bibinfo {volume} {8}},\ \bibinfo {pages} {011046} (\bibinfo {year} {2018})}\BibitemShut {NoStop}%
	\bibitem [{\citenamefont {Bohrdt}\ \emph {et~al.}(2020)\citenamefont {Bohrdt}, \citenamefont {Demler}, \citenamefont {Pollmann}, \citenamefont {Knap},\ and\ \citenamefont {Grusdt}}]{Bohrdt2020}%
	\BibitemOpen
	\bibfield  {author} {\bibinfo {author} {\bibfnamefont {A.}~\bibnamefont {Bohrdt}}, \bibinfo {author} {\bibfnamefont {E.}~\bibnamefont {Demler}}, \bibinfo {author} {\bibfnamefont {F.}~\bibnamefont {Pollmann}}, \bibinfo {author} {\bibfnamefont {M.}~\bibnamefont {Knap}},\ and\ \bibinfo {author} {\bibfnamefont {F.}~\bibnamefont {Grusdt}},\ }\bibfield  {title} {\bibinfo {title} {Parton theory of angle-resolved photoemission spectroscopy spectra in antiferromagnetic mott insulators},\ }\href {https://doi.org/10.1103/PhysRevB.102.035139} {\bibfield  {journal} {\bibinfo  {journal} {Phys. Rev. B}\ }\textbf {\bibinfo {volume} {102}},\ \bibinfo {pages} {035139} (\bibinfo {year} {2020})}\BibitemShut {NoStop}%
	\bibitem [{\citenamefont {Bloch}\ \emph {et~al.}(2012)\citenamefont {Bloch}, \citenamefont {Dalibard},\ and\ \citenamefont {Nascimb{\`e}ne}}]{Bloch2012_review}%
	\BibitemOpen
	\bibfield  {author} {\bibinfo {author} {\bibfnamefont {I.}~\bibnamefont {Bloch}}, \bibinfo {author} {\bibfnamefont {J.}~\bibnamefont {Dalibard}},\ and\ \bibinfo {author} {\bibfnamefont {S.}~\bibnamefont {Nascimb{\`e}ne}},\ }\bibfield  {title} {\bibinfo {title} {Quantum simulations with ultracold quantum gases},\ }\href@noop {} {\bibfield  {journal} {\bibinfo  {journal} {Nat. Phys.}\ }\textbf {\bibinfo {volume} {8}},\ \bibinfo {pages} {267} (\bibinfo {year} {2012})}\BibitemShut {NoStop}%
	\bibitem [{\citenamefont {Blatt}\ and\ \citenamefont {Roos}(2012)}]{Blatt2012_review}%
	\BibitemOpen
	\bibfield  {author} {\bibinfo {author} {\bibfnamefont {R.}~\bibnamefont {Blatt}}\ and\ \bibinfo {author} {\bibfnamefont {C.~F.}\ \bibnamefont {Roos}},\ }\bibfield  {title} {\bibinfo {title} {Quantum simulations with trapped ions},\ }\href@noop {} {\bibfield  {journal} {\bibinfo  {journal} {Nat. Phys.}\ }\textbf {\bibinfo {volume} {8}},\ \bibinfo {pages} {277} (\bibinfo {year} {2012})}\BibitemShut {NoStop}%
	\bibitem [{\citenamefont {Browaeys}\ and\ \citenamefont {Lahaye}(2020)}]{Browaeys2020_review}%
	\BibitemOpen
	\bibfield  {author} {\bibinfo {author} {\bibfnamefont {A.}~\bibnamefont {Browaeys}}\ and\ \bibinfo {author} {\bibfnamefont {T.}~\bibnamefont {Lahaye}},\ }\bibfield  {title} {\bibinfo {title} {Many-body physics with individually controlled rydberg atoms},\ }\href@noop {} {\bibfield  {journal} {\bibinfo  {journal} {Nat. Phys.}\ }\textbf {\bibinfo {volume} {16}},\ \bibinfo {pages} {132} (\bibinfo {year} {2020})}\BibitemShut {NoStop}%
	\bibitem [{\citenamefont {Koepsell}\ \emph {et~al.}(2019)\citenamefont {Koepsell}, \citenamefont {Vijayan}, \citenamefont {Sompet}, \citenamefont {Grusdt}, \citenamefont {Hilker}, \citenamefont {Demler}, \citenamefont {Salomon}, \citenamefont {Bloch},\ and\ \citenamefont {Gross}}]{Koepsell2019}%
	\BibitemOpen
	\bibfield  {author} {\bibinfo {author} {\bibfnamefont {J.}~\bibnamefont {Koepsell}}, \bibinfo {author} {\bibfnamefont {J.}~\bibnamefont {Vijayan}}, \bibinfo {author} {\bibfnamefont {P.}~\bibnamefont {Sompet}}, \bibinfo {author} {\bibfnamefont {F.}~\bibnamefont {Grusdt}}, \bibinfo {author} {\bibfnamefont {T.~A.}\ \bibnamefont {Hilker}}, \bibinfo {author} {\bibfnamefont {E.}~\bibnamefont {Demler}}, \bibinfo {author} {\bibfnamefont {G.}~\bibnamefont {Salomon}}, \bibinfo {author} {\bibfnamefont {I.}~\bibnamefont {Bloch}},\ and\ \bibinfo {author} {\bibfnamefont {C.}~\bibnamefont {Gross}},\ }\bibfield  {title} {\bibinfo {title} {Imaging magnetic polarons in the doped fermi--hubbard model},\ }\href {https://doi.org/10.1038/s41586-019-1463-1} {\bibfield  {journal} {\bibinfo  {journal} {Nature}\ }\textbf {\bibinfo {volume} {572}},\ \bibinfo {pages} {358} (\bibinfo {year} {2019})}\BibitemShut {NoStop}%
	\bibitem [{\citenamefont {Prichard}\ \emph {et~al.}(2024)\citenamefont {Prichard}, \citenamefont {Spar}, \citenamefont {Morera}, \citenamefont {Demler}, \citenamefont {Yan},\ and\ \citenamefont {Bakr}}]{OL_Prichard2024_Nagaoka}%
	\BibitemOpen
	\bibfield  {author} {\bibinfo {author} {\bibfnamefont {M.~L.}\ \bibnamefont {Prichard}}, \bibinfo {author} {\bibfnamefont {B.~M.}\ \bibnamefont {Spar}}, \bibinfo {author} {\bibfnamefont {I.}~\bibnamefont {Morera}}, \bibinfo {author} {\bibfnamefont {E.}~\bibnamefont {Demler}}, \bibinfo {author} {\bibfnamefont {Z.~Z.}\ \bibnamefont {Yan}},\ and\ \bibinfo {author} {\bibfnamefont {W.~S.}\ \bibnamefont {Bakr}},\ }\bibfield  {title} {\bibinfo {title} {{Directly imaging spin polarons in a kinetically frustrated Hubbard system}},\ }\href {https://doi.org/10.1038/s41586-024-07356-6} {\bibfield  {journal} {\bibinfo  {journal} {Nature}\ }\textbf {\bibinfo {volume} {629}},\ \bibinfo {pages} {323} (\bibinfo {year} {2024})}\BibitemShut {NoStop}%
	\bibitem [{\citenamefont {Lebrat}\ \emph {et~al.}(2024)\citenamefont {Lebrat}, \citenamefont {Xu}, \citenamefont {Kendrick}, \citenamefont {Kale}, \citenamefont {Gang}, \citenamefont {Seetharaman}, \citenamefont {Morera}, \citenamefont {Khatami}, \citenamefont {Demler},\ and\ \citenamefont {Greiner}}]{OL_Lebrat2024_Nagaoka}%
	\BibitemOpen
	\bibfield  {author} {\bibinfo {author} {\bibfnamefont {M.}~\bibnamefont {Lebrat}}, \bibinfo {author} {\bibfnamefont {M.}~\bibnamefont {Xu}}, \bibinfo {author} {\bibfnamefont {L.~H.}\ \bibnamefont {Kendrick}}, \bibinfo {author} {\bibfnamefont {A.}~\bibnamefont {Kale}}, \bibinfo {author} {\bibfnamefont {Y.}~\bibnamefont {Gang}}, \bibinfo {author} {\bibfnamefont {P.}~\bibnamefont {Seetharaman}}, \bibinfo {author} {\bibfnamefont {I.}~\bibnamefont {Morera}}, \bibinfo {author} {\bibfnamefont {E.}~\bibnamefont {Khatami}}, \bibinfo {author} {\bibfnamefont {E.}~\bibnamefont {Demler}},\ and\ \bibinfo {author} {\bibfnamefont {M.}~\bibnamefont {Greiner}},\ }\bibfield  {title} {\bibinfo {title} {{Observation of Nagaoka polarons in a Fermi--Hubbard quantum simulator}},\ }\href {https://doi.org/10.1038/s41586-024-07272-9} {\bibfield  {journal} {\bibinfo  {journal} {Nature}\ }\textbf {\bibinfo {volume} {629}},\ \bibinfo {pages} {317} (\bibinfo {year} {2024})}\BibitemShut {NoStop}%
	\bibitem [{\citenamefont {Koepsell}\ \emph {et~al.}(2021)\citenamefont {Koepsell}, \citenamefont {Bourgund}, \citenamefont {Sompet}, \citenamefont {Hirthe}, \citenamefont {Bohrdt}, \citenamefont {Wang}, \citenamefont {Grusdt}, \citenamefont {Demler}, \citenamefont {Salomon}, \citenamefont {Gross},\ and\ \citenamefont {Bloch}}]{Koepsell_doping_2021}%
	\BibitemOpen
	\bibfield  {author} {\bibinfo {author} {\bibfnamefont {J.}~\bibnamefont {Koepsell}}, \bibinfo {author} {\bibfnamefont {D.}~\bibnamefont {Bourgund}}, \bibinfo {author} {\bibfnamefont {P.}~\bibnamefont {Sompet}}, \bibinfo {author} {\bibfnamefont {S.}~\bibnamefont {Hirthe}}, \bibinfo {author} {\bibfnamefont {A.}~\bibnamefont {Bohrdt}}, \bibinfo {author} {\bibfnamefont {Y.}~\bibnamefont {Wang}}, \bibinfo {author} {\bibfnamefont {F.}~\bibnamefont {Grusdt}}, \bibinfo {author} {\bibfnamefont {E.}~\bibnamefont {Demler}}, \bibinfo {author} {\bibfnamefont {G.}~\bibnamefont {Salomon}}, \bibinfo {author} {\bibfnamefont {C.}~\bibnamefont {Gross}},\ and\ \bibinfo {author} {\bibfnamefont {I.}~\bibnamefont {Bloch}},\ }\bibfield  {title} {\bibinfo {title} {Microscopic evolution of doped mott insulators from polaronic metal to fermi liquid},\ }\href {https://doi.org/10.1126/science.abe7165} {\bibfield  {journal} {\bibinfo  {journal} {Science}\ }\textbf {\bibinfo {volume} {374}},\ \bibinfo {pages} {82} (\bibinfo {year}
		{2021})},\ \Eprint {https://arxiv.org/abs/https://www.science.org/doi/pdf/10.1126/science.abe7165} {https://www.science.org/doi/pdf/10.1126/science.abe7165} \BibitemShut {NoStop}%
	\bibitem [{\citenamefont {Gull}\ \emph {et~al.}(2013)\citenamefont {Gull}, \citenamefont {Parcollet},\ and\ \citenamefont {Millis}}]{Gull.2013}%
	\BibitemOpen
	\bibfield  {author} {\bibinfo {author} {\bibfnamefont {E.}~\bibnamefont {Gull}}, \bibinfo {author} {\bibfnamefont {O.}~\bibnamefont {Parcollet}},\ and\ \bibinfo {author} {\bibfnamefont {A.~J.}\ \bibnamefont {Millis}},\ }\bibfield  {title} {\bibinfo {title} {{Superconductivity and the Pseudogap in the Two-Dimensional Hubbard Model}},\ }\href {https://doi.org/10.1103/physrevlett.110.216405} {\bibfield  {journal} {\bibinfo  {journal} {Physical Review Letters}\ }\textbf {\bibinfo {volume} {110}},\ \bibinfo {pages} {216405} (\bibinfo {year} {2013})},\ \Eprint {https://arxiv.org/abs/1207.2490} {1207.2490} \BibitemShut {NoStop}%
	\bibitem [{\citenamefont {Werner}\ and\ \citenamefont {Millis}(2007)}]{Werner.2007lvq}%
	\BibitemOpen
	\bibfield  {author} {\bibinfo {author} {\bibfnamefont {P.}~\bibnamefont {Werner}}\ and\ \bibinfo {author} {\bibfnamefont {A.~J.}\ \bibnamefont {Millis}},\ }\bibfield  {title} {\bibinfo {title} {{Doping-driven Mott transition in the one-band Hubbard model}},\ }\href {https://doi.org/10.1103/physrevb.75.085108} {\bibfield  {journal} {\bibinfo  {journal} {Physical Review B}\ }\textbf {\bibinfo {volume} {75}},\ \bibinfo {pages} {085108} (\bibinfo {year} {2007})},\ \Eprint {https://arxiv.org/abs/cond-mat/0610401} {cond-mat/0610401} \BibitemShut {NoStop}%
	\bibitem [{\citenamefont {Kyung}\ \emph {et~al.}(2006)\citenamefont {Kyung}, \citenamefont {Kancharla}, \citenamefont {Sénéchal}, \citenamefont {Tremblay}, \citenamefont {Civelli},\ and\ \citenamefont {Kotliar}}]{Kyung.2006}%
	\BibitemOpen
	\bibfield  {author} {\bibinfo {author} {\bibfnamefont {B.}~\bibnamefont {Kyung}}, \bibinfo {author} {\bibfnamefont {S.~S.}\ \bibnamefont {Kancharla}}, \bibinfo {author} {\bibfnamefont {D.}~\bibnamefont {Sénéchal}}, \bibinfo {author} {\bibfnamefont {A.~M.~S.}\ \bibnamefont {Tremblay}}, \bibinfo {author} {\bibfnamefont {M.}~\bibnamefont {Civelli}},\ and\ \bibinfo {author} {\bibfnamefont {G.}~\bibnamefont {Kotliar}},\ }\bibfield  {title} {\bibinfo {title} {{Pseudogap induced by short-range spin correlations in a doped Mott insulator}},\ }\href {https://doi.org/10.1103/physrevb.73.165114} {\bibfield  {journal} {\bibinfo  {journal} {Physical Review B}\ }\textbf {\bibinfo {volume} {73}},\ \bibinfo {pages} {165114} (\bibinfo {year} {2006})},\ \Eprint {https://arxiv.org/abs/cond-mat/0502565} {cond-mat/0502565} \BibitemShut {NoStop}%
	\bibitem [{\citenamefont {Xu}\ \emph {et~al.}(2025)\citenamefont {Xu}, \citenamefont {Kendrick}, \citenamefont {Kale}, \citenamefont {Gang}, \citenamefont {Feng}, \citenamefont {Zhang}, \citenamefont {Young}, \citenamefont {Lebrat},\ and\ \citenamefont {Greiner}}]{Xu_cryogenic_2025}%
	\BibitemOpen
	\bibfield  {author} {\bibinfo {author} {\bibfnamefont {M.}~\bibnamefont {Xu}}, \bibinfo {author} {\bibfnamefont {L.~H.}\ \bibnamefont {Kendrick}}, \bibinfo {author} {\bibfnamefont {A.}~\bibnamefont {Kale}}, \bibinfo {author} {\bibfnamefont {Y.}~\bibnamefont {Gang}}, \bibinfo {author} {\bibfnamefont {C.}~\bibnamefont {Feng}}, \bibinfo {author} {\bibfnamefont {S.}~\bibnamefont {Zhang}}, \bibinfo {author} {\bibfnamefont {A.~W.}\ \bibnamefont {Young}}, \bibinfo {author} {\bibfnamefont {M.}~\bibnamefont {Lebrat}},\ and\ \bibinfo {author} {\bibfnamefont {M.}~\bibnamefont {Greiner}},\ }\bibfield  {title} {\bibinfo {title} {A neutral-atom hubbard quantum simulator in the cryogenic regime},\ }\href {https://doi.org/10.1038/s41586-025-09112-w} {\bibfield  {journal} {\bibinfo  {journal} {Nature}\ }\textbf {\bibinfo {volume} {642}},\ \bibinfo {pages} {909} (\bibinfo {year} {2025})}\BibitemShut {NoStop}%
	\bibitem [{\citenamefont {Chalopin}\ \emph {et~al.}(2026)\citenamefont {Chalopin}, \citenamefont {Bojović}, \citenamefont {Wang}, \citenamefont {Franz}, \citenamefont {Sinha}, \citenamefont {Wang}, \citenamefont {Bourgund}, \citenamefont {Obermeyer}, \citenamefont {Grusdt}, \citenamefont {Bohrdt}, \citenamefont {Pollet}, \citenamefont {Wietek}, \citenamefont {Georges}, \citenamefont {Hilker},\ and\ \citenamefont {Bloch}}]{chalopin_observation_2026}%
	\BibitemOpen
	\bibfield  {author} {\bibinfo {author} {\bibfnamefont {T.}~\bibnamefont {Chalopin}}, \bibinfo {author} {\bibfnamefont {P.}~\bibnamefont {Bojović}}, \bibinfo {author} {\bibfnamefont {S.}~\bibnamefont {Wang}}, \bibinfo {author} {\bibfnamefont {T.}~\bibnamefont {Franz}}, \bibinfo {author} {\bibfnamefont {A.}~\bibnamefont {Sinha}}, \bibinfo {author} {\bibfnamefont {Z.}~\bibnamefont {Wang}}, \bibinfo {author} {\bibfnamefont {D.}~\bibnamefont {Bourgund}}, \bibinfo {author} {\bibfnamefont {J.}~\bibnamefont {Obermeyer}}, \bibinfo {author} {\bibfnamefont {F.}~\bibnamefont {Grusdt}}, \bibinfo {author} {\bibfnamefont {A.}~\bibnamefont {Bohrdt}}, \bibinfo {author} {\bibfnamefont {L.}~\bibnamefont {Pollet}}, \bibinfo {author} {\bibfnamefont {A.}~\bibnamefont {Wietek}}, \bibinfo {author} {\bibfnamefont {A.}~\bibnamefont {Georges}}, \bibinfo {author} {\bibfnamefont {T.}~\bibnamefont {Hilker}},\ and\ \bibinfo {author} {\bibfnamefont {I.}~\bibnamefont {Bloch}},\ }\bibfield  {title} {\bibinfo {title} {Observation of
			emergent scaling of spin–charge correlations at the onset of the pseudogap},\ }\href {https://doi.org/10.1073/pnas.2525539123} {\bibfield  {journal} {\bibinfo  {journal} {Proceedings of the National Academy of Sciences}\ }\textbf {\bibinfo {volume} {123}},\ \bibinfo {pages} {e2525539123} (\bibinfo {year} {2026})}\BibitemShut {NoStop}%
	\bibitem [{\citenamefont {Kendrick}\ \emph {et~al.}(2025)\citenamefont {Kendrick}, \citenamefont {Kale}, \citenamefont {Gang}, \citenamefont {Deters}, \citenamefont {Lebrat}, \citenamefont {Young},\ and\ \citenamefont {Greiner}}]{kendrick2025pseudogapfermihubbardquantumsimulator}%
	\BibitemOpen
	\bibfield  {author} {\bibinfo {author} {\bibfnamefont {L.~H.}\ \bibnamefont {Kendrick}}, \bibinfo {author} {\bibfnamefont {A.}~\bibnamefont {Kale}}, \bibinfo {author} {\bibfnamefont {Y.}~\bibnamefont {Gang}}, \bibinfo {author} {\bibfnamefont {A.~D.}\ \bibnamefont {Deters}}, \bibinfo {author} {\bibfnamefont {M.}~\bibnamefont {Lebrat}}, \bibinfo {author} {\bibfnamefont {A.~W.}\ \bibnamefont {Young}},\ and\ \bibinfo {author} {\bibfnamefont {M.}~\bibnamefont {Greiner}},\ }\href {https://arxiv.org/abs/2509.18075} {\bibinfo {title} {Pseudogap in a fermi-hubbard quantum simulator}} (\bibinfo {year} {2025}),\ \Eprint {https://arxiv.org/abs/2509.18075} {arXiv:2509.18075 [cond-mat.quant-gas]} \BibitemShut {NoStop}%
	\bibitem [{\citenamefont {Millis}\ \emph {et~al.}(1990)\citenamefont {Millis}, \citenamefont {Monien},\ and\ \citenamefont {Pines}}]{Millis.1990}%
	\BibitemOpen
	\bibfield  {author} {\bibinfo {author} {\bibfnamefont {A.~J.}\ \bibnamefont {Millis}}, \bibinfo {author} {\bibfnamefont {H.}~\bibnamefont {Monien}},\ and\ \bibinfo {author} {\bibfnamefont {D.}~\bibnamefont {Pines}},\ }\bibfield  {title} {\bibinfo {title} {{Phenomenological model of nuclear relaxation in the normal state of YBa2Cu3O7}},\ }\href {https://doi.org/10.1103/physrevb.42.167} {\bibfield  {journal} {\bibinfo  {journal} {Physical Review B}\ }\textbf {\bibinfo {volume} {42}},\ \bibinfo {pages} {167} (\bibinfo {year} {1990})}\BibitemShut {NoStop}%
	\bibitem [{\citenamefont {Chakravarty}\ \emph {et~al.}(1989)\citenamefont {Chakravarty}, \citenamefont {Halperin},\ and\ \citenamefont {Nelson}}]{Chakravarty.1989}%
	\BibitemOpen
	\bibfield  {author} {\bibinfo {author} {\bibfnamefont {S.}~\bibnamefont {Chakravarty}}, \bibinfo {author} {\bibfnamefont {B.~I.}\ \bibnamefont {Halperin}},\ and\ \bibinfo {author} {\bibfnamefont {D.~R.}\ \bibnamefont {Nelson}},\ }\bibfield  {title} {\bibinfo {title} {{Two-dimensional quantum Heisenberg antiferromagnet at low temperatures}},\ }\href {https://doi.org/10.1103/physrevb.39.2344} {\bibfield  {journal} {\bibinfo  {journal} {Physical Review B}\ }\textbf {\bibinfo {volume} {39}},\ \bibinfo {pages} {2344} (\bibinfo {year} {1989})}\BibitemShut {NoStop}%
	\bibitem [{\citenamefont {Blumberg}\ \emph {et~al.}(1996)\citenamefont {Blumberg}, \citenamefont {Abbamonte}, \citenamefont {Klein}, \citenamefont {Lee}, \citenamefont {Ginsberg}, \citenamefont {Miller},\ and\ \citenamefont {Zibold}}]{Blumberg1996}%
	\BibitemOpen
	\bibfield  {author} {\bibinfo {author} {\bibfnamefont {G.}~\bibnamefont {Blumberg}}, \bibinfo {author} {\bibfnamefont {P.}~\bibnamefont {Abbamonte}}, \bibinfo {author} {\bibfnamefont {M.~V.}\ \bibnamefont {Klein}}, \bibinfo {author} {\bibfnamefont {W.~C.}\ \bibnamefont {Lee}}, \bibinfo {author} {\bibfnamefont {D.~M.}\ \bibnamefont {Ginsberg}}, \bibinfo {author} {\bibfnamefont {L.~L.}\ \bibnamefont {Miller}},\ and\ \bibinfo {author} {\bibfnamefont {A.}~\bibnamefont {Zibold}},\ }\bibfield  {title} {\bibinfo {title} {Resonant two-magnon raman scattering in cuprate antiferromagnetic insulators},\ }\href {https://doi.org/10.1103/PhysRevB.53.R11930} {\bibfield  {journal} {\bibinfo  {journal} {Phys. Rev. B}\ }\textbf {\bibinfo {volume} {53}},\ \bibinfo {pages} {R11930} (\bibinfo {year} {1996})}\BibitemShut {NoStop}%
	\bibitem [{\citenamefont {Devereaux}\ and\ \citenamefont {Hackl}(2007)}]{Devereaux.2007}%
	\BibitemOpen
	\bibfield  {author} {\bibinfo {author} {\bibfnamefont {T.~P.}\ \bibnamefont {Devereaux}}\ and\ \bibinfo {author} {\bibfnamefont {R.}~\bibnamefont {Hackl}},\ }\bibfield  {title} {\bibinfo {title} {{Inelastic light scattering from correlated electrons}},\ }\href {https://doi.org/10.1103/revmodphys.79.175} {\bibfield  {journal} {\bibinfo  {journal} {Reviews of Modern Physics}\ }\textbf {\bibinfo {volume} {79}},\ \bibinfo {pages} {175} (\bibinfo {year} {2007})},\ \Eprint {https://arxiv.org/abs/cond-mat/0607554} {cond-mat/0607554} \BibitemShut {NoStop}%
	\bibitem [{\citenamefont {Coldea}\ \emph {et~al.}(2001)\citenamefont {Coldea}, \citenamefont {Hayden}, \citenamefont {Aeppli}, \citenamefont {Perring}, \citenamefont {Frost}, \citenamefont {Mason}, \citenamefont {Cheong},\ and\ \citenamefont {Fisk}}]{Coldea.2001}%
	\BibitemOpen
	\bibfield  {author} {\bibinfo {author} {\bibfnamefont {R.}~\bibnamefont {Coldea}}, \bibinfo {author} {\bibfnamefont {S.~M.}\ \bibnamefont {Hayden}}, \bibinfo {author} {\bibfnamefont {G.}~\bibnamefont {Aeppli}}, \bibinfo {author} {\bibfnamefont {T.~G.}\ \bibnamefont {Perring}}, \bibinfo {author} {\bibfnamefont {C.~D.}\ \bibnamefont {Frost}}, \bibinfo {author} {\bibfnamefont {T.~E.}\ \bibnamefont {Mason}}, \bibinfo {author} {\bibfnamefont {S.-W.}\ \bibnamefont {Cheong}},\ and\ \bibinfo {author} {\bibfnamefont {Z.}~\bibnamefont {Fisk}},\ }\bibfield  {title} {\bibinfo {title} {{Spin Waves and Electronic Interactions in La2CuO4}},\ }\href {https://doi.org/10.1103/physrevlett.86.5377} {\bibfield  {journal} {\bibinfo  {journal} {Physical Review Letters}\ }\textbf {\bibinfo {volume} {86}},\ \bibinfo {pages} {5377} (\bibinfo {year} {2001})},\ \Eprint {https://arxiv.org/abs/cond-mat/0006384} {cond-mat/0006384} \BibitemShut {NoStop}%
	\bibitem [{\citenamefont {Headings}\ \emph {et~al.}(2010)\citenamefont {Headings}, \citenamefont {Hayden}, \citenamefont {Coldea},\ and\ \citenamefont {Perring}}]{Headings.2010}%
	\BibitemOpen
	\bibfield  {author} {\bibinfo {author} {\bibfnamefont {N.~S.}\ \bibnamefont {Headings}}, \bibinfo {author} {\bibfnamefont {S.~M.}\ \bibnamefont {Hayden}}, \bibinfo {author} {\bibfnamefont {R.}~\bibnamefont {Coldea}},\ and\ \bibinfo {author} {\bibfnamefont {T.~G.}\ \bibnamefont {Perring}},\ }\bibfield  {title} {\bibinfo {title} {{Anomalous High-Energy Spin Excitations in the High-Tc Superconductor-Parent Antiferromagnet La2CuO4}},\ }\href {https://doi.org/10.1103/physrevlett.105.247001} {\bibfield  {journal} {\bibinfo  {journal} {Physical Review Letters}\ }\textbf {\bibinfo {volume} {105}},\ \bibinfo {pages} {247001} (\bibinfo {year} {2010})},\ \Eprint {https://arxiv.org/abs/1009.2915} {1009.2915} \BibitemShut {NoStop}%
	\bibitem [{\citenamefont {Yamada}\ \emph {et~al.}(1989)\citenamefont {Yamada}, \citenamefont {Kakurai}, \citenamefont {Endoh}, \citenamefont {Thurston}, \citenamefont {Kastner}, \citenamefont {Birgeneau}, \citenamefont {Shirane}, \citenamefont {Hidaka},\ and\ \citenamefont {Murakami}}]{Yamada.1989}%
	\BibitemOpen
	\bibfield  {author} {\bibinfo {author} {\bibfnamefont {K.}~\bibnamefont {Yamada}}, \bibinfo {author} {\bibfnamefont {K.}~\bibnamefont {Kakurai}}, \bibinfo {author} {\bibfnamefont {Y.}~\bibnamefont {Endoh}}, \bibinfo {author} {\bibfnamefont {T.~R.}\ \bibnamefont {Thurston}}, \bibinfo {author} {\bibfnamefont {M.~A.}\ \bibnamefont {Kastner}}, \bibinfo {author} {\bibfnamefont {R.~J.}\ \bibnamefont {Birgeneau}}, \bibinfo {author} {\bibfnamefont {G.}~\bibnamefont {Shirane}}, \bibinfo {author} {\bibfnamefont {Y.}~\bibnamefont {Hidaka}},\ and\ \bibinfo {author} {\bibfnamefont {T.}~\bibnamefont {Murakami}},\ }\bibfield  {title} {\bibinfo {title} {{Spin dynamics in the two-dimensional quantum antiferromagnet La2CuO4}},\ }\href {https://doi.org/10.1103/physrevb.40.4557} {\bibfield  {journal} {\bibinfo  {journal} {Physical Review B}\ }\textbf {\bibinfo {volume} {40}},\ \bibinfo {pages} {4557} (\bibinfo {year} {1989})}\BibitemShut {NoStop}%
	\bibitem [{\citenamefont {Hayden}\ \emph {et~al.}(1991)\citenamefont {Hayden}, \citenamefont {Aeppli}, \citenamefont {Osborn}, \citenamefont {Taylor}, \citenamefont {Perring}, \citenamefont {Cheong},\ and\ \citenamefont {Fisk}}]{Hayden.1991}%
	\BibitemOpen
	\bibfield  {author} {\bibinfo {author} {\bibfnamefont {S.~M.}\ \bibnamefont {Hayden}}, \bibinfo {author} {\bibfnamefont {G.}~\bibnamefont {Aeppli}}, \bibinfo {author} {\bibfnamefont {R.}~\bibnamefont {Osborn}}, \bibinfo {author} {\bibfnamefont {A.~D.}\ \bibnamefont {Taylor}}, \bibinfo {author} {\bibfnamefont {T.~G.}\ \bibnamefont {Perring}}, \bibinfo {author} {\bibfnamefont {S.-W.}\ \bibnamefont {Cheong}},\ and\ \bibinfo {author} {\bibfnamefont {Z.}~\bibnamefont {Fisk}},\ }\bibfield  {title} {\bibinfo {title} {{High-energy spin waves in La2CuO4}},\ }\href {https://doi.org/10.1103/physrevlett.67.3622} {\bibfield  {journal} {\bibinfo  {journal} {Physical Review Letters}\ }\textbf {\bibinfo {volume} {67}},\ \bibinfo {pages} {3622} (\bibinfo {year} {1991})}\BibitemShut {NoStop}%
	\bibitem [{\citenamefont {Canali}\ and\ \citenamefont {Girvin}(1992)}]{Canali.1992}%
	\BibitemOpen
	\bibfield  {author} {\bibinfo {author} {\bibfnamefont {C.~M.}\ \bibnamefont {Canali}}\ and\ \bibinfo {author} {\bibfnamefont {S.~M.}\ \bibnamefont {Girvin}},\ }\bibfield  {title} {\bibinfo {title} {{Theory of Raman scattering in layered cuprate materials}},\ }\href {https://doi.org/10.1103/physrevb.45.7127} {\bibfield  {journal} {\bibinfo  {journal} {Physical Review B}\ }\textbf {\bibinfo {volume} {45}},\ \bibinfo {pages} {7127} (\bibinfo {year} {1992})}\BibitemShut {NoStop}%
	\bibitem [{Note1()}]{Note1}%
	\BibitemOpen
	\bibinfo {note} {The slope $a$ in Eq. \protect \eqref {Eq:Universal_Scaling} does have a weak dependence on the Hamiltonian parameters $t$ and $U$, entering at orders $t/U$ and higher; in the strong-coupling regime $t/U \ll 1$ this is however unimportant.}\BibitemShut {Stop}%
	\bibitem [{\citenamefont {Oguchi}(1960)}]{Oguchi.1960}%
	\BibitemOpen
	\bibfield  {author} {\bibinfo {author} {\bibfnamefont {T.}~\bibnamefont {Oguchi}},\ }\bibfield  {title} {\bibinfo {title} {{Theory of Spin-Wave Interactions in Ferro- and Antiferromagnetism}},\ }\href {https://doi.org/10.1103/physrev.117.117} {\bibfield  {journal} {\bibinfo  {journal} {Physical Review}\ }\textbf {\bibinfo {volume} {117}},\ \bibinfo {pages} {117} (\bibinfo {year} {1960})}\BibitemShut {NoStop}%
	\bibitem [{\citenamefont {Sugai}\ \emph {et~al.}(2003)\citenamefont {Sugai}, \citenamefont {Suzuki}, \citenamefont {Takayanagi}, \citenamefont {Hosokawa},\ and\ \citenamefont {Hayamizu}}]{sugai_bimagnon_2003}%
	\BibitemOpen
	\bibfield  {author} {\bibinfo {author} {\bibfnamefont {S.}~\bibnamefont {Sugai}}, \bibinfo {author} {\bibfnamefont {H.}~\bibnamefont {Suzuki}}, \bibinfo {author} {\bibfnamefont {Y.}~\bibnamefont {Takayanagi}}, \bibinfo {author} {\bibfnamefont {T.}~\bibnamefont {Hosokawa}},\ and\ \bibinfo {author} {\bibfnamefont {N.}~\bibnamefont {Hayamizu}},\ }\bibfield  {title} {\bibinfo {title} {Carrier-density-dependent momentum shift of the coherent peak and the lo phonon mode in p-type high-${T}_{c}$ superconductors},\ }\href {https://doi.org/10.1103/PhysRevB.68.184504} {\bibfield  {journal} {\bibinfo  {journal} {Phys. Rev. B}\ }\textbf {\bibinfo {volume} {68}},\ \bibinfo {pages} {184504} (\bibinfo {year} {2003})}\BibitemShut {NoStop}%
	\bibitem [{\citenamefont {Ammon}\ \emph {et~al.}(1995)\citenamefont {Ammon}, \citenamefont {Troyer},\ and\ \citenamefont {Tsunetsugu}}]{Troyer1995}%
	\BibitemOpen
	\bibfield  {author} {\bibinfo {author} {\bibfnamefont {B.}~\bibnamefont {Ammon}}, \bibinfo {author} {\bibfnamefont {M.}~\bibnamefont {Troyer}},\ and\ \bibinfo {author} {\bibfnamefont {H.}~\bibnamefont {Tsunetsugu}},\ }\bibfield  {title} {\bibinfo {title} {Effect of the three-site hopping term on the t-j model},\ }\href {https://doi.org/10.1103/PhysRevB.52.629} {\bibfield  {journal} {\bibinfo  {journal} {Phys. Rev. B}\ }\textbf {\bibinfo {volume} {52}},\ \bibinfo {pages} {629} (\bibinfo {year} {1995})}\BibitemShut {NoStop}%
	\bibitem [{\citenamefont {Nagaoka}(1966)}]{Nagaoka1966}%
	\BibitemOpen
	\bibfield  {author} {\bibinfo {author} {\bibfnamefont {Y.}~\bibnamefont {Nagaoka}},\ }\bibfield  {title} {\bibinfo {title} {Ferromagnetism in a narrow, almost half-filled $s$ band},\ }\href {https://doi.org/10.1103/PhysRev.147.392} {\bibfield  {journal} {\bibinfo  {journal} {Phys. Rev.}\ }\textbf {\bibinfo {volume} {147}},\ \bibinfo {pages} {392} (\bibinfo {year} {1966})}\BibitemShut {NoStop}%
	\bibitem [{\citenamefont {{Thouless}}(1965)}]{Thouless1965}%
	\BibitemOpen
	\bibfield  {author} {\bibinfo {author} {\bibfnamefont {D.~J.}\ \bibnamefont {{Thouless}}},\ }\bibfield  {title} {\bibinfo {title} {{Exchange in solid $^{3}$He and the Heisenberg Hamiltonian}},\ }\href {https://doi.org/10.1088/0370-1328/86/5/301} {\bibfield  {journal} {\bibinfo  {journal} {Proceedings of the Physical Society}\ }\textbf {\bibinfo {volume} {86}},\ \bibinfo {pages} {893} (\bibinfo {year} {1965})}\BibitemShut {NoStop}%
	\bibitem [{\citenamefont {Kan\'asz-Nagy}\ \emph {et~al.}(2017)\citenamefont {Kan\'asz-Nagy}, \citenamefont {Lovas}, \citenamefont {Grusdt}, \citenamefont {Greif}, \citenamefont {Greiner},\ and\ \citenamefont {Demler}}]{NagyLovas2017}%
	\BibitemOpen
	\bibfield  {author} {\bibinfo {author} {\bibfnamefont {M.}~\bibnamefont {Kan\'asz-Nagy}}, \bibinfo {author} {\bibfnamefont {I.}~\bibnamefont {Lovas}}, \bibinfo {author} {\bibfnamefont {F.}~\bibnamefont {Grusdt}}, \bibinfo {author} {\bibfnamefont {D.}~\bibnamefont {Greif}}, \bibinfo {author} {\bibfnamefont {M.}~\bibnamefont {Greiner}},\ and\ \bibinfo {author} {\bibfnamefont {E.~A.}\ \bibnamefont {Demler}},\ }\bibfield  {title} {\bibinfo {title} {Quantum correlations at infinite temperature: The dynamical nagaoka effect},\ }\href {https://doi.org/10.1103/PhysRevB.96.014303} {\bibfield  {journal} {\bibinfo  {journal} {Phys. Rev. B}\ }\textbf {\bibinfo {volume} {96}},\ \bibinfo {pages} {014303} (\bibinfo {year} {2017})}\BibitemShut {NoStop}%
	\bibitem [{\citenamefont {Haerter}\ and\ \citenamefont {Shastry}(2005)}]{Haerter-Shastry}%
	\BibitemOpen
	\bibfield  {author} {\bibinfo {author} {\bibfnamefont {J.~O.}\ \bibnamefont {Haerter}}\ and\ \bibinfo {author} {\bibfnamefont {B.~S.}\ \bibnamefont {Shastry}},\ }\bibfield  {title} {\bibinfo {title} {Kinetic antiferromagnetism in the triangular lattice},\ }\href {https://doi.org/10.1103/PhysRevLett.95.087202} {\bibfield  {journal} {\bibinfo  {journal} {Phys. Rev. Lett.}\ }\textbf {\bibinfo {volume} {95}},\ \bibinfo {pages} {087202} (\bibinfo {year} {2005})}\BibitemShut {NoStop}%
	\bibitem [{\citenamefont {Morera}\ \emph {et~al.}(2023)\citenamefont {Morera}, \citenamefont {Kan\'asz-Nagy}, \citenamefont {Smolenski}, \citenamefont {Ciorciaro}, \citenamefont {Imamo\ifmmode~\breve{g}\else \u{g}\fi{}lu},\ and\ \citenamefont {Demler}}]{Morera_High_temperature}%
	\BibitemOpen
	\bibfield  {author} {\bibinfo {author} {\bibfnamefont {I.}~\bibnamefont {Morera}}, \bibinfo {author} {\bibfnamefont {M.}~\bibnamefont {Kan\'asz-Nagy}}, \bibinfo {author} {\bibfnamefont {T.}~\bibnamefont {Smolenski}}, \bibinfo {author} {\bibfnamefont {L.}~\bibnamefont {Ciorciaro}}, \bibinfo {author} {\bibfnamefont {A.}~\bibnamefont {Imamo\ifmmode~\breve{g}\else \u{g}\fi{}lu}},\ and\ \bibinfo {author} {\bibfnamefont {E.}~\bibnamefont {Demler}},\ }\bibfield  {title} {\bibinfo {title} {High-temperature kinetic magnetism in triangular lattices},\ }\href {https://doi.org/10.1103/PhysRevResearch.5.L022048} {\bibfield  {journal} {\bibinfo  {journal} {Phys. Rev. Res.}\ }\textbf {\bibinfo {volume} {5}},\ \bibinfo {pages} {L022048} (\bibinfo {year} {2023})}\BibitemShut {NoStop}%
	\bibitem [{\citenamefont {Ciorciaro}\ \emph {et~al.}(2023)\citenamefont {Ciorciaro}, \citenamefont {Smole{\'{n}}ski}, \citenamefont {Morera}, \citenamefont {Kiper}, \citenamefont {Hiestand}, \citenamefont {Kroner}, \citenamefont {Zhang}, \citenamefont {Watanabe}, \citenamefont {Taniguchi}, \citenamefont {Demler},\ and\ \citenamefont {{\.{I}}mamo{\u{g}}lu}}]{TMD_Ciorciaro2023}%
	\BibitemOpen
	\bibfield  {author} {\bibinfo {author} {\bibfnamefont {L.}~\bibnamefont {Ciorciaro}}, \bibinfo {author} {\bibfnamefont {T.}~\bibnamefont {Smole{\'{n}}ski}}, \bibinfo {author} {\bibfnamefont {I.}~\bibnamefont {Morera}}, \bibinfo {author} {\bibfnamefont {N.}~\bibnamefont {Kiper}}, \bibinfo {author} {\bibfnamefont {S.}~\bibnamefont {Hiestand}}, \bibinfo {author} {\bibfnamefont {M.}~\bibnamefont {Kroner}}, \bibinfo {author} {\bibfnamefont {Y.}~\bibnamefont {Zhang}}, \bibinfo {author} {\bibfnamefont {K.}~\bibnamefont {Watanabe}}, \bibinfo {author} {\bibfnamefont {T.}~\bibnamefont {Taniguchi}}, \bibinfo {author} {\bibfnamefont {E.}~\bibnamefont {Demler}},\ and\ \bibinfo {author} {\bibfnamefont {A.}~\bibnamefont {{\.{I}}mamo{\u{g}}lu}},\ }\bibfield  {title} {\bibinfo {title} {Kinetic magnetism in triangular moir{\'e} materials},\ }\href {https://doi.org/10.1038/s41586-023-06633-0} {\bibfield  {journal} {\bibinfo  {journal} {Nature}\ }\textbf {\bibinfo {volume} {623}},\ \bibinfo {pages} {509} (\bibinfo {year}
		{2023})}\BibitemShut {NoStop}%
	\bibitem [{\citenamefont {Prichard}\ \emph {et~al.}(2025)\citenamefont {Prichard}, \citenamefont {Ba}, \citenamefont {Morera}, \citenamefont {Spar}, \citenamefont {Huse}, \citenamefont {Demler},\ and\ \citenamefont {Bakr}}]{Prichard_MagnonPolaron}%
	\BibitemOpen
	\bibfield  {author} {\bibinfo {author} {\bibfnamefont {M.~L.}\ \bibnamefont {Prichard}}, \bibinfo {author} {\bibfnamefont {Z.}~\bibnamefont {Ba}}, \bibinfo {author} {\bibfnamefont {I.}~\bibnamefont {Morera}}, \bibinfo {author} {\bibfnamefont {B.~M.}\ \bibnamefont {Spar}}, \bibinfo {author} {\bibfnamefont {D.~A.}\ \bibnamefont {Huse}}, \bibinfo {author} {\bibfnamefont {E.}~\bibnamefont {Demler}},\ and\ \bibinfo {author} {\bibfnamefont {W.~S.}\ \bibnamefont {Bakr}},\ }\bibfield  {title} {\bibinfo {title} {Magnon-polarons in the fermi--hubbard model},\ }\href {https://doi.org/10.1038/s41567-025-03004-6} {\bibfield  {journal} {\bibinfo  {journal} {Nature Physics}\ }\textbf {\bibinfo {volume} {21}},\ \bibinfo {pages} {1548} (\bibinfo {year} {2025})}\BibitemShut {NoStop}%
	\bibitem [{\citenamefont {Radovskaia}\ \emph {et~al.}(2026)\citenamefont {Radovskaia}, \citenamefont {Andrei}, \citenamefont {Hortensius}, \citenamefont {Mikhaylovskiy}, \citenamefont {Citro}, \citenamefont {Chattopadhyay}, \citenamefont {Na}, \citenamefont {Ivanov}, \citenamefont {Demler}, \citenamefont {Kimel}, \citenamefont {Caviglia},\ and\ \citenamefont {Afanasiev}}]{Radovskaia2026}%
	\BibitemOpen
	\bibfield  {author} {\bibinfo {author} {\bibfnamefont {V.}~\bibnamefont {Radovskaia}}, \bibinfo {author} {\bibfnamefont {R.}~\bibnamefont {Andrei}}, \bibinfo {author} {\bibfnamefont {J.~R.}\ \bibnamefont {Hortensius}}, \bibinfo {author} {\bibfnamefont {R.~V.}\ \bibnamefont {Mikhaylovskiy}}, \bibinfo {author} {\bibfnamefont {R.}~\bibnamefont {Citro}}, \bibinfo {author} {\bibfnamefont {S.}~\bibnamefont {Chattopadhyay}}, \bibinfo {author} {\bibfnamefont {M.~X.}\ \bibnamefont {Na}}, \bibinfo {author} {\bibfnamefont {B.~A.}\ \bibnamefont {Ivanov}}, \bibinfo {author} {\bibfnamefont {E.}~\bibnamefont {Demler}}, \bibinfo {author} {\bibfnamefont {A.~V.}\ \bibnamefont {Kimel}}, \bibinfo {author} {\bibfnamefont {A.~D.}\ \bibnamefont {Caviglia}},\ and\ \bibinfo {author} {\bibfnamefont {D.}~\bibnamefont {Afanasiev}},\ }\bibfield  {title} {\bibinfo {title} {Photoengineering the magnon spectrum in an insulating antiferromagnet},\ }\bibfield  {journal} {\bibinfo  {journal} {Nature Physics}\ }\href
	{https://doi.org/10.1038/s41567-026-03230-6} {10.1038/s41567-026-03230-6} (\bibinfo {year} {2026})\BibitemShut {NoStop}%
	\bibitem [{Note2()}]{Note2}%
	\BibitemOpen
	\bibinfo {note} {Ignoring the difference between $J_\delta $ and $J_0$, as well as the $U/t$ dependence of $g(0)$, yields Eq. \protect \eqref {Eq:Universal_Scaling} with a constant slope $a = g(0) / 2(1+r_0)$. Taking these two effects into account will introduce corrections to the slope $a$ at order $t/U$ and above, as mentioned before.}\BibitemShut {Stop}%
	\bibitem [{\citenamefont {Fleury}\ and\ \citenamefont {Loudon}(1968)}]{Fleury.1968}%
	\BibitemOpen
	\bibfield  {author} {\bibinfo {author} {\bibfnamefont {P.~A.}\ \bibnamefont {Fleury}}\ and\ \bibinfo {author} {\bibfnamefont {R.}~\bibnamefont {Loudon}},\ }\bibfield  {title} {\bibinfo {title} {{Scattering of Light by One- and Two-Magnon Excitations}},\ }\href {https://doi.org/10.1103/physrev.166.514} {\bibfield  {journal} {\bibinfo  {journal} {Physical Review}\ }\textbf {\bibinfo {volume} {166}},\ \bibinfo {pages} {514} (\bibinfo {year} {1968})}\BibitemShut {NoStop}%
	\bibitem [{\citenamefont {Shraiman}\ and\ \citenamefont {Siggia}(1989)}]{Shraiman.1989}%
	\BibitemOpen
	\bibfield  {author} {\bibinfo {author} {\bibfnamefont {B.~I.}\ \bibnamefont {Shraiman}}\ and\ \bibinfo {author} {\bibfnamefont {E.~D.}\ \bibnamefont {Siggia}},\ }\bibfield  {title} {\bibinfo {title} {{Spiral phase of a doped quantum antiferromagnet}},\ }\href {https://doi.org/10.1103/physrevlett.62.1564} {\bibfield  {journal} {\bibinfo  {journal} {Physical Review Letters}\ }\textbf {\bibinfo {volume} {62}},\ \bibinfo {pages} {1564} (\bibinfo {year} {1989})}\BibitemShut {NoStop}%
	\bibitem [{\citenamefont {Auerbach}\ and\ \citenamefont {Larson}(1991{\natexlab{b}})}]{Auerbach.1991}%
	\BibitemOpen
	\bibfield  {author} {\bibinfo {author} {\bibfnamefont {A.}~\bibnamefont {Auerbach}}\ and\ \bibinfo {author} {\bibfnamefont {B.~E.}\ \bibnamefont {Larson}},\ }\bibfield  {title} {\bibinfo {title} {{Doped antiferromagnet: The instability of homogeneous magnetic phases}},\ }\href {https://doi.org/10.1103/physrevb.43.7800} {\bibfield  {journal} {\bibinfo  {journal} {Physical Review B}\ }\textbf {\bibinfo {volume} {43}},\ \bibinfo {pages} {7800} (\bibinfo {year} {1991}{\natexlab{b}})}\BibitemShut {NoStop}%
	\bibitem [{\citenamefont {Yamada}\ \emph {et~al.}(1998)\citenamefont {Yamada}, \citenamefont {Lee}, \citenamefont {Kurahashi}, \citenamefont {Wada}, \citenamefont {Wakimoto}, \citenamefont {Ueki}, \citenamefont {Kimura}, \citenamefont {Endoh}, \citenamefont {Hosoya}, \citenamefont {Shirane}, \citenamefont {Birgeneau}, \citenamefont {Greven}, \citenamefont {Kastner},\ and\ \citenamefont {Kim}}]{Yamada.1998}%
	\BibitemOpen
	\bibfield  {author} {\bibinfo {author} {\bibfnamefont {K.}~\bibnamefont {Yamada}}, \bibinfo {author} {\bibfnamefont {C.~H.}\ \bibnamefont {Lee}}, \bibinfo {author} {\bibfnamefont {K.}~\bibnamefont {Kurahashi}}, \bibinfo {author} {\bibfnamefont {J.}~\bibnamefont {Wada}}, \bibinfo {author} {\bibfnamefont {S.}~\bibnamefont {Wakimoto}}, \bibinfo {author} {\bibfnamefont {S.}~\bibnamefont {Ueki}}, \bibinfo {author} {\bibfnamefont {H.}~\bibnamefont {Kimura}}, \bibinfo {author} {\bibfnamefont {Y.}~\bibnamefont {Endoh}}, \bibinfo {author} {\bibfnamefont {S.}~\bibnamefont {Hosoya}}, \bibinfo {author} {\bibfnamefont {G.}~\bibnamefont {Shirane}}, \bibinfo {author} {\bibfnamefont {R.~J.}\ \bibnamefont {Birgeneau}}, \bibinfo {author} {\bibfnamefont {M.}~\bibnamefont {Greven}}, \bibinfo {author} {\bibfnamefont {M.~A.}\ \bibnamefont {Kastner}},\ and\ \bibinfo {author} {\bibfnamefont {Y.~J.}\ \bibnamefont {Kim}},\ }\bibfield  {title} {\bibinfo {title} {{Doping dependence of the spatially modulated dynamical spin
				correlations and the superconducting-transition temperature in La2-xSrxCuO4}},\ }\href {https://doi.org/10.1103/physrevb.57.6165} {\bibfield  {journal} {\bibinfo  {journal} {Physical Review B}\ }\textbf {\bibinfo {volume} {57}},\ \bibinfo {pages} {6165} (\bibinfo {year} {1998})}\BibitemShut {NoStop}%
	\bibitem [{\citenamefont {Brunner}\ \emph {et~al.}(2000)\citenamefont {Brunner}, \citenamefont {Assaad},\ and\ \citenamefont {Muramatsu}}]{monte_carlo_single_hole_z}%
	\BibitemOpen
	\bibfield  {author} {\bibinfo {author} {\bibfnamefont {M.}~\bibnamefont {Brunner}}, \bibinfo {author} {\bibfnamefont {F.~F.}\ \bibnamefont {Assaad}},\ and\ \bibinfo {author} {\bibfnamefont {A.}~\bibnamefont {Muramatsu}},\ }\bibfield  {title} {\bibinfo {title} {Single-hole dynamics in the $t\ensuremath{-}j$ model on a square lattice},\ }\href {https://doi.org/10.1103/PhysRevB.62.15480} {\bibfield  {journal} {\bibinfo  {journal} {Phys. Rev. B}\ }\textbf {\bibinfo {volume} {62}},\ \bibinfo {pages} {15480} (\bibinfo {year} {2000})}\BibitemShut {NoStop}%
	\bibitem [{\citenamefont {Mishchenko}\ \emph {et~al.}(2001)\citenamefont {Mishchenko}, \citenamefont {Prokof'ev},\ and\ \citenamefont {Svistunov}}]{mischchenko_delta_function_peak}%
	\BibitemOpen
	\bibfield  {author} {\bibinfo {author} {\bibfnamefont {A.~S.}\ \bibnamefont {Mishchenko}}, \bibinfo {author} {\bibfnamefont {N.~V.}\ \bibnamefont {Prokof'ev}},\ and\ \bibinfo {author} {\bibfnamefont {B.~V.}\ \bibnamefont {Svistunov}},\ }\bibfield  {title} {\bibinfo {title} {Single-hole spectral function and spin-charge separation in the $t\ensuremath{-}j$ model},\ }\href {https://doi.org/10.1103/PhysRevB.64.033101} {\bibfield  {journal} {\bibinfo  {journal} {Phys. Rev. B}\ }\textbf {\bibinfo {volume} {64}},\ \bibinfo {pages} {033101} (\bibinfo {year} {2001})}\BibitemShut {NoStop}%
	\bibitem [{\citenamefont {Sheng}\ \emph {et~al.}(1996{\natexlab{a}})\citenamefont {Sheng}, \citenamefont {Chen},\ and\ \citenamefont {Weng}}]{phase_string_1}%
	\BibitemOpen
	\bibfield  {author} {\bibinfo {author} {\bibfnamefont {D.~N.}\ \bibnamefont {Sheng}}, \bibinfo {author} {\bibfnamefont {Y.~C.}\ \bibnamefont {Chen}},\ and\ \bibinfo {author} {\bibfnamefont {Z.~Y.}\ \bibnamefont {Weng}},\ }\bibfield  {title} {\bibinfo {title} {Phase string effect in a doped antiferromagnet},\ }\href {https://doi.org/10.1103/PhysRevLett.77.5102} {\bibfield  {journal} {\bibinfo  {journal} {Phys. Rev. Lett.}\ }\textbf {\bibinfo {volume} {77}},\ \bibinfo {pages} {5102} (\bibinfo {year} {1996}{\natexlab{a}})}\BibitemShut {NoStop}%
	\bibitem [{\citenamefont {Sheng}\ \emph {et~al.}(1996{\natexlab{b}})\citenamefont {Sheng}, \citenamefont {Chen},\ and\ \citenamefont {Weng}}]{phase_string_2}%
	\BibitemOpen
	\bibfield  {author} {\bibinfo {author} {\bibfnamefont {D.~N.}\ \bibnamefont {Sheng}}, \bibinfo {author} {\bibfnamefont {Y.~C.}\ \bibnamefont {Chen}},\ and\ \bibinfo {author} {\bibfnamefont {Z.~Y.}\ \bibnamefont {Weng}},\ }\bibfield  {title} {\bibinfo {title} {Phase string effect in a doped antiferromagnet},\ }\href {https://doi.org/10.1103/PhysRevLett.77.5102} {\bibfield  {journal} {\bibinfo  {journal} {Phys. Rev. Lett.}\ }\textbf {\bibinfo {volume} {77}},\ \bibinfo {pages} {5102} (\bibinfo {year} {1996}{\natexlab{b}})}\BibitemShut {NoStop}%
	\bibitem [{\citenamefont {Pekker}\ \emph {et~al.}(2011)\citenamefont {Pekker}, \citenamefont {Babadi}, \citenamefont {Sensarma}, \citenamefont {Zinner}, \citenamefont {Pollet}, \citenamefont {Zwierlein},\ and\ \citenamefont {Demler}}]{pekker_babadi_instabilities}%
	\BibitemOpen
	\bibfield  {author} {\bibinfo {author} {\bibfnamefont {D.}~\bibnamefont {Pekker}}, \bibinfo {author} {\bibfnamefont {M.}~\bibnamefont {Babadi}}, \bibinfo {author} {\bibfnamefont {R.}~\bibnamefont {Sensarma}}, \bibinfo {author} {\bibfnamefont {N.}~\bibnamefont {Zinner}}, \bibinfo {author} {\bibfnamefont {L.}~\bibnamefont {Pollet}}, \bibinfo {author} {\bibfnamefont {M.~W.}\ \bibnamefont {Zwierlein}},\ and\ \bibinfo {author} {\bibfnamefont {E.}~\bibnamefont {Demler}},\ }\bibfield  {title} {\bibinfo {title} {Competition between pairing and ferromagnetic instabilities in ultracold fermi gases near feshbach resonances},\ }\href {https://doi.org/10.1103/PhysRevLett.106.050402} {\bibfield  {journal} {\bibinfo  {journal} {Phys. Rev. Lett.}\ }\textbf {\bibinfo {volume} {106}},\ \bibinfo {pages} {050402} (\bibinfo {year} {2011})}\BibitemShut {NoStop}%
	\bibitem [{\citenamefont {Yuzbashyan}\ \emph {et~al.}(2006)\citenamefont {Yuzbashyan}, \citenamefont {Tsyplyatyev},\ and\ \citenamefont {Altshuler}}]{Yuzbashyan_instability}%
	\BibitemOpen
	\bibfield  {author} {\bibinfo {author} {\bibfnamefont {E.~A.}\ \bibnamefont {Yuzbashyan}}, \bibinfo {author} {\bibfnamefont {O.}~\bibnamefont {Tsyplyatyev}},\ and\ \bibinfo {author} {\bibfnamefont {B.~L.}\ \bibnamefont {Altshuler}},\ }\bibfield  {title} {\bibinfo {title} {Relaxation and persistent oscillations of the order parameter in fermionic condensates},\ }\href {https://doi.org/10.1103/PhysRevLett.96.097005} {\bibfield  {journal} {\bibinfo  {journal} {Phys. Rev. Lett.}\ }\textbf {\bibinfo {volume} {96}},\ \bibinfo {pages} {097005} (\bibinfo {year} {2006})}\BibitemShut {NoStop}%
	\bibitem [{\citenamefont {Barankov}\ \emph {et~al.}(2004)\citenamefont {Barankov}, \citenamefont {Levitov},\ and\ \citenamefont {Spivak}}]{Barankov_instability}%
	\BibitemOpen
	\bibfield  {author} {\bibinfo {author} {\bibfnamefont {R.~A.}\ \bibnamefont {Barankov}}, \bibinfo {author} {\bibfnamefont {L.~S.}\ \bibnamefont {Levitov}},\ and\ \bibinfo {author} {\bibfnamefont {B.~Z.}\ \bibnamefont {Spivak}},\ }\bibfield  {title} {\bibinfo {title} {Collective rabi oscillations and solitons in a time-dependent bcs pairing problem},\ }\href {https://doi.org/10.1103/PhysRevLett.93.160401} {\bibfield  {journal} {\bibinfo  {journal} {Phys. Rev. Lett.}\ }\textbf {\bibinfo {volume} {93}},\ \bibinfo {pages} {160401} (\bibinfo {year} {2004})}\BibitemShut {NoStop}%
	\bibitem [{\citenamefont {Abrikosov}\ \emph {et~al.}(2012)\citenamefont {Abrikosov}, \citenamefont {Gorkov}, \citenamefont {Dzyaloshinski},\ and\ \citenamefont {Silverman}}]{abrikosov2012methods}%
	\BibitemOpen
	\bibfield  {author} {\bibinfo {author} {\bibfnamefont {A.}~\bibnamefont {Abrikosov}}, \bibinfo {author} {\bibfnamefont {L.}~\bibnamefont {Gorkov}}, \bibinfo {author} {\bibfnamefont {I.}~\bibnamefont {Dzyaloshinski}},\ and\ \bibinfo {author} {\bibfnamefont {R.}~\bibnamefont {Silverman}},\ }\href {https://books.google.ch/books?id=JYTCAgAAQBAJ} {\emph {\bibinfo {title} {Methods of Quantum Field Theory in Statistical Physics}}},\ Dover Books on Physics\ (\bibinfo  {publisher} {Dover Publications},\ \bibinfo {year} {2012})\BibitemShut {NoStop}%
	\bibitem [{\citenamefont {White}\ and\ \citenamefont {Scalapino}(1998)}]{WhiteScalapinoStripes1998}%
	\BibitemOpen
	\bibfield  {author} {\bibinfo {author} {\bibfnamefont {S.~R.}\ \bibnamefont {White}}\ and\ \bibinfo {author} {\bibfnamefont {D.~J.}\ \bibnamefont {Scalapino}},\ }\bibfield  {title} {\bibinfo {title} {Density matrix renormalization group study of the striped phase in the 2d $\mathit{t}\ensuremath{-}\mathit{J}$ model},\ }\href {https://doi.org/10.1103/PhysRevLett.80.1272} {\bibfield  {journal} {\bibinfo  {journal} {Phys. Rev. Lett.}\ }\textbf {\bibinfo {volume} {80}},\ \bibinfo {pages} {1272} (\bibinfo {year} {1998})}\BibitemShut {NoStop}%
	\bibitem [{\citenamefont {White}\ and\ \citenamefont {Scalapino}(2003)}]{WhiteScalapino2003}%
	\BibitemOpen
	\bibfield  {author} {\bibinfo {author} {\bibfnamefont {S.~R.}\ \bibnamefont {White}}\ and\ \bibinfo {author} {\bibfnamefont {D.~J.}\ \bibnamefont {Scalapino}},\ }\bibfield  {title} {\bibinfo {title} {Stripes on a 6-leg hubbard ladder},\ }\href {https://doi.org/10.1103/PhysRevLett.91.136403} {\bibfield  {journal} {\bibinfo  {journal} {Phys. Rev. Lett.}\ }\textbf {\bibinfo {volume} {91}},\ \bibinfo {pages} {136403} (\bibinfo {year} {2003})}\BibitemShut {NoStop}%
	\bibitem [{\citenamefont {Hager}\ \emph {et~al.}(2005)\citenamefont {Hager}, \citenamefont {Wellein}, \citenamefont {Jeckelmann},\ and\ \citenamefont {Fehske}}]{Hager2005}%
	\BibitemOpen
	\bibfield  {author} {\bibinfo {author} {\bibfnamefont {G.}~\bibnamefont {Hager}}, \bibinfo {author} {\bibfnamefont {G.}~\bibnamefont {Wellein}}, \bibinfo {author} {\bibfnamefont {E.}~\bibnamefont {Jeckelmann}},\ and\ \bibinfo {author} {\bibfnamefont {H.}~\bibnamefont {Fehske}},\ }\bibfield  {title} {\bibinfo {title} {Stripe formation in doped hubbard ladders},\ }\href {https://doi.org/10.1103/PhysRevB.71.075108} {\bibfield  {journal} {\bibinfo  {journal} {Phys. Rev. B}\ }\textbf {\bibinfo {volume} {71}},\ \bibinfo {pages} {075108} (\bibinfo {year} {2005})}\BibitemShut {NoStop}%
	\bibitem [{\citenamefont {Chang}\ and\ \citenamefont {Zhang}(2010)}]{ShiweiStripes2010}%
	\BibitemOpen
	\bibfield  {author} {\bibinfo {author} {\bibfnamefont {C.-C.}\ \bibnamefont {Chang}}\ and\ \bibinfo {author} {\bibfnamefont {S.}~\bibnamefont {Zhang}},\ }\bibfield  {title} {\bibinfo {title} {Spin and charge order in the doped hubbard model: Long-wavelength collective modes},\ }\href {https://doi.org/10.1103/PhysRevLett.104.116402} {\bibfield  {journal} {\bibinfo  {journal} {Phys. Rev. Lett.}\ }\textbf {\bibinfo {volume} {104}},\ \bibinfo {pages} {116402} (\bibinfo {year} {2010})}\BibitemShut {NoStop}%
	\bibitem [{\citenamefont {Zheng}\ \emph {et~al.}(2017)\citenamefont {Zheng}, \citenamefont {Chung}, \citenamefont {Corboz}, \citenamefont {Ehlers}, \citenamefont {Qin}, \citenamefont {Noack}, \citenamefont {Shi}, \citenamefont {White}, \citenamefont {Zhang},\ and\ \citenamefont {Chan}}]{Bo-XiaoStripes2017}%
	\BibitemOpen
	\bibfield  {author} {\bibinfo {author} {\bibfnamefont {B.-X.}\ \bibnamefont {Zheng}}, \bibinfo {author} {\bibfnamefont {C.-M.}\ \bibnamefont {Chung}}, \bibinfo {author} {\bibfnamefont {P.}~\bibnamefont {Corboz}}, \bibinfo {author} {\bibfnamefont {G.}~\bibnamefont {Ehlers}}, \bibinfo {author} {\bibfnamefont {M.-P.}\ \bibnamefont {Qin}}, \bibinfo {author} {\bibfnamefont {R.~M.}\ \bibnamefont {Noack}}, \bibinfo {author} {\bibfnamefont {H.}~\bibnamefont {Shi}}, \bibinfo {author} {\bibfnamefont {S.~R.}\ \bibnamefont {White}}, \bibinfo {author} {\bibfnamefont {S.}~\bibnamefont {Zhang}},\ and\ \bibinfo {author} {\bibfnamefont {G.~K.-L.}\ \bibnamefont {Chan}},\ }\bibfield  {title} {\bibinfo {title} {Stripe order in the underdoped region of the two-dimensional hubbard model},\ }\href {https://doi.org/10.1126/science.aam7127} {\bibfield  {journal} {\bibinfo  {journal} {Science}\ }\textbf {\bibinfo {volume} {358}},\ \bibinfo {pages} {1155} (\bibinfo {year} {2017})},\ \Eprint
	{https://arxiv.org/abs/https://www.science.org/doi/pdf/10.1126/science.aam7127} {https://www.science.org/doi/pdf/10.1126/science.aam7127} \BibitemShut {NoStop}%
	\bibitem [{\citenamefont {Ehlers}\ \emph {et~al.}(2017)\citenamefont {Ehlers}, \citenamefont {White},\ and\ \citenamefont {Noack}}]{WhiteHybridStripes2017}%
	\BibitemOpen
	\bibfield  {author} {\bibinfo {author} {\bibfnamefont {G.}~\bibnamefont {Ehlers}}, \bibinfo {author} {\bibfnamefont {S.~R.}\ \bibnamefont {White}},\ and\ \bibinfo {author} {\bibfnamefont {R.~M.}\ \bibnamefont {Noack}},\ }\bibfield  {title} {\bibinfo {title} {Hybrid-space density matrix renormalization group study of the doped two-dimensional hubbard model},\ }\href {https://doi.org/10.1103/PhysRevB.95.125125} {\bibfield  {journal} {\bibinfo  {journal} {Phys. Rev. B}\ }\textbf {\bibinfo {volume} {95}},\ \bibinfo {pages} {125125} (\bibinfo {year} {2017})}\BibitemShut {NoStop}%
	\bibitem [{\citenamefont {Huang}\ \emph {et~al.}(2018)\citenamefont {Huang}, \citenamefont {Mendl}, \citenamefont {Jiang}, \citenamefont {Moritz},\ and\ \citenamefont {Devereaux}}]{HuangStripes2018}%
	\BibitemOpen
	\bibfield  {author} {\bibinfo {author} {\bibfnamefont {E.~W.}\ \bibnamefont {Huang}}, \bibinfo {author} {\bibfnamefont {C.~B.}\ \bibnamefont {Mendl}}, \bibinfo {author} {\bibfnamefont {H.-C.}\ \bibnamefont {Jiang}}, \bibinfo {author} {\bibfnamefont {B.}~\bibnamefont {Moritz}},\ and\ \bibinfo {author} {\bibfnamefont {T.~P.}\ \bibnamefont {Devereaux}},\ }\bibfield  {title} {\bibinfo {title} {Stripe order from the perspective of the hubbard model},\ }\href {https://doi.org/10.1038/s41535-018-0097-0} {\bibfield  {journal} {\bibinfo  {journal} {npj Quantum Materials}\ }\textbf {\bibinfo {volume} {3}},\ \bibinfo {pages} {22} (\bibinfo {year} {2018})}\BibitemShut {NoStop}%
	\bibitem [{\citenamefont {Jiang}\ \emph {et~al.}(2020)\citenamefont {Jiang}, \citenamefont {Zaanen}, \citenamefont {Devereaux},\ and\ \citenamefont {Jiang}}]{Thom4leg2020}%
	\BibitemOpen
	\bibfield  {author} {\bibinfo {author} {\bibfnamefont {Y.-F.}\ \bibnamefont {Jiang}}, \bibinfo {author} {\bibfnamefont {J.}~\bibnamefont {Zaanen}}, \bibinfo {author} {\bibfnamefont {T.~P.}\ \bibnamefont {Devereaux}},\ and\ \bibinfo {author} {\bibfnamefont {H.-C.}\ \bibnamefont {Jiang}},\ }\bibfield  {title} {\bibinfo {title} {Ground state phase diagram of the doped hubbard model on the four-leg cylinder},\ }\href {https://doi.org/10.1103/PhysRevResearch.2.033073} {\bibfield  {journal} {\bibinfo  {journal} {Phys. Rev. Res.}\ }\textbf {\bibinfo {volume} {2}},\ \bibinfo {pages} {033073} (\bibinfo {year} {2020})}\BibitemShut {NoStop}%
	\bibitem [{\citenamefont {Wietek}\ \emph {et~al.}(2021)\citenamefont {Wietek}, \citenamefont {He}, \citenamefont {White}, \citenamefont {Georges},\ and\ \citenamefont {Stoudenmire}}]{Wietek.2021}%
	\BibitemOpen
	\bibfield  {author} {\bibinfo {author} {\bibfnamefont {A.}~\bibnamefont {Wietek}}, \bibinfo {author} {\bibfnamefont {Y.-Y.}\ \bibnamefont {He}}, \bibinfo {author} {\bibfnamefont {S.~R.}\ \bibnamefont {White}}, \bibinfo {author} {\bibfnamefont {A.}~\bibnamefont {Georges}},\ and\ \bibinfo {author} {\bibfnamefont {E.~M.}\ \bibnamefont {Stoudenmire}},\ }\bibfield  {title} {\bibinfo {title} {{Stripes, Antiferromagnetism, and the Pseudogap in the Doped Hubbard Model at Finite Temperature}},\ }\href {https://doi.org/10.1103/physrevx.11.031007} {\bibfield  {journal} {\bibinfo  {journal} {Physical Review X}\ }\textbf {\bibinfo {volume} {11}},\ \bibinfo {pages} {031007} (\bibinfo {year} {2021})},\ \Eprint {https://arxiv.org/abs/2009.10736} {2009.10736} \BibitemShut {NoStop}%
	\bibitem [{\citenamefont {Qin}\ \emph {et~al.}(2020)\citenamefont {Qin}, \citenamefont {Chung}, \citenamefont {Shi}, \citenamefont {Vitali}, \citenamefont {Hubig}, \citenamefont {Schollw\"ock}, \citenamefont {White},\ and\ \citenamefont {Zhang}}]{ShiweiAbsence2020}%
	\BibitemOpen
	\bibfield  {author} {\bibinfo {author} {\bibfnamefont {M.}~\bibnamefont {Qin}}, \bibinfo {author} {\bibfnamefont {C.-M.}\ \bibnamefont {Chung}}, \bibinfo {author} {\bibfnamefont {H.}~\bibnamefont {Shi}}, \bibinfo {author} {\bibfnamefont {E.}~\bibnamefont {Vitali}}, \bibinfo {author} {\bibfnamefont {C.}~\bibnamefont {Hubig}}, \bibinfo {author} {\bibfnamefont {U.}~\bibnamefont {Schollw\"ock}}, \bibinfo {author} {\bibfnamefont {S.~R.}\ \bibnamefont {White}},\ and\ \bibinfo {author} {\bibfnamefont {S.}~\bibnamefont {Zhang}} (\bibinfo {collaboration} {Simons Collaboration on the Many-Electron Problem}),\ }\bibfield  {title} {\bibinfo {title} {Absence of superconductivity in the pure two-dimensional hubbard model},\ }\href {https://doi.org/10.1103/PhysRevX.10.031016} {\bibfield  {journal} {\bibinfo  {journal} {Phys. Rev. X}\ }\textbf {\bibinfo {volume} {10}},\ \bibinfo {pages} {031016} (\bibinfo {year} {2020})}\BibitemShut {NoStop}%
	\bibitem [{\citenamefont {Miyashita}(1988)}]{miyashita_thermodynamic_1988}%
	\BibitemOpen
	\bibfield  {author} {\bibinfo {author} {\bibfnamefont {S.}~\bibnamefont {Miyashita}},\ }\bibfield  {title} {\bibinfo {title} {Thermodynamic {Properties} of {Spin} 1/2 {Antiferromagnetic} {Heisenberg} {Model} on the {Square} {Lattice}},\ }\href {https://doi.org/10.1143/JPSJ.57.1934} {\bibfield  {journal} {\bibinfo  {journal} {Journal of the Physical Society of Japan}\ }\textbf {\bibinfo {volume} {57}},\ \bibinfo {pages} {1934} (\bibinfo {year} {1988})}\BibitemShut {NoStop}%
	\bibitem [{\citenamefont {Gomez-Santos}\ \emph {et~al.}(1989)\citenamefont {Gomez-Santos}, \citenamefont {Joannopoulos},\ and\ \citenamefont {Negele}}]{gomez-santos_monte_1989}%
	\BibitemOpen
	\bibfield  {author} {\bibinfo {author} {\bibfnamefont {G.}~\bibnamefont {Gomez-Santos}}, \bibinfo {author} {\bibfnamefont {J.~D.}\ \bibnamefont {Joannopoulos}},\ and\ \bibinfo {author} {\bibfnamefont {J.~W.}\ \bibnamefont {Negele}},\ }\bibfield  {title} {\bibinfo {title} {Monte {Carlo} study of the quantum spin-(1/2 {Heisenberg} antiferromagnet on the square lattice},\ }\href {https://doi.org/10.1103/PhysRevB.39.4435} {\bibfield  {journal} {\bibinfo  {journal} {Physical Review B}\ }\textbf {\bibinfo {volume} {39}},\ \bibinfo {pages} {4435} (\bibinfo {year} {1989})}\BibitemShut {NoStop}%
	\bibitem [{\citenamefont {Makivić}\ and\ \citenamefont {Ding}(1991)}]{makivic_two-dimensional_1991}%
	\BibitemOpen
	\bibfield  {author} {\bibinfo {author} {\bibfnamefont {M.~S.}\ \bibnamefont {Makivić}}\ and\ \bibinfo {author} {\bibfnamefont {H.-Q.}\ \bibnamefont {Ding}},\ }\bibfield  {title} {\bibinfo {title} {Two-dimensional spin-1/2 {Heisenberg} antiferromagnet: {A} quantum {Monte} {Carlo} study},\ }\href {https://doi.org/10.1103/PhysRevB.43.3562} {\bibfield  {journal} {\bibinfo  {journal} {Physical Review B}\ }\textbf {\bibinfo {volume} {43}},\ \bibinfo {pages} {3562} (\bibinfo {year} {1991})}\BibitemShut {NoStop}%
	\bibitem [{\citenamefont {Kim}\ and\ \citenamefont {Troyer}(1998)}]{kim_low_1998}%
	\BibitemOpen
	\bibfield  {author} {\bibinfo {author} {\bibfnamefont {J.-K.}\ \bibnamefont {Kim}}\ and\ \bibinfo {author} {\bibfnamefont {M.}~\bibnamefont {Troyer}},\ }\bibfield  {title} {\bibinfo {title} {Low {Temperature} {Behavior} and {Crossovers} of the {Square} {Lattice} {Quantum} {Heisenberg} {Antiferromagnet}},\ }\href {https://doi.org/10.1103/PhysRevLett.80.2705} {\bibfield  {journal} {\bibinfo  {journal} {Physical Review Letters}\ }\textbf {\bibinfo {volume} {80}},\ \bibinfo {pages} {2705} (\bibinfo {year} {1998})}\BibitemShut {NoStop}%
	\bibitem [{\citenamefont {Okabe}\ and\ \citenamefont {Kikuchi}(1988)}]{okabe_quantum_1988}%
	\BibitemOpen
	\bibfield  {author} {\bibinfo {author} {\bibfnamefont {Y.}~\bibnamefont {Okabe}}\ and\ \bibinfo {author} {\bibfnamefont {M.}~\bibnamefont {Kikuchi}},\ }\bibfield  {title} {\bibinfo {title} {Quantum {Monte} {Carlo} {Simulation} of the {Spin} 1/2 {XXZ} {Model} on the {Square} {Lattice}},\ }\href {https://doi.org/10.1143/JPSJ.57.4351} {\bibfield  {journal} {\bibinfo  {journal} {Journal of the Physical Society of Japan}\ }\textbf {\bibinfo {volume} {57}},\ \bibinfo {pages} {4351} (\bibinfo {year} {1988})}\BibitemShut {NoStop}%
	\bibitem [{\citenamefont {Tallon}\ and\ \citenamefont {Loram}(2001)}]{pseudogap_temperature_doping_dependence}%
	\BibitemOpen
	\bibfield  {author} {\bibinfo {author} {\bibfnamefont {J.}~\bibnamefont {Tallon}}\ and\ \bibinfo {author} {\bibfnamefont {J.}~\bibnamefont {Loram}},\ }\bibfield  {title} {\bibinfo {title} {The doping dependence of t* – what is the real high-tc phase diagram?},\ }\href {https://doi.org/https://doi.org/10.1016/S0921-4534(00)01524-0} {\bibfield  {journal} {\bibinfo  {journal} {Physica C: Superconductivity}\ }\textbf {\bibinfo {volume} {349}},\ \bibinfo {pages} {53} (\bibinfo {year} {2001})}\BibitemShut {NoStop}%
	\bibitem [{\citenamefont {Chalopin}\ \emph {et~al.}(2024)\citenamefont {Chalopin}, \citenamefont {Bojović}, \citenamefont {Wang}, \citenamefont {Franz}, \citenamefont {Sinha}, \citenamefont {Wang}, \citenamefont {Bourgund}, \citenamefont {Obermeyer}, \citenamefont {Grusdt}, \citenamefont {Bohrdt}, \citenamefont {Pollet}, \citenamefont {Wietek}, \citenamefont {Georges}, \citenamefont {Hilker},\ and\ \citenamefont {Bloch}}]{chalopin2024probingmagneticoriginpseudogap}%
	\BibitemOpen
	\bibfield  {author} {\bibinfo {author} {\bibfnamefont {T.}~\bibnamefont {Chalopin}}, \bibinfo {author} {\bibfnamefont {P.}~\bibnamefont {Bojović}}, \bibinfo {author} {\bibfnamefont {S.}~\bibnamefont {Wang}}, \bibinfo {author} {\bibfnamefont {T.}~\bibnamefont {Franz}}, \bibinfo {author} {\bibfnamefont {A.}~\bibnamefont {Sinha}}, \bibinfo {author} {\bibfnamefont {Z.}~\bibnamefont {Wang}}, \bibinfo {author} {\bibfnamefont {D.}~\bibnamefont {Bourgund}}, \bibinfo {author} {\bibfnamefont {J.}~\bibnamefont {Obermeyer}}, \bibinfo {author} {\bibfnamefont {F.}~\bibnamefont {Grusdt}}, \bibinfo {author} {\bibfnamefont {A.}~\bibnamefont {Bohrdt}}, \bibinfo {author} {\bibfnamefont {L.}~\bibnamefont {Pollet}}, \bibinfo {author} {\bibfnamefont {A.}~\bibnamefont {Wietek}}, \bibinfo {author} {\bibfnamefont {A.}~\bibnamefont {Georges}}, \bibinfo {author} {\bibfnamefont {T.}~\bibnamefont {Hilker}},\ and\ \bibinfo {author} {\bibfnamefont {I.}~\bibnamefont {Bloch}},\ }\href {https://arxiv.org/abs/2412.17801} {\bibinfo {title}
		{Probing the magnetic origin of the pseudogap using a fermi-hubbard quantum simulator}} (\bibinfo {year} {2024}),\ \Eprint {https://arxiv.org/abs/2412.17801} {arXiv:2412.17801 [cond-mat.str-el]} \BibitemShut {NoStop}%
	\bibitem [{Note3()}]{Note3}%
	\BibitemOpen
	\bibinfo {note} {This step is not strictly necessary: one may keep the full exchange terms \protect \eqref {eq:exchange_components_full} and use them to write diagrams. When the legs corresponding to holes, originating from a single vertex, are contracted with each other, the result is equivalent to taking the approximate form \protect \eqref {eq:exchange_components_decoupled}. On the other hand, having multiple such vertices and joining their hole legs to each other would give rise to higher-order terms, which we will ignore.}\BibitemShut {Stop}%
	\bibitem [{Note4()}]{Note4}%
	\BibitemOpen
	\bibinfo {note} {The linear fit returns a small, nonzero intercept - and so the dahsed line in Fig.~\ref {fig:sm_afm_doping}~(b) does not go exactly through the origin. This is likely an artifact of our frequency grid discretization.}\BibitemShut {Stop}%
	\bibitem [{Note5()}]{Note5}%
	\BibitemOpen
	\bibinfo {note} {From the definition \protect \eqref {eq:magnon_rpa_pauli_coefficients}, the time-ordered self-energies $\Pi ^0_{\protect \mathbf k}(\omega )$ and $\Pi ^1_{\protect \mathbf k}(\omega )$ are even functions of frequency. However, they are not analytic: the imaginary parts start as $|\omega |$. We can make the functions analytic near the origin by working with the retarded versions, i.e. flipping the sign of the imaginary part at negative frequencies. Then the real parts are even in $\omega $, while the imaginary ones are odd, yielding the expansion from the text. Of course, this mapping does not change the $\omega >0$ values of the self-energy.}\BibitemShut {Stop}%
\end{thebibliography}
\end{document}